\documentclass[sigconf]{acmart}

\newtheorem{definition}{Definition}
\newtheorem{remark}{Remark}
\usepackage{algorithm}
\usepackage{algorithmic}
\usepackage{amsthm,amsmath}
\usepackage{enumitem}
\usepackage{multirow}
\usepackage{multicol}
\usepackage{tcolorbox}
\usepackage{bm}
\usepackage{subfigure}
\usepackage{graphicx}
\usepackage{soul}

\AtBeginDocument{%
  \providecommand\BibTeX{{%
    \normalfont B\kern-0.5em{\scshape i\kern-0.25em b}\kern-0.8em\TeX}}}

\setcopyright{acmcopyright}
\copyrightyear{2018}
\acmYear{2018}
\acmDOI{XXXXXXX.XXXXXXX}

\acmConference[Conference acronym 'XX]{Make sure to enter the correct
  conference title from your rights confirmation emai}{June 03--05,
  2018}{Woodstock, NY}

\acmPrice{15.00}
\acmISBN{978-1-4503-XXXX-X/18/06}

\newcommand{\OURS}{BHE}

\newcommand{\Eupper}{E}
\newcommand{\Elower}{e}

\newcommand{\Rupper}{R}
\newcommand{\Rlower}{r}

\newcommand{\recomender}{h}
\newcommand{\recomendersub}{f}

\begin{document}

\title{Exploring and Exploiting Data Heterogeneity in Recommendation}

\author{Zimu Wang}
\affiliation{%
  \institution{\small{Tsinghua University}
  \country{\small{Beijing, China}}
  }
}
\email{14317593@qq.com}

\author{Jiashuo Liu}
\affiliation{%
  \institution{\small{Tsinghua University}
  \country{\small{Beijing, China}}
  }
}
\email{liujiashuo77@gmail.com}

\author{Hao Zou}
\affiliation{%
  \institution{\small{Tsinghua University}
  \country{\small{Beijing, China}}
  }
}
\email{zouh18@mails.tsinghua.edu.cn}

\author{Xingxuan Zhang}
\affiliation{%
  \institution{\small{Tsinghua University}
  \country{\small{Beijing, China}}
  }
}
\email{xingxuanzhang@hotmail.com}

\author{Yue He}
\affiliation{%
  \institution{\small{Tsinghua University}
  \country{\small{Beijing, China}}
  }
}
\email{heyue18@mails.tsinghua.edu.cn}

\author{Dongxu Liang}
\affiliation{%
  \institution{\small{Kuaishou China}
  \country{\small{Beijing, China}}
  }
}
\email{liangdongxu@kuaishou.com}

\author{Peng Cui$^{\dagger}$}
\affiliation{%
  \institution{\small{Tsinghua University}
  \country{\small{Beijing, China}}
  }
}
\email{cuip@tsinghua.edu.cn}

\begin{abstract}
Massive amounts of data are the foundation of data-driven recommendation models. As an inherent nature of big data, data heterogeneity widely exists in real-world recommendation systems. It reflects the differences in the properties among sub-populations. Ignoring the heterogeneity in recommendation data could limit the performance of recommendation models, hurt the sub-populational robustness, and make the models misled by biases. However, data heterogeneity has not attracted substantial attention in the recommendation community. Therefore, it inspires us to adequately explore and exploit heterogeneity for solving the above problems and assisting data analysis. In this work, we focus on exploring two representative categories of heterogeneity in recommendation data that is the heterogeneity of prediction mechanism and covariate distribution and propose an algorithm that explores the heterogeneity through a bilevel clustering method. Furthermore, the uncovered heterogeneity is exploited for two purposes in recommendation scenarios which are prediction with multiple sub-models and supporting debias. Extensive experiments on real-world data validate the existence of heterogeneity in recommendation data and the effectiveness of exploring and exploiting data heterogeneity in recommendation.
\end{abstract}



\maketitle

\section{Introduction}


 Big data is often created by aggregating many data sources corresponding to different sub-populations, leading to widespread data heterogeneity, since each sub-population may exhibit some unique characteristics that other sub-populations do not possess. Previous literature\cite{kearns2018preventing,wagner1982simpson,fan2014challenges} has pointed out that ignoring the data heterogeneity will significantly hurt the generalization performance, damage the sub-populational robustness, and make the models misled by biases. Exploiting the unique characteristics of each sub-populations can help the machine learning system provide more targeted services and unveil commonality across sub-populations\cite{fan2014challenges, he2017learning}.

Modern data-driven recommendation systems also face significant data heterogeneity. The heterogeneity may come from many sources. For instance, the data is collected from different cities, and the consumption data consists of new and old users. 
Correspondingly, the heterogeneity can be studied by dividing sub-population from many perspectives 
depending on the important features considered (e.g. age, item popularity, and which city the data is from).

However, no matter from which perspective we study the heterogeneity of recommendation data, the heterogeneity can be reflected in the joint distributions $P(Y, U, V)$ among different sub-populations, where $Y$ indicates the scores rated by the users to the items and $U$ and $V$ indicate the profiles of users and items, respectively. Since it is common practice to decompose the joint distribution into prediction mechanism $P(Y | U, V)$ and covariate distribution $P(U, V)$, we, therefore, argue to systematically and intrinsically study the data heterogeneity in recommendation from these two perspectives which reflect the rating pattern and the interaction pattern respectively\cite{wang2022invariant, chen2021autodebias, chen2020bias}.
Different sub-populations with heterogeneity of prediction mechanism may have unique rating patterns. Hence, ignoring the heterogeneity of prediction mechanism in recommendation data could limit the performance of recommenders. 
For instance, the rating pattern of data during a sales promotion could be significantly different from ordinary days. During a sales promotion, users' preference for the expensive item they like could surge, while the preference for the cheap substitution could decrease.
Moreover, the heterogeneity of covariate distribution implies the imbalance among each sub-population. Ignoring heterogeneity could make the recommendation models focus on obvious patterns of major sub-population and neglect the patterns of minor sub-population. This could also lead to poor performance and damage sub-populational robustness. 
As a result, it is necessary to explicitly exploit the prediction mechanism heterogeneity in recommendation appropriately for dealing with the above problems.
To be concrete, the heterogeneity of prediction mechanism can be leveraged to train exclusive sub-models for different sub-populations to resolve the distinction of rating patterns among sub-populations. As for the imbalance problem, the current mainstream methods attempt to debias the data based on the propensity score. However, due to the heterogeneity of both prediction mechanism and covariate distribution, the propensity score also faces heterogeneity. As a result, the propensity score could also be estimated more accurately by exploiting heterogeneity of both prediction mechanism and covariate distribution.

\begin{figure} [t]
 \subfigure[Compactness on Yelp] {
     \includegraphics[width=0.475\linewidth]{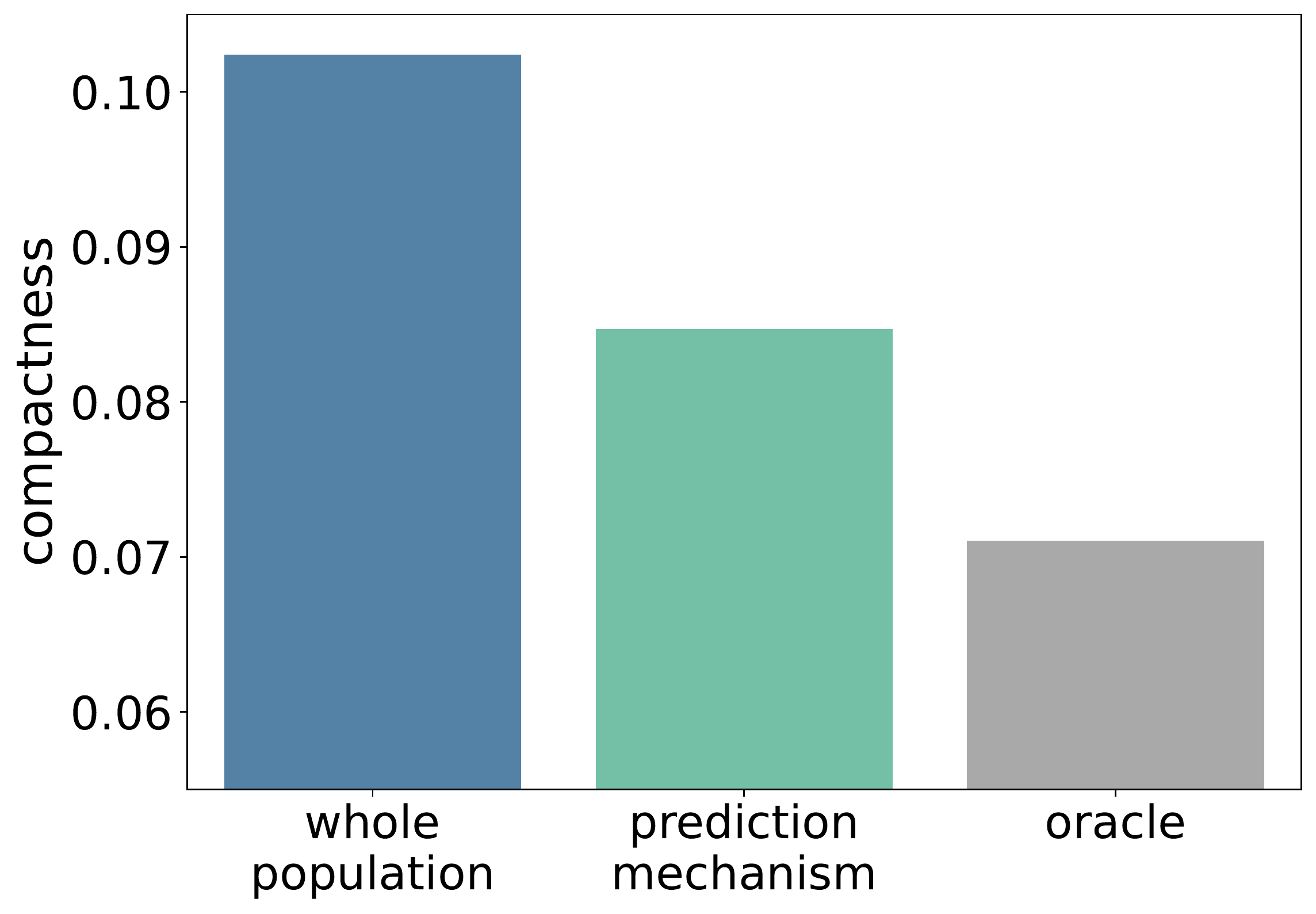}
 }
 \subfigure[Compactness on MovieLens-1M] {
     \includegraphics[width=0.475\linewidth]{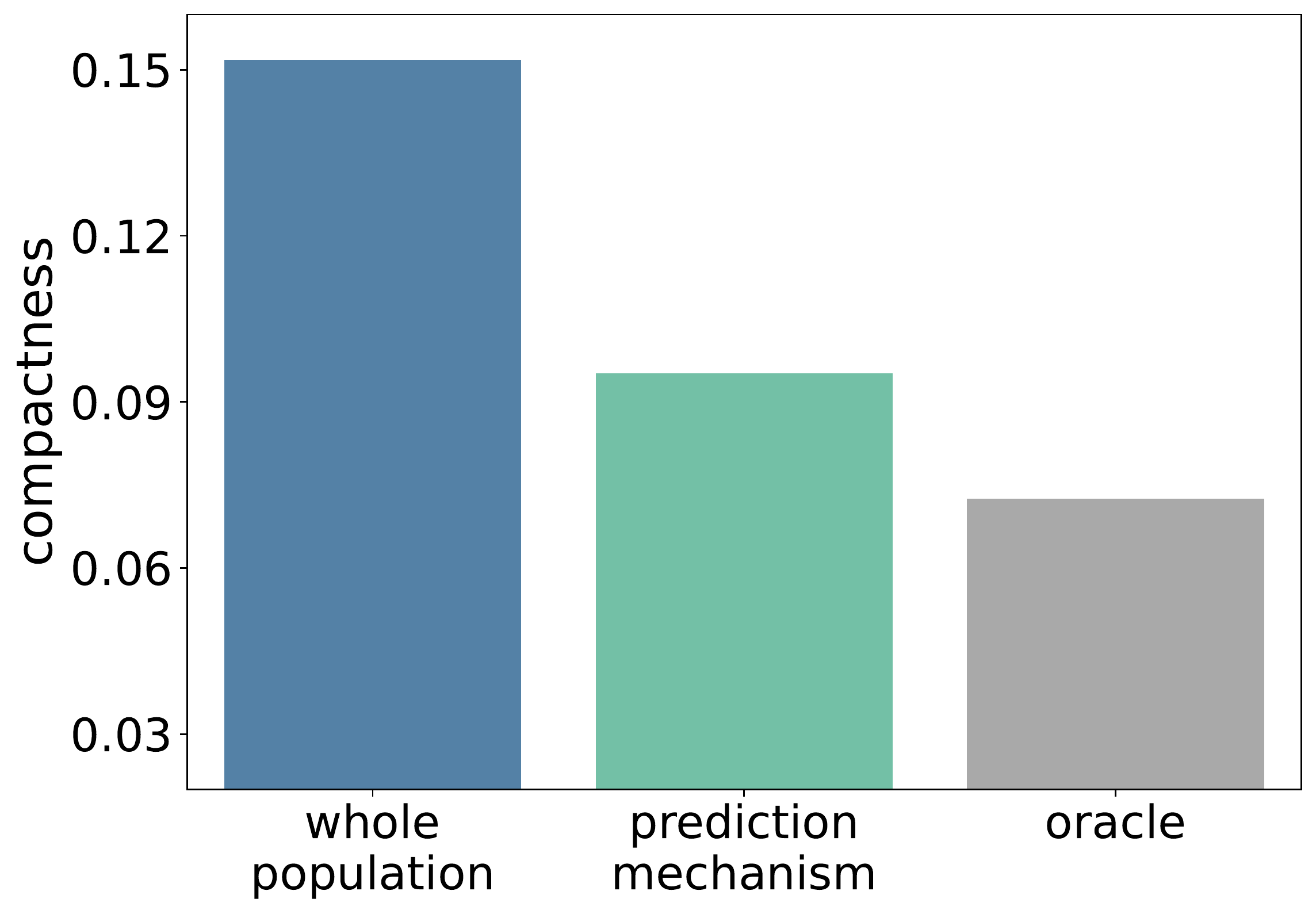}
 }
 \vskip -0.2in
 \caption{The compactness of covariate values is improved after clustering according to the prediction mechanism. A lower compactness value indicates more aggregated sub-populations.}
 \label{E_R_coupling}
 \vskip -0.2in
\end{figure}

However, when we attempt to predict new samples outside the training datasets, there is no ground truth rating $Y$ for them and this absence of rating brings a challenge to the determination of which sub-population the samples are from. 
Fortunately, due to the unique selection mechanism of recommendation systems\cite{schnabel2016recommendations}, the heterogeneity of the prediction mechanism and covariate distribution are aligned, which means there is a coupling between them. This presents an opportunity for us to infer the sub-population based on only covariates.
To empirically verify the assumption, we also conduct 
experiments on real-world datasets. 
As shown in Figure\ref{E_R_coupling}, the sub-populations divided by prediction mechanism enjoy significantly better compactness\footnote{The compactness of (sub-)population is defined as the average euclidean distance between the center of (sub-)population and each point. And the compactness of multiple sub-population is the average value of each sub-population compactness.}  in terms of covariate values (i.e. the metric is smaller which means the covariates are more aggregated in sub-population than the whole population\cite{calinski1974dendrite}). 
So these sub-populations are also likely to be obtained by dividing the population based on covariate distributions instead of prediction mechanism. This phenomenon indicates that the heterogeneity of prediction mechanism and covariate distribution are aligned and coupled to some extent and support our idea.

In this work, we focus on exploring the heterogeneity of both prediction mechanism and covariate distribution in recommendation data.
To achieve our goals, we propose a novel bilevel clustering method named \textbf{B}ilevel \textbf{H}eterogeneity \textbf{E}xploration(\OURS) for hierarchically exploring heterogeneity in recommendation data. 
We further exploit them for the two purposes above which are prediction with multiple sub-models and supporting debias.
The sufficient experimental results demonstrate that the heterogeneity uncovered by our method can contribute to better performance and sub-populational robustness of recommendation model.

In summary, the main contributions are as follows:
\begin{itemize}[leftmargin=0.3cm]
    \item We investigate the novel problem of exploring and exploiting the heterogeneity in recommendation data.
    
    

    \item We propose a novel bilevel clustering method (\OURS) to hierarchically explore data heterogeneity and two approaches to exploit the heterogeneity explored.

    
    \item We conduct extensive experiments on real-world data to validate the importance of heterogeneity and the improvement brought by exploiting heterogeneity.
    
\end{itemize}

\section{Related Works}
\subsection{Heterogeneity in Big Data}
Big data is an important support for machine learning applications. With the development of the industry, the scale of datasets is gradually expanding. With the expansion of data scale, heterogeneity, an inherent property of big data, has attracted attention from the academic community\cite{fan2014challenges, he2017learning}. Heterogeneity refers to any diversity inside data, and academic research focuses mainly on data generation mechanisms, sub-populations, and data sources. Neglecting heterogeneity can lead to high error rates on minor sub-populations, which are unacceptable in high-risk applications(e.g. autonomous driving\cite{breitenstein2020systematization}, medicine \cite{dzobo2018not} , and finance\cite{challen2019artificial}).

There is currently no consensus in the academic community on the definition of data heterogeneity. \cite{li1995definition} defines ecology heterogeneity from the view of the property and complexity of the system. In economics, \cite{rosenbaum2005heterogeneity} claims the unit heterogeneity in observational studies is the uncertainty of the potential outcome.

More recently, a discussion of data heterogeneity has kicked off in the causal inference and robust learning community. Given training data collected from multiple distributions, \cite{arjovsky2019invariant} proposes a learning paradigm to estimate invariant correlations across these distributions. 
\cite{duchi2018learning} proposes to assign a higher weight to the sub-population in which the model performs worst to learn a model providing good performance against perturbations. 
\cite{liu2021kernelized, liu2021heterogeneous} propose to explore data heterogeneity to find out the invariant variables so that they can improve the model generalization. 

\subsection{Bias in Recommendation}
In recent years, debiasing has become a research foci in the community of recommendation\cite{chen2020bias, saito2022unbiased, bonner2018causal, liu2020general}. Studies on bias in recommendation mainly consider the data heterogeneity caused by the recommendation system itself, such as exposure mechanism and sales strategy. The recommendation community currently defines many kinds of biases(e.g. exposure bias\cite{liu2021mitigating, chen2018modeling},  position bias\cite{collins2018study, joachims2007evaluating}, and popularity bias\cite{abdollahpouri2020multi, abdollahpouri2019unfairness, abdollahpouri2020connection}). They essentially emphasize the data imbalance problem in each sub-population. The goal of debiasing is to mitigate the consequences of the model being dominated by major sub-populations. However, these works do not explicitly model and explore the heterogeneity.

The current mainstream debiasing methods are based on Inverse Propensity Score(IPS), of which propensity score is essential to IPS-based methods. The most common approach to estimating propensity scores in the recommendation community is to estimate that based on the observed scores using the Naive Bayes Estimation. A series of methods based on propensity score are proposed: 1) IPS\cite{schnabel2016recommendations} is the most basic IPS-based method. This method uses the inverse of the propensity score as the weight of the observed sample to adjust for the biased distribution of the observed data. This method has a wide influence on the recommendation community because of its simplicity and effectiveness. 2) To solve the problem of large variance of traditional IPS-based methods, self-normalized IPS(SNIPS)\cite{schnabel2016recommendations} which uses the self-normalized inverse propensity score as the weight of each sample is proposed. 


\subsection{Heterogeneity in Recommendation}
Currently, data heterogeneity has not attracted much attention in the recommendation field. \cite{he2022causpref} points out that recommendation data may be collected from several sub-populations with different distributions. 
It mainly considers the robustness of the distribution shift. 
However, it does not explicitly define and exploit heterogeneity. 
\cite{wang2022invariant} achieves 
general debiasing by exploring heterogeneity in the recommendation and learning preference that is invariant across sub-populations. 
It does not explicitly define and model the heterogeneity in recommendation data.
\cite{kim2022fedgpo, maeng2022towards} consider the unfairness caused by the unique mechanism of federated learning and the heterogeneity of client devices. They rely on pre-specified client devices heterogeneity and neglect how the heterogeneity of prediction mechanism and covariate distribution affects recommendation.
\section{PROBLEM STATEMENT AND METHOD}
In this section, we first introduce the definition of data heterogeneity in recommendation scenarios, which is sufficient to cover most situations in practice. We then introduce the proposed method for hierarchically exploring data heterogeneity in recommendation data called \textbf{B}ilevel \textbf{H}eterogeneity \textbf{E}xploration(\OURS). Finally, as mentioned above, we present two approaches for exploiting data heterogeneity. We use capital letters (e.g., $
U$), lowercase letters (e.g., \textbf{$u$}), and calligraphic font letters (e.g., $\mathcal{U}$) to denote a variable, its specific value, and sample space, respectively. Commonly used symbols are shown in Table\ref{table1}. In this paper, we use sub-population and environment(short as env) interchangeably.

\begin{table}           
\caption{Notation}
\label{table1}
\begin{tabular}{cl}
\toprule 
Notation & Annotation\\
\midrule
$U$ & The user profile(e.g. user id, age and occupation).\\
$V$ & The item profile(e.g. item id, category and price).\\
$Y$ & The user feedback on the item in observed data.\\
$O$ & Indicate whether an interaction is observed.\\
$\Eupper$ & Sub-population with respect to $P(Y|U, V)$.\\
$\Rupper$ & Sub-population with respect to $P(U, V)$.\\
\bottomrule
\end{tabular}
\vskip -0.1in
\end{table}

\subsection{Data Heterogeneity in Recommendation}
The joint distribution $P(Y,U,V)$ can be naturally decomposed into $P(Y|U,V)$ and $P(U,V)$.
As $P(Y|U,V)$ and $P(U,V)$ are essential in recommendation, we define the heterogeneity of recommendation on their levels respectively.

\noindent\rule[+0pt]{\linewidth}{0.1em}
\begin{definition}\label{definition1} 
\vspace{-5pt}
The observed training data $D=\{D_{e,r}\}_{e,r\in\mathcal{E}\times \mathcal R}$ where $D_{\Elower,\Rlower} \coloneqq {\{(y^{\Elower,\Rlower}_j, u^{\Elower,\Rlower}_j, v^{\Elower,\Rlower}_j)\}}$ is collected from heterogeneous environments $(\Elower, \Rlower) \in \mathcal{\Eupper} \times \mathcal{\Rupper}$. 
$\Eupper$ and $\Rupper$ reflect the heterogeneity in the recommendation data from different perspectives. 
\begin{itemize}[leftmargin=0.3cm]
    \item The sample $\{(y_j^{\Elower,*}, u_j^{\Elower,*}, v_j^{\Elower,*})\}$ from the environment $\Elower$(no matter which $\Rlower$ it belongs to) has the prediction mechanism $g_{\Elower}(\cdot): \mathcal{U} \times \mathcal{V} \rightarrow \mathcal{Y}$ of environment ${\Elower}$. $\forall {\Elower}_1, {\Elower}_2  \in \mathcal{\Eupper}$, ${\Elower}_1 \neq {\Elower}_2 \Leftrightarrow g_{\Elower_1}(\cdot) \neq g_{\Elower_2}(\cdot)$. As a result, 
    ${\Elower}_1 \neq {\Elower}_2 \Leftrightarrow P_{\Elower_1}(Y|U,V) \neq P_{\Elower_2}(Y|U,V)$.
    
    \item The samples $\{(y_j^{*,\Rlower}, u_j^{*,\Rlower}, v_j^{*,\Rlower})\}$ from the environment $\Rlower$ follows the specific covariate distribution $P_{\Rlower}(U,V)$. $\forall \Rlower_1, \Rlower_2  \in \mathcal{\Rupper}$, ${\Rlower}_1 \neq {\Rlower}_2 \Leftrightarrow P_{\Rlower_1}(U,V) \neq P_{\Rlower_2}(U,V)$.
    
\end{itemize}
\textbf{Each sample has two environment labels corresponding to two different kinds of heterogeneity($\Eupper$ and $\Rupper$).}
\vspace{-2pt}
\end{definition}
\noindent\rule[+5pt]{\linewidth}{0.1em}

The selection mechanism prevalent in recommender systems is that users are more inclined to interact with items they are interested in\cite{schnabel2016recommendations, chen2020bias, wang2020information}. Formally, $O \not \! \perp \!\!\! \perp Y | (U, V)$, where $O$ is the indicator variable, which indicates whether an interaction between $u$ and $v$ is observed. Due to $O$ being often affected by $Y$, the heterogeneity of the prediction mechanism $P(Y|U, V)$ and covariate distribution $P(U, V)$ may be aligned and coupled, that is, $\forall \Elower_1, \Elower_2  \in \mathcal{\Eupper}$, ${\Elower}_1 \neq {\Elower}_2 \Leftrightarrow P_{\Elower_1, \Rlower}(U,V) \neq P_{\Elower_2, \Rlower}(U,V)$. This is consistent with the results in Figure\ref{E_R_coupling}, which makes it difficult to directly explore $\Eupper$ kind heterogeneity and $\Rupper$ kind heterogeneity independently.

We present the following algorithms to explore the heterogeneity in the recommendation data and exploit the heterogeneity to improve recommenders. 




\subsection{Exploring Recommendation Data Heterogeneity}
\begin{figure*} [t]
\centering 
\includegraphics[width=0.9\linewidth]{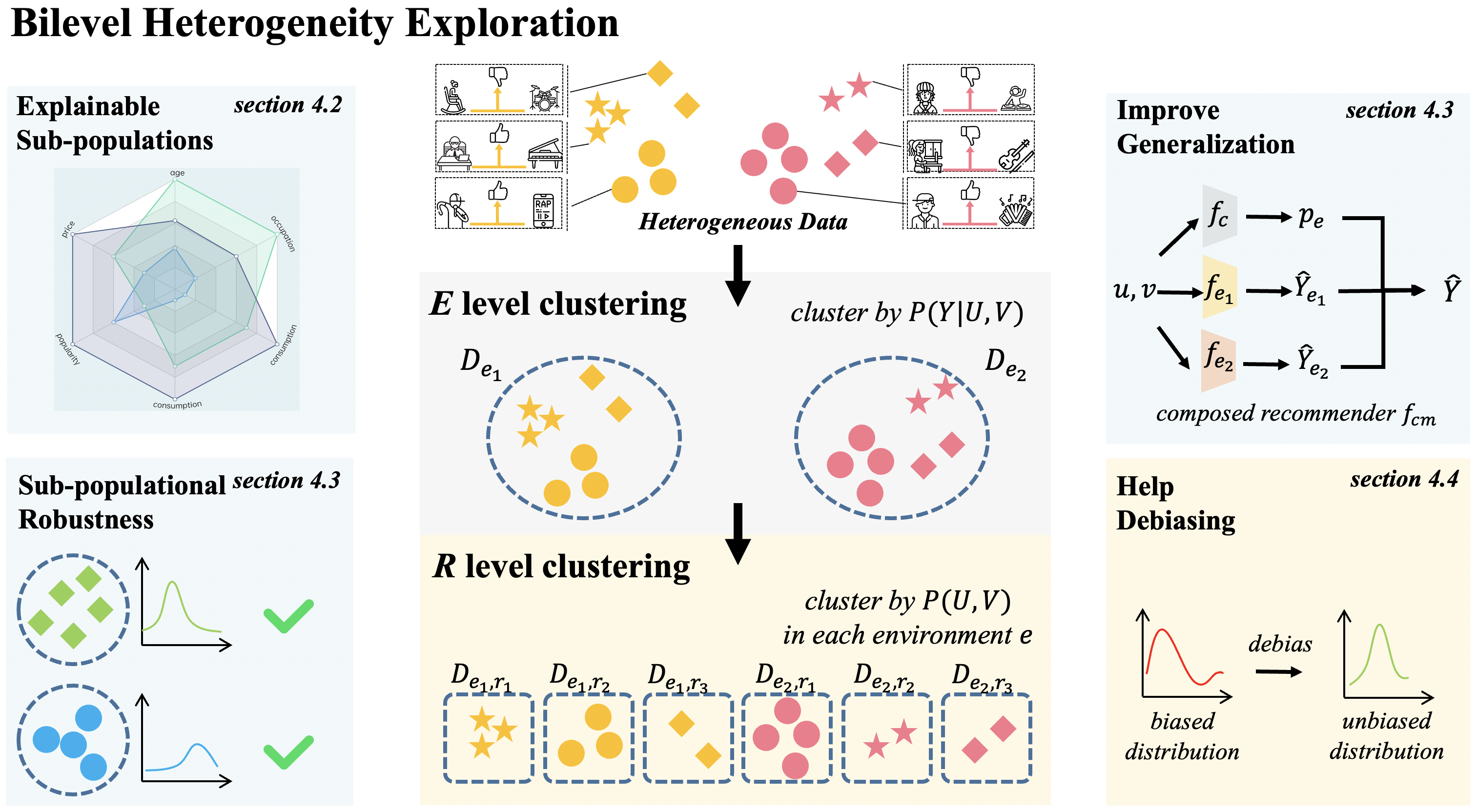}
\centering
\caption {The middle sub-part shows the framework of \OURS~ and how it proceeds. \OURS~ first clusters the observation data with respect to the prediction mechanism. Then based on the results of $\Eupper$ level clustering, \OURS~ explores heterogeneity in covariate distributions among sub-populations. The surrounding four sub-parts indicate the application scenarios and advantages of \OURS~. Clustering can obtain explainable sub-populations(Section \ref{imapct_and_explainable}), enhance the generalization of recommendation models(Section \ref{exp_better_generalization_and_sub_popu}), improve the sub-populational robustness(Section \ref{exp_better_generalization_and_sub_popu}), and assist debiasing tasks(Section \ref{exp_better_debias}).}
\label{fg2}
 \vskip -0.05in
\end{figure*}

To explore recommendation data heterogeneity, we propose a bilevel heterogeneity exploration (\OURS) framework, which consists of two cluster-like levels. The framework of \OURS~ is shown in Figure\ref{fg2}.

\textbf{Prediction Mechanism Heterogeneity\bm{$(\Eupper)$}.}  The major difference between $\Eupper$ and $\Rupper$ is that $\Eupper$ reflects the heterogeneity of prediction mechanism(or conditional distribution $P(Y|U, V)$). To explore the latent $\Eupper$ kind heterogeneity inside data, \OURS~ introduces a cluster-like method to explore the heterogeneity of $P(Y|U, V)$. In detail, given a $(u, v)$ pair, we infer its feedback $\hat{y}_{u,v,\Elower}$ under each environment $\Elower$ and select the environment ${\Elower}^{*}_{u,v}$ corresponding to the result closest to $y_{u,v}$(the true feedback observed). 
Then, we construct the sub-dataset $D_{\Elower}$ with the samples belonging to $\Elower$ for $\forall e \in \mathcal{E}$. 

\textbf{Covariate Distribution Heterogeneity\bm{$(\Rupper)$}.} Given $\Eupper$ kind environment ${\Elower}$, the samples in $D_{\Elower}$ all belong to $\Elower$. Therefore, the heterogeneity over $P_{\Elower}(U, V)$ is only controlled by environment $\Rupper$. To explore  $\Rupper$ kind heterogeneity, \OURS~ performs clustering at the $P_{\Elower}(U,V)$ level. Finally, \OURS~ splits them into multiple $D_{\Elower, \Rlower}$ according to which $\Elower$ and $\Rlower$ the samples belong.

\OURS~ performs the above two stages in sequence to explore heterogeneity in recommendation data hierarchically.

\subsubsection{Explore Prediction Mechanism Heterogeneity}

In order to explore the latent $\Eupper$ kind heterogeneity, \OURS~ generates environment labels based on the heterogeneity of $P(Y|U,V)$.
The goal of environment generation is 
\begin{equation}
    \hat{\Eupper} = \arg \max_{\Eupper} \log P(Y, U, V|\Eupper)
\end{equation}

To solve this problem, we design an EM-liked algorithm. EM algorithm maximizes both $\log P(Y, X|Z)$ and $\log P(Y, X|\theta)$ in iterations, where $Z$ is the latent variable, $X$ is the covariates, and $\theta$ is the model parameters. Corresponding to our problem, the environment $\Eupper$ is the latent variable $Z$ and $(U, V)$ are the covariates $X$. In the M step, \OURS~ uses multiple recommendation models to fit the corresponding environment's training data. In the E step, we adjust the environment labels assigned to each sample according to models trained in the M step. Although we use an EM-liked algorithm to explore heterogeneity, our final goal is not  $\theta$ which is the goal of the EM algorithm but finding the latent environment $\Eupper$. \\
\textbf{(1) M Step:} \quad In the M step, we update the model parameters $\theta$, given training data $D$ and the environment label $e$ of each sample. In our problem, we construct $|\mathcal{\Eupper}|$ mutually independent factorization-based recommendation models(e.g. MF\cite{koren2008factorization} and FM\cite{rendle2010factorization}), which is donated as $\recomender_{\Elower}$ and parameterized by $\theta_{\Elower}$ to fit the conditional distribution $P_{\Elower}(Y|U,V)$ of the corresponding environment $e$ respectively. The $\theta$ of EM algorithm is composed as $\theta := \{\theta_{\Elower}|\Elower \in  \mathcal{\Eupper} \}$. We fit the models to observed interactions across $|\mathcal{\Eupper}|$ environments by optimizing the following loss function. 
\begin{equation}
\label{loss_M_step}
\mathcal{L}_{M} = \sum_{\Elower\in\mathcal{\Eupper}}\frac{1}{|D_{\Elower}|} \sum_{(u,v) \in D_{\Elower}}  \ell_{rec}(\hat{y}_{u,v,\Elower}, y_{u,v}) 
\end{equation}
where $\hat{y}_{u,v,\Elower}$ denotes the feedback predicted by $\recomender_{\Elower}$, $y_{u,v}$ denotes the truth feedback in observed data, and $\ell_{rec}$ is the loss function for recommendation tasks, such as MSE and BCE. By optimizing Eq(\ref{loss_M_step}), \OURS~ updates $\theta$ in the M step.\\
\textbf{(2) E Step:} \quad In the E step, we adjust each sample's assigned environment labels based on the distance between the sample and each cluster center. This distance is calculated as the  sample loss of each recommendation model $\recomender_{\Elower}$:
\begin{equation}
\label{e_label_assign}
\Elower_{u,v} \leftarrow \arg \min_{\Elower\in\mathcal{\Eupper}}\{\ell_{rec}(\hat{y}_{u,v,\Elower}, y_{u,v}) \}
\end{equation}
where, $\Elower_{u,v}$ is the $\Eupper$ environment label assigned to sample $(u, v)$.

In summary, \OURS~ iteratively proceeds M step and E step until $\Eupper$ environments are explored. 
The distance between samples and the environment $\Elower$ is denoted as $G_{\Elower}$.

\subsubsection{Explore Covariate Distribution Heterogeneity}

As described above, the heterogeneity of $P_{\Elower}(u, v)$ is only controlled by environment $\Rupper$, given $\Eupper$ kind environment $\Elower$. Therefore, we can directly explore the heterogeneity of $P_{\Elower}(U, V)$ to generate $\Rupper$ environments. Specifically, \OURS~ uses k-means clustering\cite{hamerly2003learning} to explore $\Rupper$ kind heterogeneity under given environment $\Elower$.:
\begin{equation}
\label{k_means_raw}
\arg \min_{\Rupper} \sum_{\Rlower \in \mathcal{\Rupper}} \frac{1}{|\mathcal{\Rupper}|} \sum_{(u,v)\in D_{\Elower, \Rlower}} || \textbf{x}_{u,v} - \bar{\textbf{x}}_{\Elower, \Rlower}||^2
\end{equation}
where $\textbf{x}_{u, v}$ denotes the raw features vector of user $u$ and item $v$  (including but not limited to user id, gender, item id, and price) and $\bar{\textbf{x}}_{\Elower, \Rlower}$ denotes the mean of $\textbf{x}_{u,v}$ in $D_{\Elower, \Rlower}$.

Since raw features are sometimes scarce(only user/item id is available) and very sparse in recommendation scenarios, clustering directly at the raw feature space may not perform well. Therefore, we turn to the embedding space learned by $\recomender_{\Elower}$. Factorization-based models map raw features to embeddings, specifically, MF maps user/item id to its corresponding embedding, and FM maps each dimension of raw features to its corresponding embedding. A straightforward method is to concatenate the embeddings mapped. We denote the concatenated vector as $\textbf{a}_{u,v}$. Then, we can rewrite the Eq(\ref{k_means_raw}) as follows.
\begin{equation}
\label{k_means_embedding}
\arg \min_{\Rupper} \sum_{\Rlower \in \mathcal{\Rupper}} \frac{1}{|\mathcal{\Rupper}|} \sum_{(u,v)\in D_{\Elower, \Rlower}} || \textbf{a}_{u,v} - \bar{\textbf{a}}_{\Elower, \Rlower}||^2
\end{equation}

Finally, \OURS~ output the $\Rupper$ environment labels and the distances of samples to each $\Rupper$ environment $\Rlower$ ($|| \textbf{a}_{u,v} - \bar{\textbf{a}}_{\Elower, \Rlower}||^2$) denoted as $G_{\Elower, \Rlower}$, given $\Eupper$ environment  $\Elower$.

\begin{remark}
Since \OURS~ performs on the observed data, the heterogeneity of $P(Y|U,V)$ and $P(U,V)$ explored by \OURS~ are essentially those of $P(Y|U,V, O=1)$ and $P(U,V|O=1)$.
\end{remark}

\subsection{Exploiting Heterogeneity}
The heterogeneity of recommendation data has great potential. In this part, we propose two approaches to exploit the heterogeneity of recommendation data.

\subsubsection{Exploit Heterogeneity with Multiple Sub-Models}
\label{exploit_multi}
As mentioned above, there is heterogeneity in the prediction mechanisms in the recommendation data. In  $\Eupper$ environments, there are different prediction mechanisms. Roughly fitting all prediction mechanisms with a single model may limit model performance. Naturally, we propose to train the corresponding model on the data of each environment $\Elower$. However, recommendation data is quite sparse, and dividing it into multiple $D_{\Elower}$ exacerbates the sparsity. Fitting a model on an overly sparse $D_{\Elower}$ can hurt model performance. Therefore, we construct a weighted training set $D_{\Elower}^w := \{(u,v,y_{u,v},w^{tr}_{u,v, \Elower})| (u,v)\in D_{\Elower} \}$ corresponding to the environment $\Elower$ based on $G_{\Elower}$ (the distances from the samples to environment $\Elower$), where $w^{tr}_{u,v, \Elower}$ is the weight of sample in environment $\Elower$. $w^{tr}_{u,v, \Elower}$ is estimated as:
\begin{equation}
\label{sample_weight_e}
w^{tr}_{u,v, \Elower} = \text{Softmax}_{\Elower}([-d_{u,v,\Elower_1},...,-d_{u,v,\Elower_{|\mathcal{\Eupper}|}}])
\end{equation}
where $d_{u,v,\Elower} := \ell_{rec}(\hat{y}_{u,v,\Elower})$, and $\text{Softmax}_{\Elower}(\cdot)$ means the $\Elower$th dimension of Softmax's output. For each $\Elower$ we construct a corresponding recommender $\recomendersub_{\Elower}$, it can be any mainstream recommender. We optimize each $\recomendersub_{\Elower}$ by minimizing object function:
\begin{equation}
\label{loss_sub_model}
\mathcal{L}_{\Elower} = \sum_{(u,v) \in D_{e}^{w}}  \ell_{rec}(\hat{y}_{u,v,\Elower}, y_{u,v}) ~ w^{tr}_{u,v, \Elower}
\end{equation}
where $\hat{y}_{u,v,\Elower}$ is the predicted feedback of $\recomendersub_{\Elower}$.

In the test phase, we cannot directly infer $\Eupper$ because there is no information on $Y$.
Recall that $\Eupper$ can reflect not only the heterogeneity of prediction mechanism but also the heterogeneity of covariate distribution, due to the coupling between them shown in Figure\ref{E_R_coupling}. 
We jointly train a factorization-based recommender $\recomender_{emb}$ and a classifier $\recomendersub_{c}(\cdot):\mathcal{Q}\rightarrow\mathcal{\Eupper}$, where $\mathcal{Q}$ is the embedding space of input raw feature. $\recomendersub_{c}$ use the trained embeddings of $\recomender_{emb}$ as input to predict an $(u,v)$ pair belong to which environment $\Elower$. We jointly optimize both $\recomender_{emb}$ and $\recomendersub_{c}$ by minimizing:
\begin{equation}
\label{loss_classifier}
\mathcal{L}_{emb\_c} = \sum_{(u,v) \in D}  \ell_{rec}(\hat{y}_{u,v}^{emb}, y_{u,v}) + \text{CrossEntropy}(\hat{\Elower}_{u,v},\Elower_{u,v} )
\end{equation}
For a test sample $(u,v)$, we first use each $\recomendersub_{\Elower}$ predict feedback $\hat{y}_{u,v,\Elower}$. Then we use $\recomendersub_{c}$ to calculate the probability $p_{u, v, \Elower}$ that $(u,v)$ belongs to each $\Elower$. Ultimately, the weighted sum of each $\hat{y}_{u,v,\Elower}$ with $p_{u, v, \Elower}$ as the weight is the predicted result $\hat{y}_{u,v}$: 
\begin{equation}
\label{final_pred}
\hat{y}_{u,v} = \sum_{\Elower \in \mathcal{\Eupper}} \hat{y}_{u,v,\Elower}~p_{u, v, \Elower} 
\end{equation}
We combine $\recomendersub_{c}$, $\recomender_{emb}$ and each $\recomendersub_{\Elower}$ to become a composed recommender denoted as $\recomendersub_{cm}$.

\subsubsection{Support Debiasing Task}
\label{method_debias}
Debiasing the data is an important problem in the field of recommendation. The current mainstream debiasing method is based on the inverse propensity score(IPS). The propensity score in the recommendation scenario is estimated as the probability that the user/item pair is observed (i.e. $O_{u,v}=1$) given the user/item pair and score  since the raw features of users and items can hardly be accessible. The mainstream methods calculate the probability conditional on the rating score as the approximation\cite{schnabel2016recommendations}: 
\begin{equation}
\label{traditional_propensity_estimate}
P(O_{u,v}=1|Y_{u,v} = y) = \frac{P(Y_{u,v}=y|O_{u,v}=1)P(O_{u,v}=1)}{P(Y_{u,v}=y)}
\end{equation}
This is equivalent to regarding the variables of user/item as constant across the environments.

Due to the heterogeneity on user/item variable distribution for each environment, the Eq(\ref{traditional_propensity_estimate}) can not be an accurate approximation of propensity score for the samples in different environments. It is remarkable that the propensity score function and its compositions also involve heterogeneity. For example, $P_{\Elower, \Rlower}(Y|O)$ are different among the environments due to the variation of the r.h.s in the following equation.

\begin{equation}
\label{why_our_ips_better}
\begin{aligned} 
& \quad P_{\Elower, \Rlower}(Y=y|O=1) \\
&=\sum_{(u,v)} P_{\Elower, \Rlower}(Y=y|U=u,V=v,O=1) ~ P_{\Elower, \Rlower}(U=u,V=v|O=1)
\end{aligned}
\end{equation}
Therefore, we propose to learn the specific propensity score for each environment $P_{\Elower, \Rlower}(Y=y|O=1)$, which can be a better approximation for the samples in the corresponding environment.

Based on the Bayesian theorem, we have
\begin{equation}
P_{\Elower, \Rlower}(O_{u,v}=1|Y_{u,v} = y) =\frac{P_{\Elower, \Rlower}(Y=y|O=1)P_{\Elower, \Rlower}(O=1)}{P_{\Elower, \Rlower}(Y=y)}.
\end{equation}
Since the term $P_{\Elower, \Rlower}(O=1)$ and  $P_{\Elower, \Rlower}(Y=y)$ is intractable in common practice, we resort to compute the following as substitute
\begin{equation}
\label{our_estimate_propensity_method}
P_{\Elower, \Rlower}(O_{u,v}=1|Y_{u,v} = y) \simeq \frac{P_{\Elower, \Rlower}(Y=y|O=1)P(O=1)}{P(Y=y)}.
\end{equation}
Given the estimated propensity score $p_{u,v}$ of each sample$(u,v)$, we train a debiased recommender according to the objective functions of IPS and SNIPS(Eq(\ref{ips_loss}) and Eq(\ref{snips_loss}), respectively).
\begin{equation}
\label{ips_loss}
\mathcal{L}_{IPS} = \frac{1}{|D|} \sum_{(u,v)\in D} \frac{\ell_{rec}(\hat{y}_{u,v}, y_{u,v})}{p_{u,v}}
\end{equation}
\begin{equation}
\label{snips_loss}
\mathcal{L}_{SNIPS} = \frac{1}{\sum_{(u,v)\in D} \frac{1}{p_{u,v}}} \sum_{(u,v)\in D} \frac{\ell_{rec}(\hat{y}_{u,v}, y_{u,v})}{p_{u,v}}
\end{equation}



\begin{figure*}[t!]
 \subfigure[Showcase on Yelp] {
 \label{cross_case_yelp}
     \includegraphics[width=0.235\linewidth]{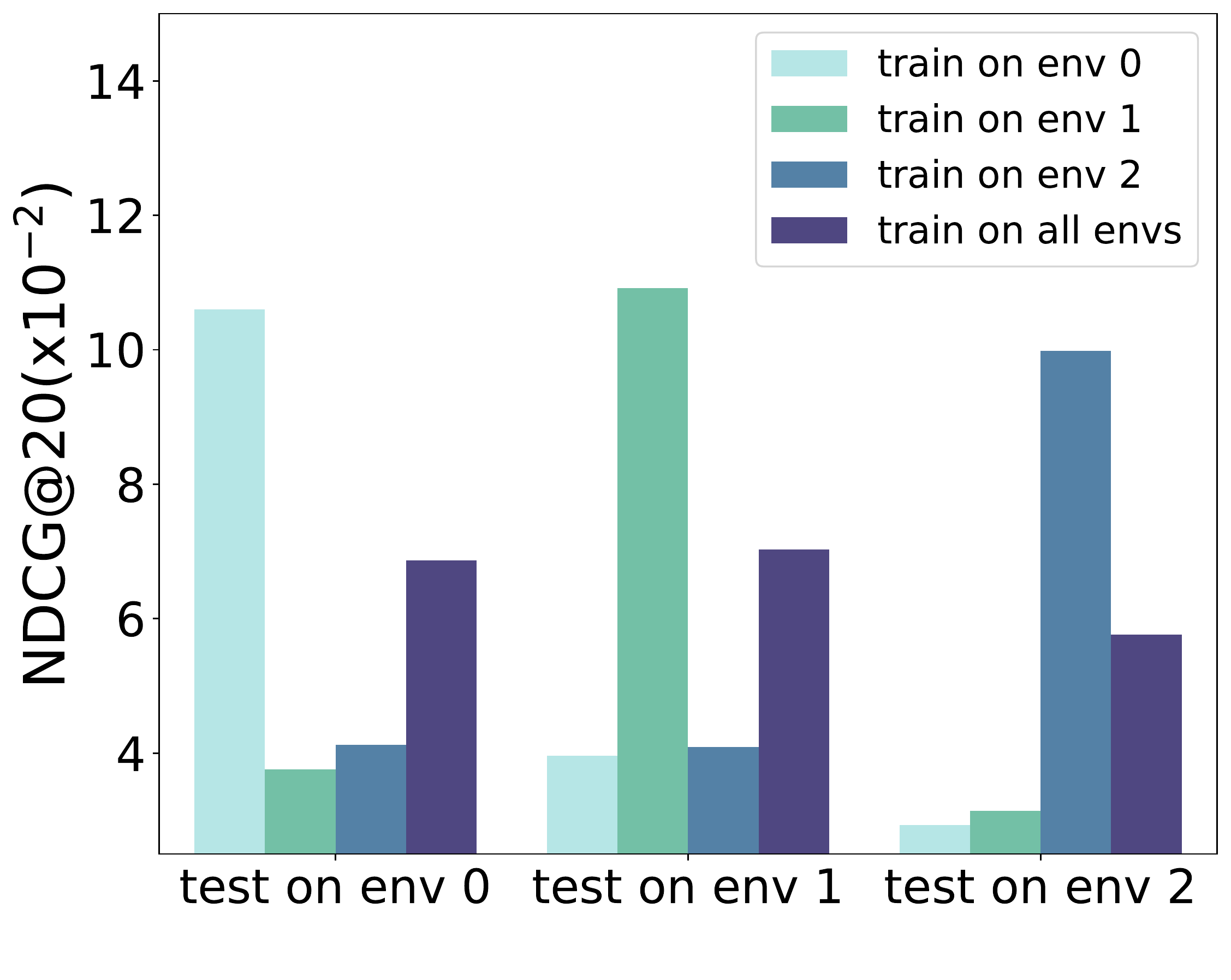}
 }
  \subfigure[Average on Yelp] {
 \label{cross_mean_yelp}
     \includegraphics[width=0.235\linewidth]{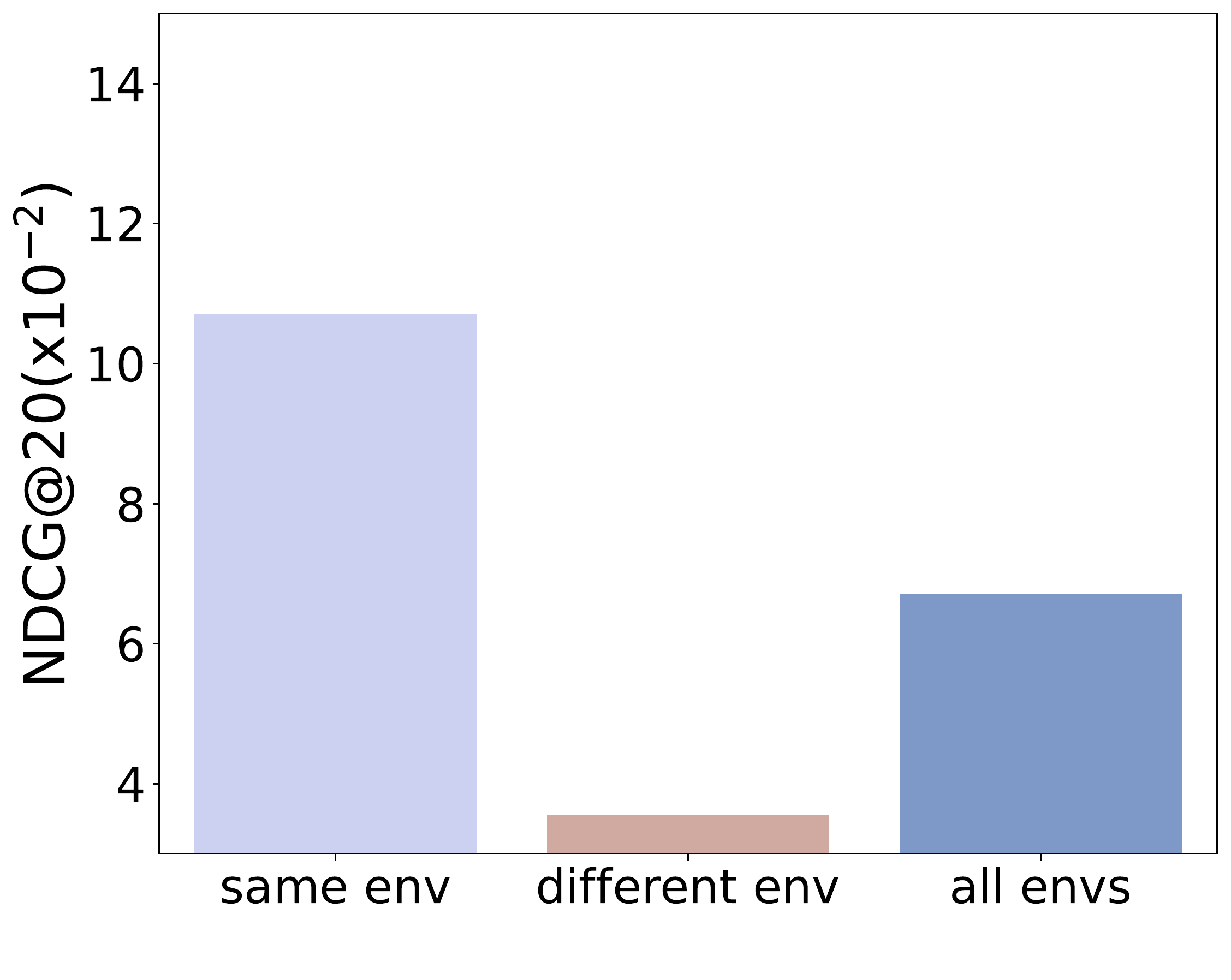}
 }
 \subfigure[Showcase on MovieLens-1M] {
 \label{cross_case_ml}
     \includegraphics[width=0.235\linewidth]{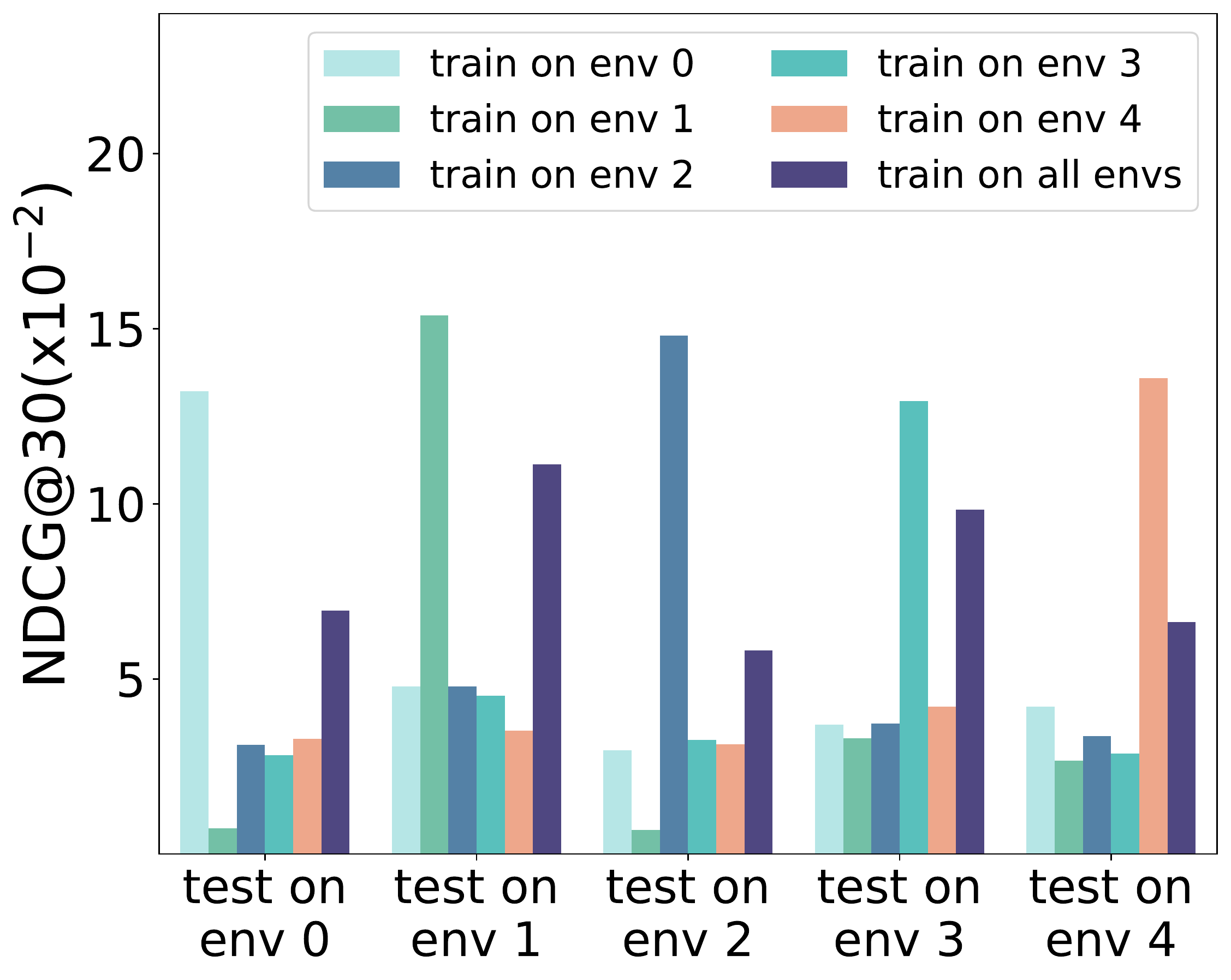}
 }
 \subfigure[Average on MovieLens-1M] {
 \label{cross_mean_ml}
     \includegraphics[width=0.235\linewidth]{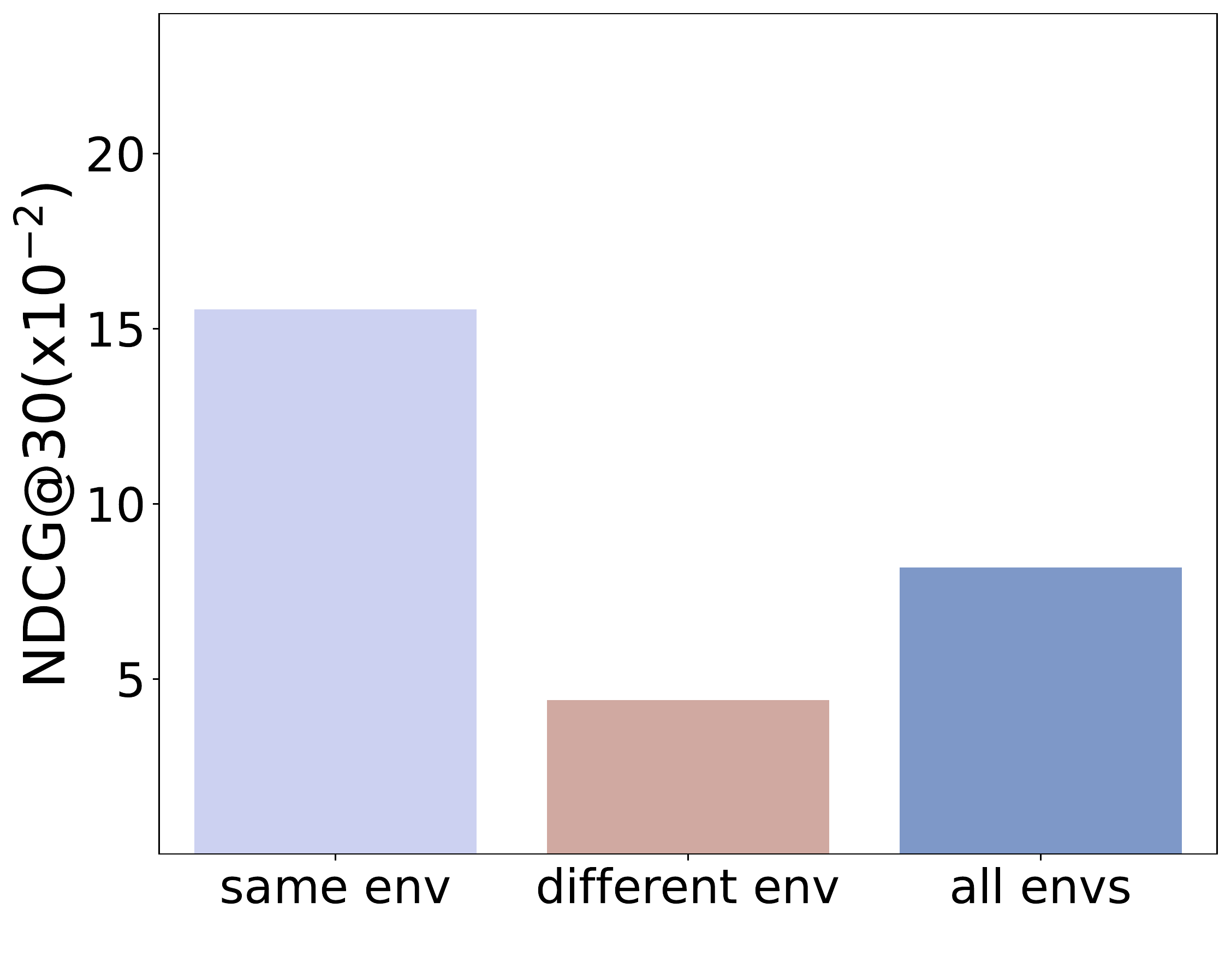}
 }
 \vskip -0.1in
 \caption{The heterogeneity of recommendation data is consistent with cognition and significantly impacts the recommenders. When the model is trained and evaluated in different environments, the performance will drop sharply.   On the other hand, when the model is trained and evaluated in the same environment, it outperforms other training settings. The backbone is FM.}
 \label{cross_fig}
 \vskip -0.1in
\end{figure*}


\section{Experiments}

We conduct extensive experiments on real-world datasets to demonstrate significant heterogeneity in recommendation data and evaluate the proposed methods. Our experiments aim to answer the following questions.
\begin{itemize}[leftmargin=0.3cm]
    \item \textbf{RQ1:} Does the heterogeneity in recommendation fit our cognition and be explainable?
    \item \textbf{RQ2:} Does exploiting heterogeneity explored by \OURS~ promote better generalization and sub-populational robustness?
    \item \textbf{RQ3:} Does exploiting heterogeneity improve debiasing?
    \item \textbf{RQ4:} How do the environment numbers($|\mathcal{E}|$ and $|\mathcal{R}|$) influence \OURS?
\end{itemize}

\subsection{Experimental Setting}
In this part, we detail the datasets and backbones we used. The details of the metrics we used are in Appendix~\ref{app_metrics}.
\subsubsection{Dataset} We conduct experiments on the following datasets.
\quad\\
\textbf{Yelp}\footnote{\url{https://www.yelp.com/dataset}} \& \textbf{MovieLens-1M}\footnote{\url{https://grouplens.org/datasets/movielens/1m/}}. Yelp consists of user reviews of brick-and-mortar restaurants, and MovieLens-1M consists of user ratings of movies. Both datasets contain sufficient profiles(including but not limited to the user's age, number of users' fans, and category of item). Most of these raw features are discrete categorical features(e.g., gender),  and we discretize continuous features(e.g., number of fans) into categorical features. The interaction between the user and the item is that the user rates the item (rate 1-5). We treat interactions with scores $\geq$ 4 as positive samples  and the rest as negative samples. We random sample 20\% of all data as test data, and  the rest as training data. \\
\textbf{Yahoo}\footnote{\url{https://webscope.sandbox.yahoo.com/catalog.php?datatype=r\&did=3}} \& \textbf{Coat}\footnote{\url{https://www.cs.cornell.edu/~schnabts/mnar/}}. Both datasets consist of a biased dataset of normal user interactions, and an unbiased uniform dataset collected by a random logging strategy. The interaction between the user and the item is that the user rates the item (rate 1-5). We randomly sample 5\% of the uniform dataset to estimate the propensity score and the remaining 95\% as test data. We treat interactions with scores $\geq$ 3 as positive samples  and the rest as negative samples. 

\subsubsection{Backbones} In the setting of experiments conducted on the datasets with sufficient raw features(Yelp and MovieLens-1M), we use FM\cite{rendle2010factorization} and NFM\cite{he2017neural} as backbones, since they are the representatives of linear and nonlinear recommenders using raw features, respectively. In the debiasing setting, we follow the settings of related studies\cite{wang2020information, schnabel2016recommendations}. Therefore, we do not use raw features and use MF\cite{koren2008factorization} and NCF\cite{he2017NCF} as backbones.

\begin{figure}[t]
\includegraphics[width=1.0\linewidth]{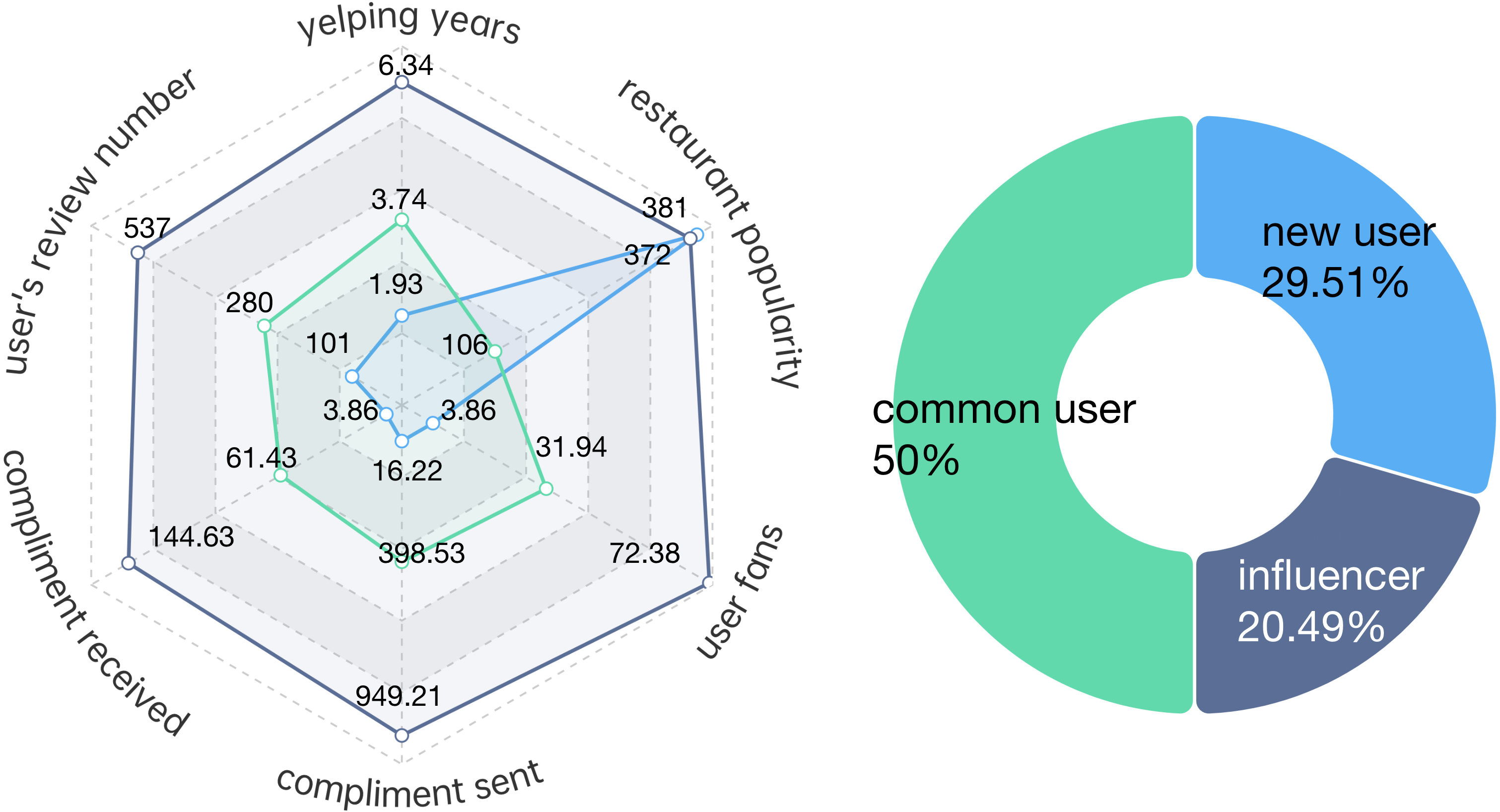}
\centering 
\caption {\OURS~ explores three typical sub-populations from the reviews data of Yelp. Each sub-population represents the pattern of a kind of users(new user, common user, and influencer). The left radar chart shows their attribute characteristics, and the right ring chart shows their proportions.}
\label{view_feature}
 \vskip -0.15in
\end{figure}

\begin{table*}[t]

\centering
\caption{Overall performance of exploiting heterogeneity for better generalization. \OURS~ achieves the best performance with a remarkable improvement, which demonstrates the superiority of exploiting the heterogeneity explored by \OURS.}
\begin{tabular}{c|c|ccc|ccc|ccc|ccc}
    \midrule
    \multicolumn{2}{c|}{Dataset} & \multicolumn{6}{c|}{Yelp}                     & \multicolumn{6}{c}{MovieLens-1M} \\
    \midrule
    \multicolumn{2}{c|}{Metric} & \multicolumn{3}{c|}{$NDCG(\times 10^{-2})$} & \multicolumn{3}{c|}{$Recall(\times 10^{-2})$} & \multicolumn{3}{c|}{$NDCG(\times 10^{-2})$} & \multicolumn{3}{c}{$Recall(\times 10^{-2})$} \\
    \midrule
    Backbone & Env & top-10 & top-20 & top-30 & top-10 & top-20 & top-30 & top-20 & top-30 & top-40 & top-20 & top-30 & top-40 \\
    \midrule
    \multirow{6}[8]{*}{FM} & None  & 4.99  & 6.52  & 7.74  & 5.53  & 9.62  & 13.12  & 7.86  & 8.29  & 8.83  & 7.12  & 9.34  & 11.25  \\
\cmidrule{2-14}          & cluster-user & 5.64  & 7.20  & 8.44  & 6.09  & 10.33  & 13.85  & 9.56  & 9.94  & 10.45  & 8.19  & 10.67  & 12.75  \\
          & cluster-item & 5.57  & 7.08  & 8.31  & 6.07  & 10.18  & 13.64  & 10.97  & 11.33  & 11.80  & 9.08  & 11.80  & 13.88  \\
\cmidrule{2-14}          & raw feature & 5.37  & 6.87  & 8.09  & 5.87  & 9.91  & 13.37  & 7.67  & 8.10  & 8.63  & 6.57  & 8.68  & 10.53  \\
          & embedding & 4.97  & 6.42  & 7.60  & 5.50  & 9.39  & 12.72  & 7.40  & 7.91  & 8.45  & 6.76  & 8.98  & 10.87  \\
\cmidrule{2-14}          & \OURS & \textbf{6.74 } & \textbf{8.24 } & \textbf{9.30 } & \textbf{7.32 } & \textbf{11.64 } & \textbf{14.72 } & \textbf{12.14 } & \textbf{12.37 } & \textbf{12.79 } & \textbf{10.84 } & \textbf{13.47 } & \textbf{15.46 } \\
    \midrule
    \multirow{6}[8]{*}{NFM} & None  & 9.95  & 14.01  & 17.03  & 12.07  & 22.48  & 30.90  & 11.11  & 11.30  & 11.73  & 8.82  & 11.24  & 13.26  \\
\cmidrule{2-14}          & cluster-user & 14.44  & 18.36  & 21.26  & 16.63  & 26.99  & 35.05  & 12.03  & 12.34  & 12.88  & 9.67  & 12.53  & 14.85  \\
          & cluster-item & 13.34  & 16.84  & 19.44  & 15.08  & 24.52  & 31.87  & 13.39  & 13.56  & 13.97  & 10.63  & 13.36  & 15.56  \\
\cmidrule{2-14}          & raw feature & 10.60  & 14.47  & 17.43  & 12.46  & 22.59  & 30.93  & 11.25  & 11.55  & 12.07  & 9.78  & 12.50  & 14.71  \\
          & embedding & 10.33  & 13.93  & 16.52  & 11.78  & 21.12  & 28.34  & 10.82  & 11.18  & 11.72  & 8.89  & 11.46  & 13.65  \\
\cmidrule{2-14}          & \OURS & \textbf{18.07 } & \textbf{22.57 } & \textbf{25.76 } & \textbf{20.10 } & \textbf{32.23 } & \textbf{41.22 } & \textbf{14.61 } & \textbf{14.86 } & \textbf{15.41 } & \textbf{11.69 } & \textbf{14.91 } & \textbf{17.53 } \\
    \bottomrule
    \end{tabular}%
    \label{labeling_func_table}
  \vskip -0.05in
\end{table*}

\subsection{RQ1: Heterogeneity Explored by \OURS~ is Practical and Explainable}
\label{imapct_and_explainable}

In this subsection, we first verify that the prediction mechanism heterogeneity $\Eupper$ explored by \OURS~ reflects the prediction mechanism heterogeneity in recommendation data which has a significant impact on the recommendation models. Secondly, we verify the assumption that the heterogeneity of prediction mechanism and covariate distribution are aligned. Thirdly, we analyze the explainable sub-populations explored by \OURS. The datasets we used in this subsection are Yelp and MovieLens-1M.

\textbf{Significant Impact on Models.} 
Without loss of generality, we categorized Yelp and MovieLens-1M into three and five environments respectively based on the $\Eupper$ explored by \OURS. We train FM models separately in each environment and evaluate them in each environment. Also, we train an FM model in all environments and evaluate it in each environment as a benchmark. More details about the above methods are shown in Appendix~\ref{app_baselines}.

The results are shown in Figure\ref{cross_fig}.
Figure\ref{cross_case_yelp} and Figure\ref{cross_case_ml} are two samples from multiple runs on Yelp and MovieLens-1M respectively, and Figure\ref{cross_mean_yelp} and Figure\ref{cross_mean_ml} are the average results on Yelp and MovieLens-1M respectively. From the results, we can conclude the following points: 
\begin{itemize}[leftmargin=0.3cm]
    \item When the model is trained and evaluated in different environments, the performance will drop sharply or even collapse. 
    \item However, when the model is trained and evaluated in the same environment, it outperforms other training settings.
    \item Results show that the prediction mechanism among the environments explored by \OURS~ could be significantly different and has a huge impact on recommenders. 
    \item Mishandling the data heterogeneity could limit the performance of recommenders in practice, while properly exploiting the unique statistical characteristics and patterns in different environments could improve the generalization of recommenders.
\end{itemize}
We could also get consistent conclusions when using NFM as the backbone, see Appendix~\ref{app_cross_nfm} for details.

\textbf{Prediction Mechanism and Covariate Distribution are Aligned.} 
We use compactness as an indicator (details are shown in Appendix~\ref{app_metrics}) to measure how aggregated covariates are in sub-populations. 
A lower compactness value indicates more aggregated sub-populations.
We use the cluster results of k-means which divide the population by covariate distribution, as the oracle benchmark.
The results are shown in Figure\ref{E_R_coupling}. 

Compared to the whole population, the compactness of sub-populations divided by prediction mechanism is significantly improved.
It shows that these sub-populations also help unveil the heterogeneity of covariate distribution.
Therefore, the heterogeneity of prediction mechanism and covariate distribution are aligned and coupled to some extent.

\textbf{Explainability.}
We analyze the explainability of sub-populations explored on Yelp, first. 
From the three sub-populations explored by \OURS~ we unveil that each of them represents the behaviors and consumption habits of different types of users: new users, common users, and influencers. From the results shown in Figure\ref{view_feature}, we could observe that:



\begin{itemize}[leftmargin=0.3cm]
    \item \textbf{new users} are unfamiliar with the platform and do not know which restaurants are suitable for them, so they prefer to choose the very popular restaurants. In addition, they are not integrated into the social circle of the platform so there is a lack of communication with other users.
    \item The \textbf{common user}'s reviews account for the majority. These users have been on the platform for a long time and have formed their unique tastes instead of blindly following the popularity. These users have their social circles and interact significantly more than newcomers.
    \item In the internet era, some senior users become \textbf{influencers} on the platform. They prefer well-known restaurants and provide many reviews which be widely complimented. They are in the foci of social networks with a great number of fans. 
\end{itemize}

The behavior patterns and proportions of different categories of people are crucial to the platform’s strategic decision-making. The results also show that the heterogeneity of the prediction mechanism and the covariate distribution is aligned to some extent. \OURS~ also unveils explainable sub-populations on MovieLens-1M, see Appendix~\ref{app_explainable} for details.

\begin{figure*}[t]
 \subfigure[FM-BHE on MovieLens-1M(item popularity)] {
 \label{minor_group_ml_item}
     \includegraphics[width=0.23\linewidth]{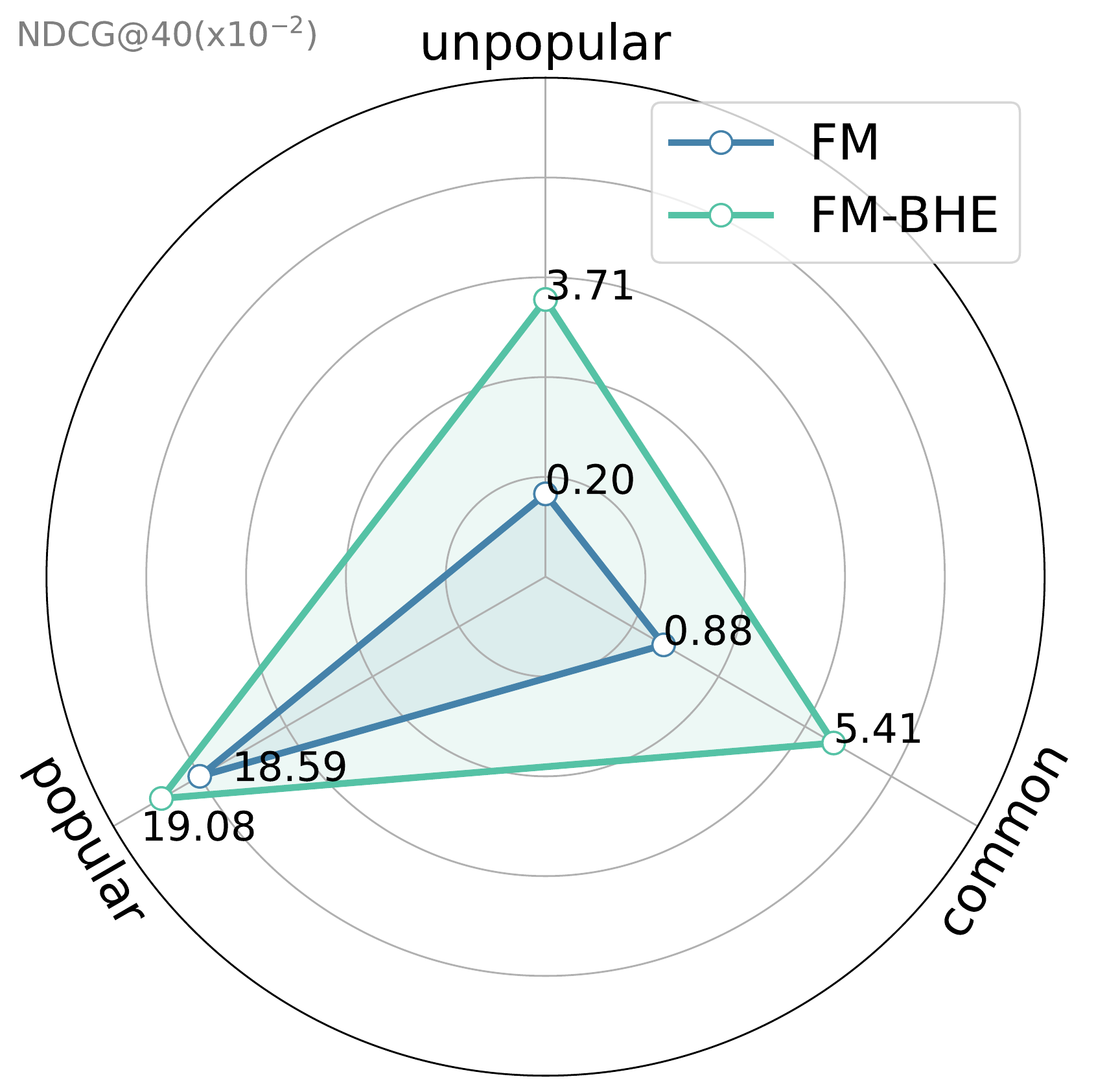}
 }
 \subfigure[FM-BHE on MovieLens-1M(user age)] {
 \label{minor_group_ml_user}
     \includegraphics[width=0.23\linewidth]{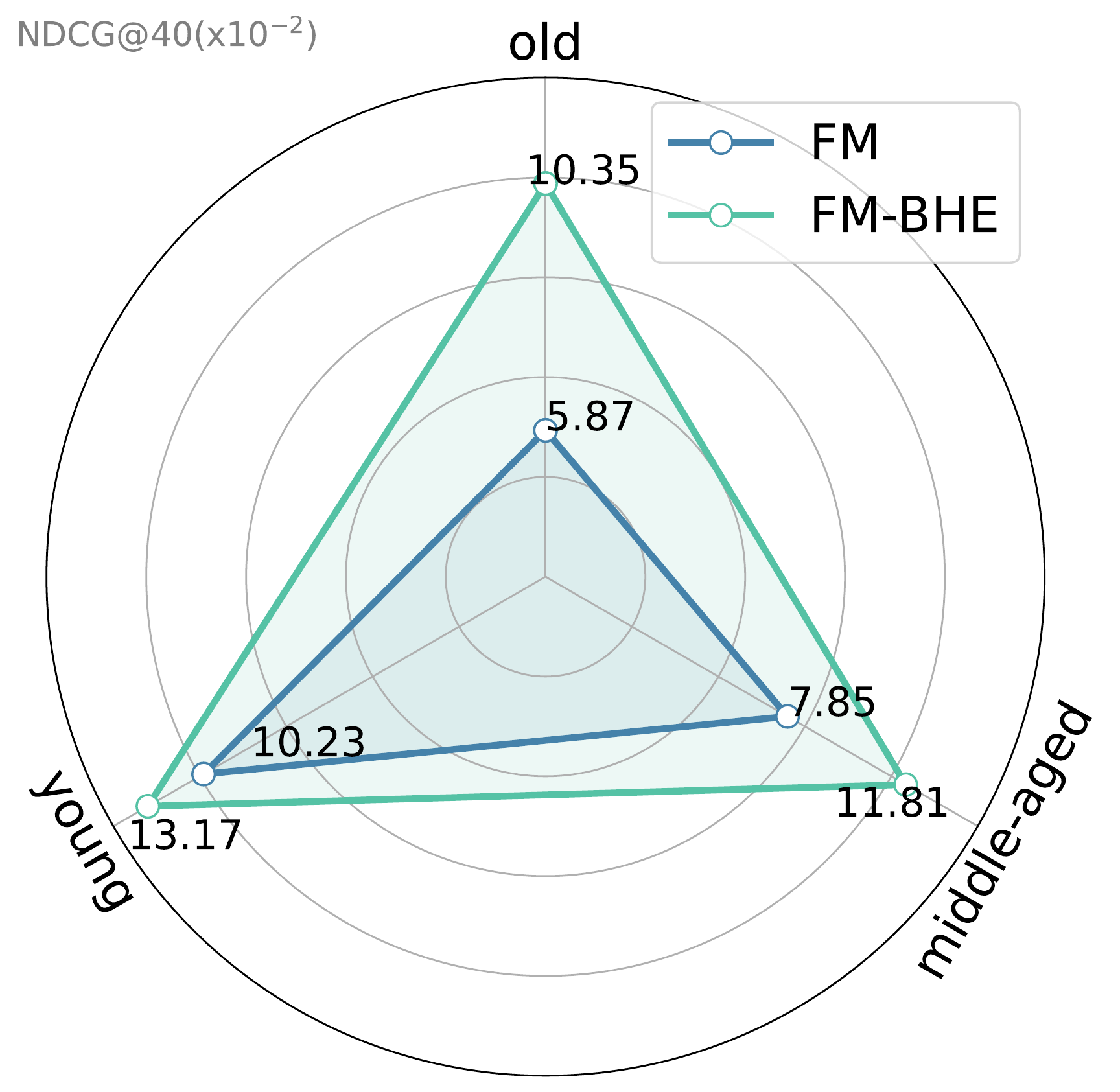}
 }
  \subfigure[FM-BHE on Yelp(item popularity)] {
 \label{minor_group_yelp_item}
     \includegraphics[width=0.23\linewidth]{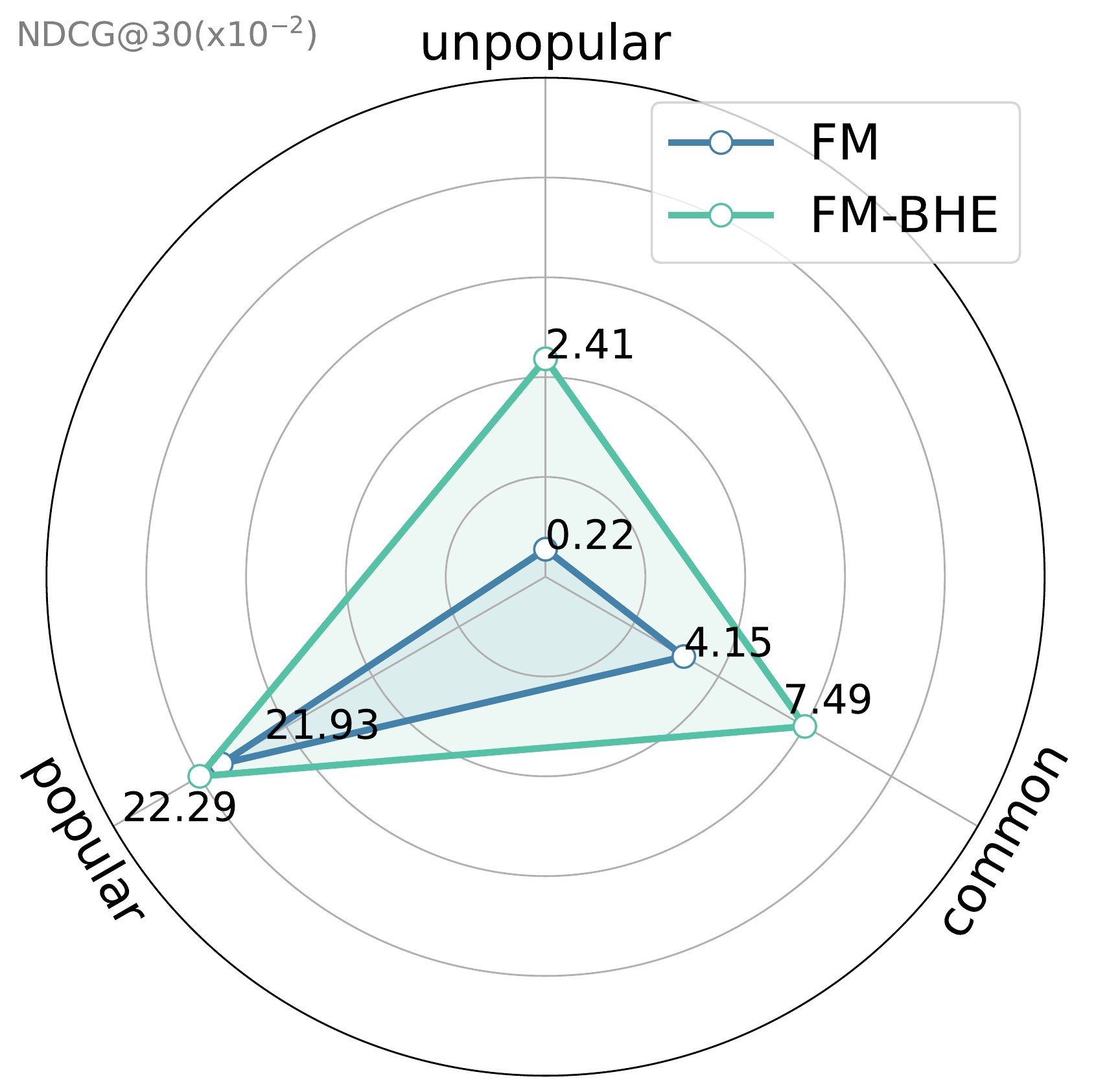}
 }
  \subfigure[FM-BHE on Yelp(user fans number)] {
 \label{minor_group_yelp_user}
     \includegraphics[width=0.23\linewidth]{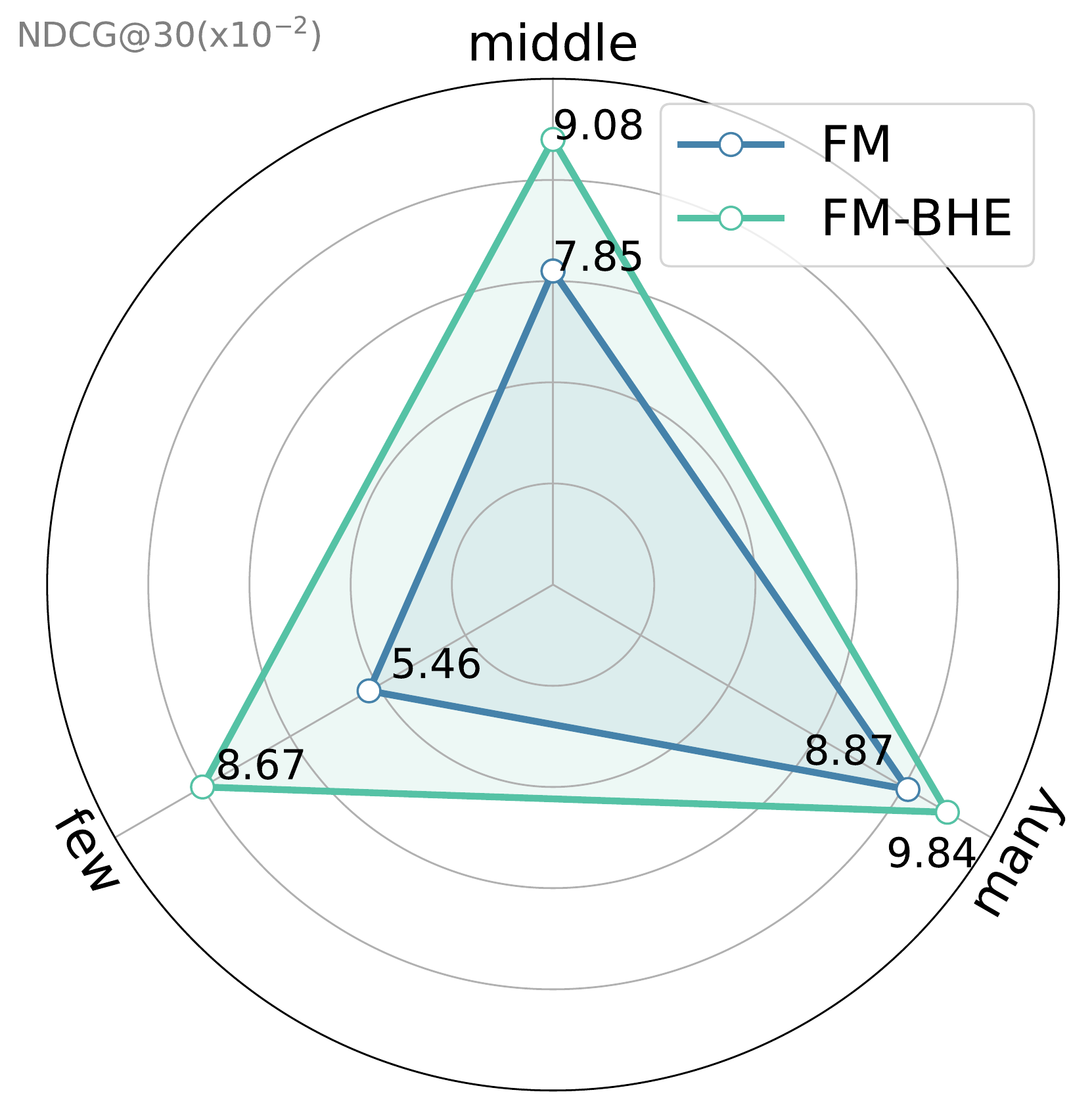}
 }
 
 \vskip -0.2in
 \caption{Exploiting heterogeneity explored by BHE makes the recommender better serve each sub-population. We divide the evaluation results into several sub-populations according to the attributes of item and user(e.g., popularity, age, and fans number). Compared with FM, \OURS~ has improved in each sub-population, and the improvement in minor sub-populations (e.g., unpopular, old, and few fans number) is more significant.}
 \label{minor_group}
 
 \vskip -0.2in
\end{figure*}

\begin{figure} [ht!]
 \subfigure[MF as backbone on Yahoo] {
     \includegraphics[width=0.475\linewidth]{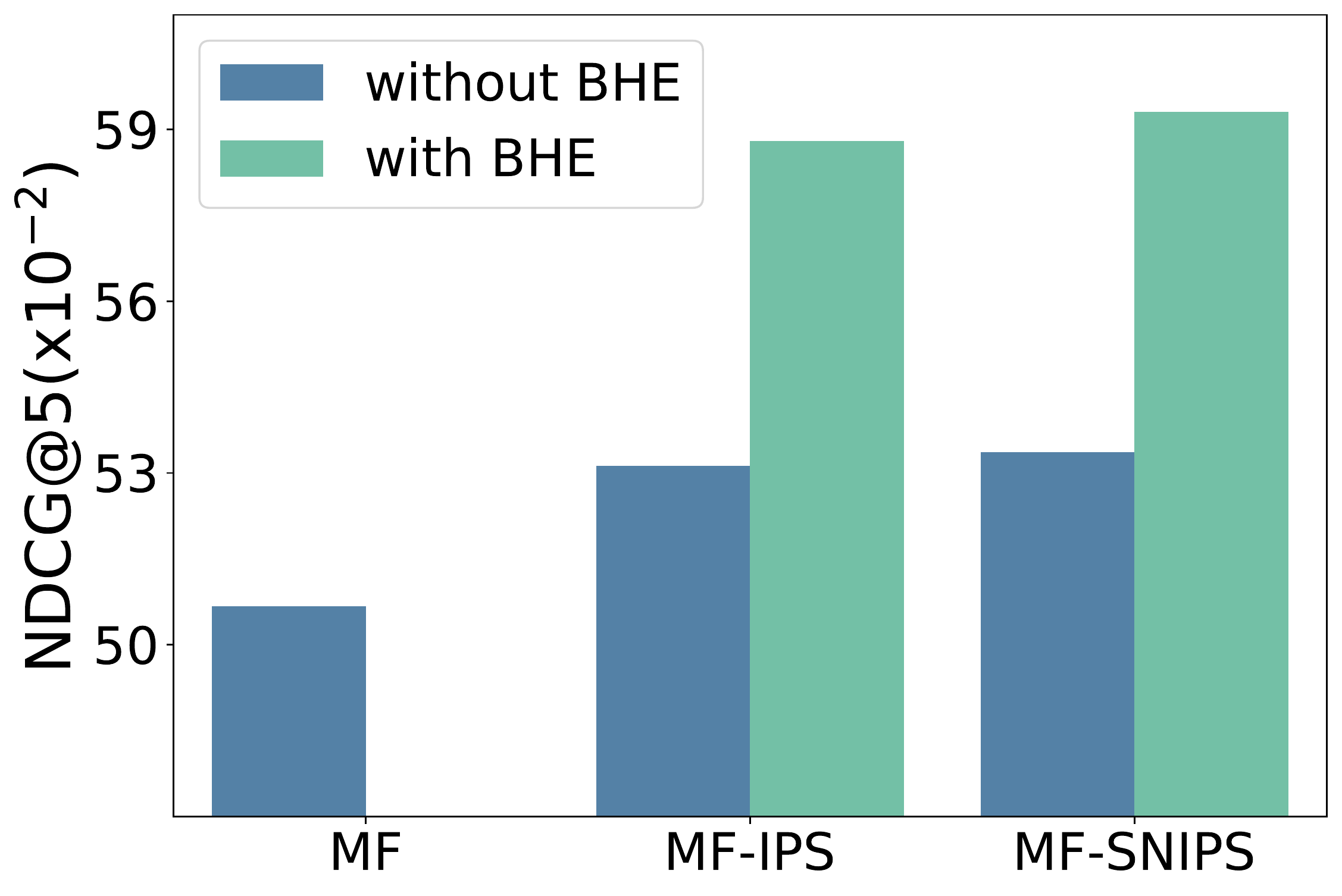}
 }
 \subfigure[NCF as backbone on Yahoo] {
     \includegraphics[width=0.475\linewidth]{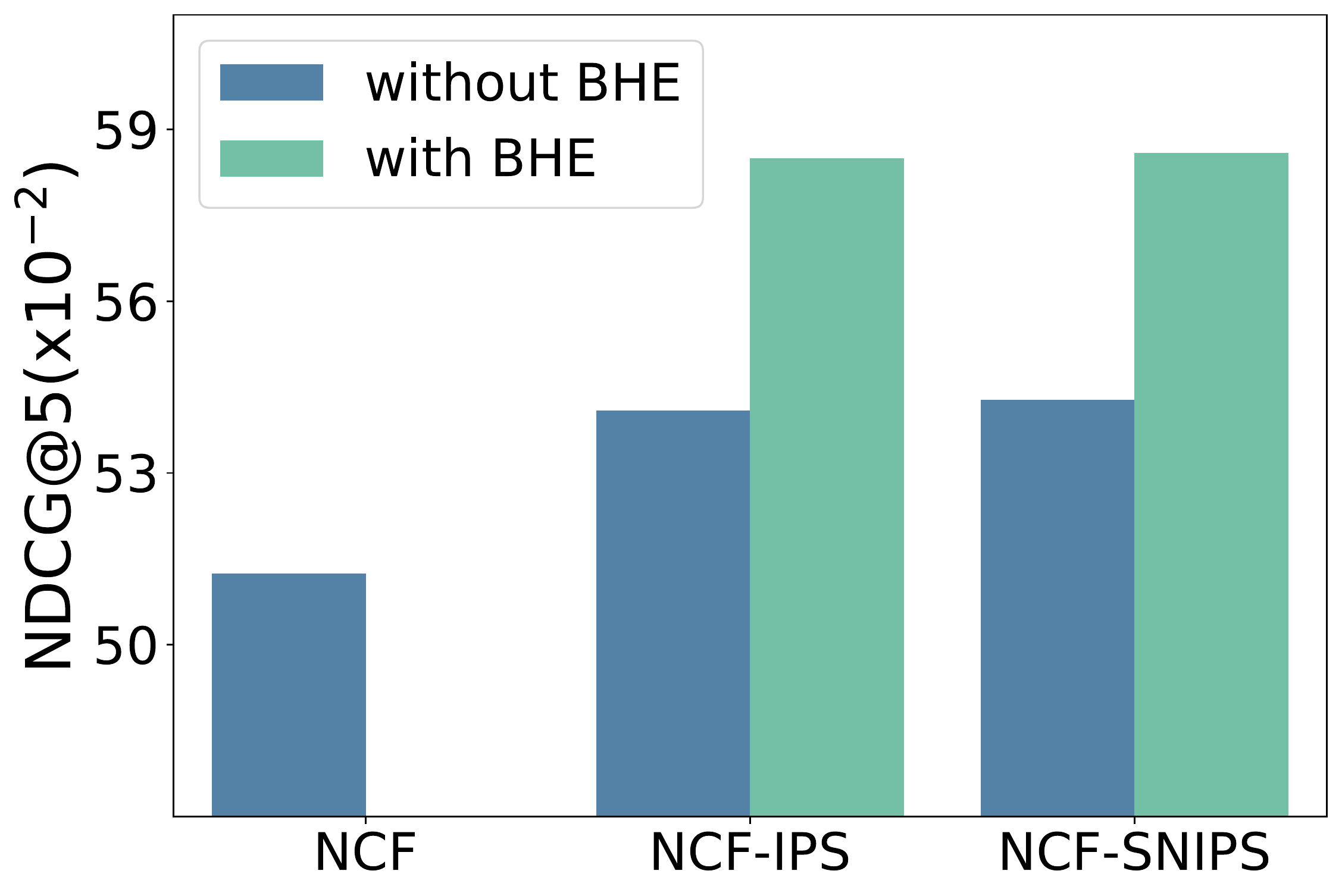}
 }
  \subfigure[MF as backbone on Coat] {
     \includegraphics[width=0.475\linewidth]{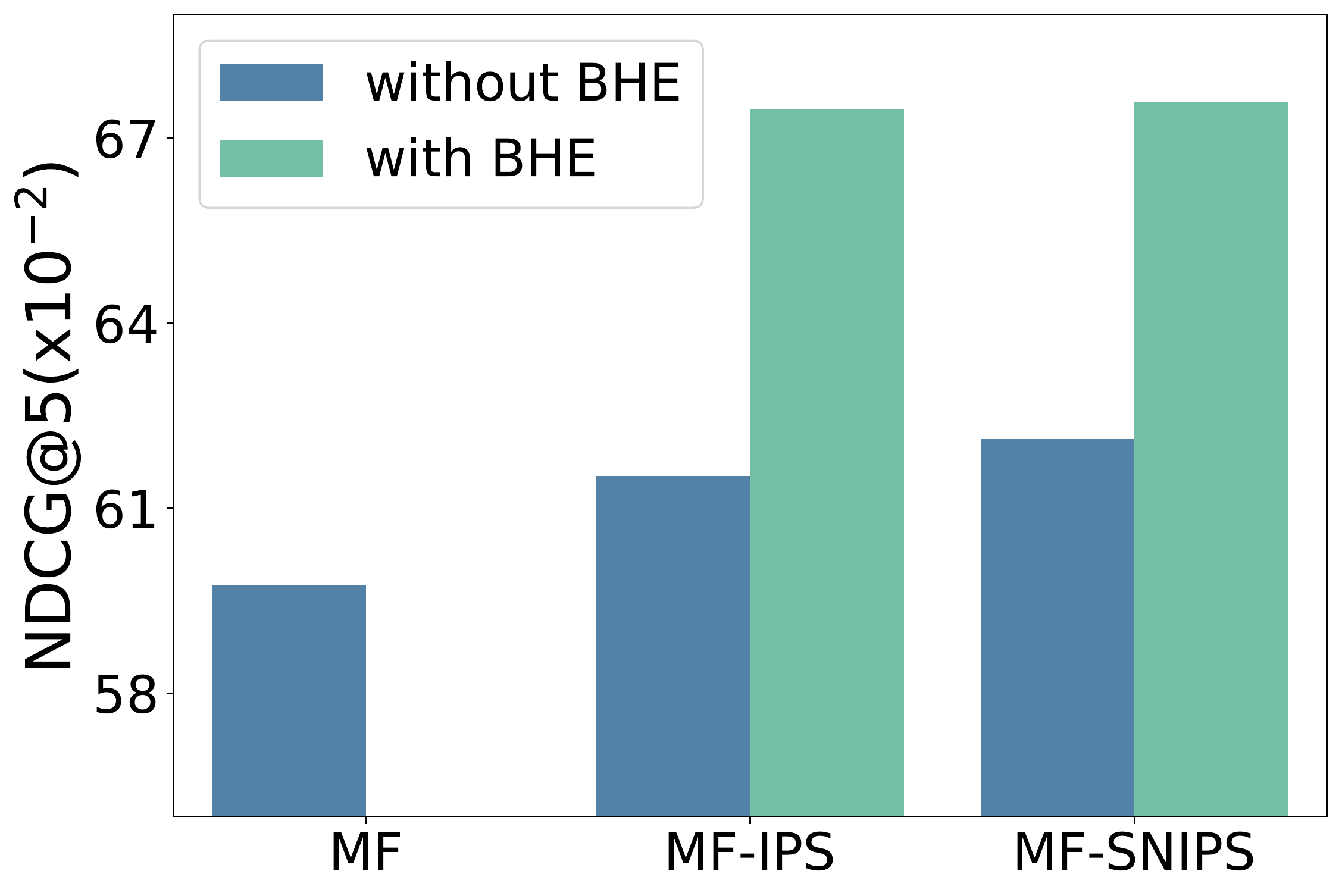}
 }
 \subfigure[NCF as backbone on Coat] {
     \includegraphics[width=0.475\linewidth]{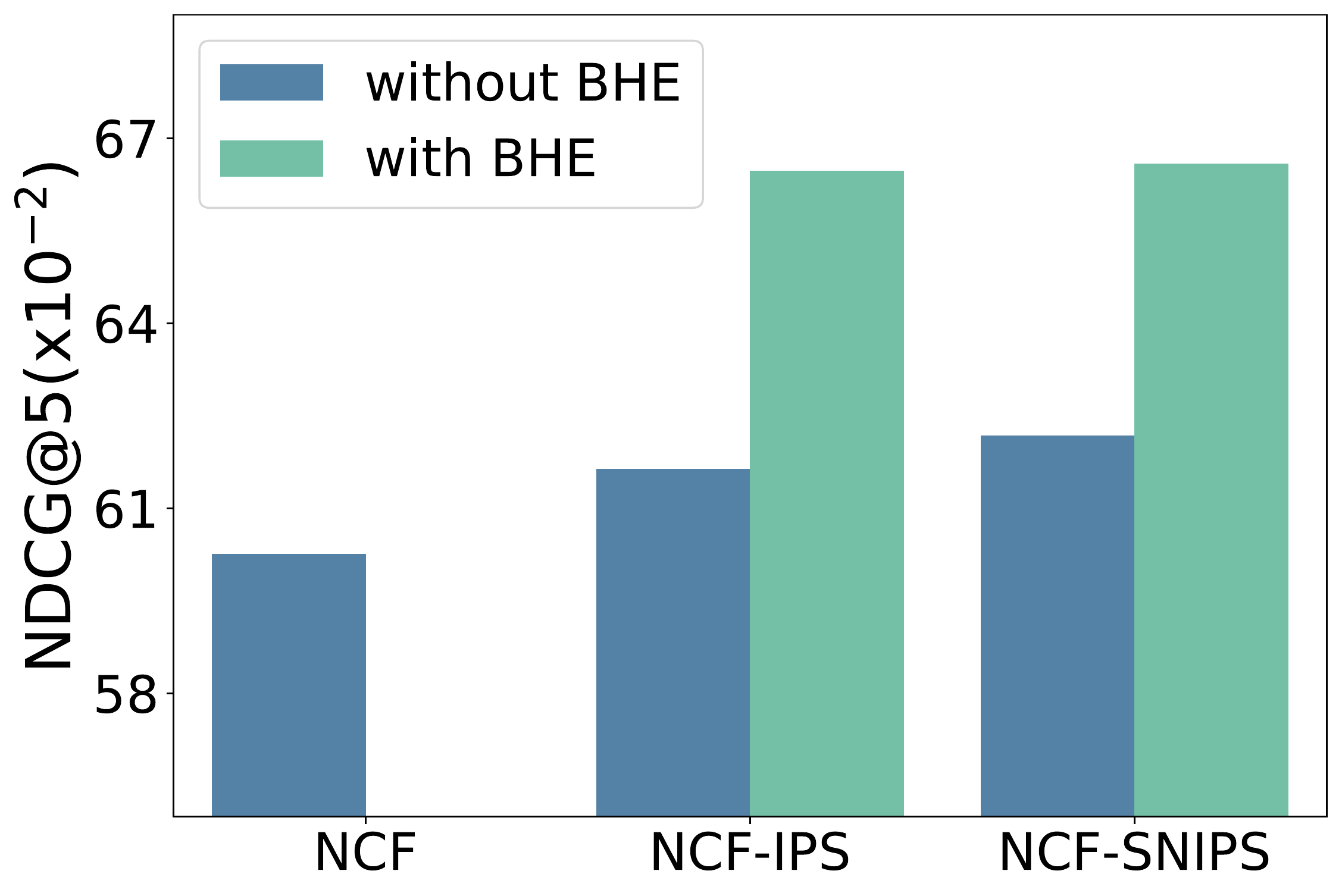}
 }
 \vskip -0.2in
 \caption{Performance of exploiting heterogeneity for better debiasing. Since \OURS~ needs to be combined with IPS-based methods for better debiasing, it cannot directly improve backbones(MF and NCF). With the support of heterogeneity explored by \OURS, each backbone and debiasing method achieves significant improvement.}
 \label{debias_main_fig}
 \vskip -0.2in
\end{figure}

\begin{figure} [t]
 \subfigure[MF as backbone on Yahoo] {
     \includegraphics[width=0.475\linewidth]{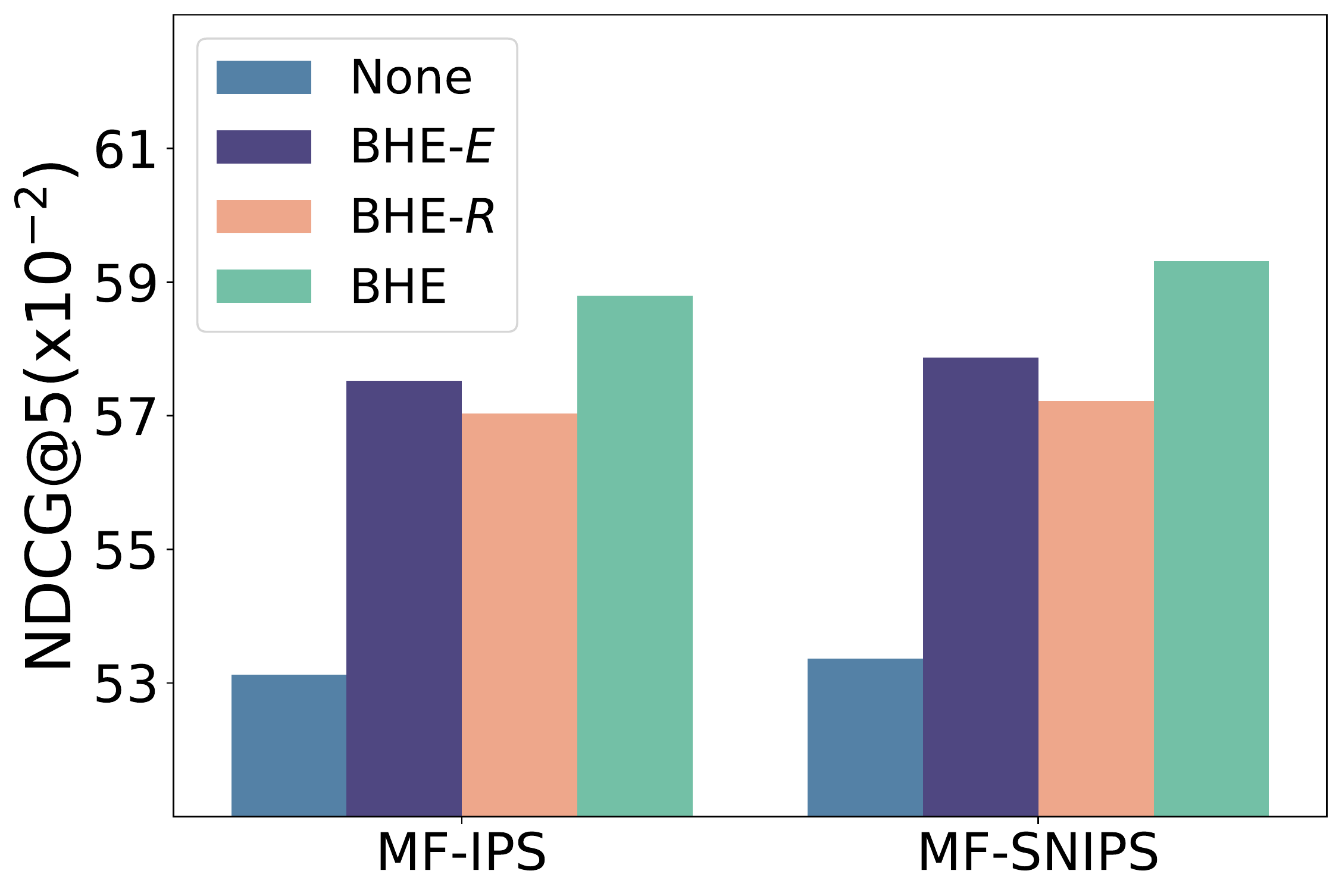}
 }
 \subfigure[NCF as backbone on Yahoo] {
     \includegraphics[width=0.475\linewidth]{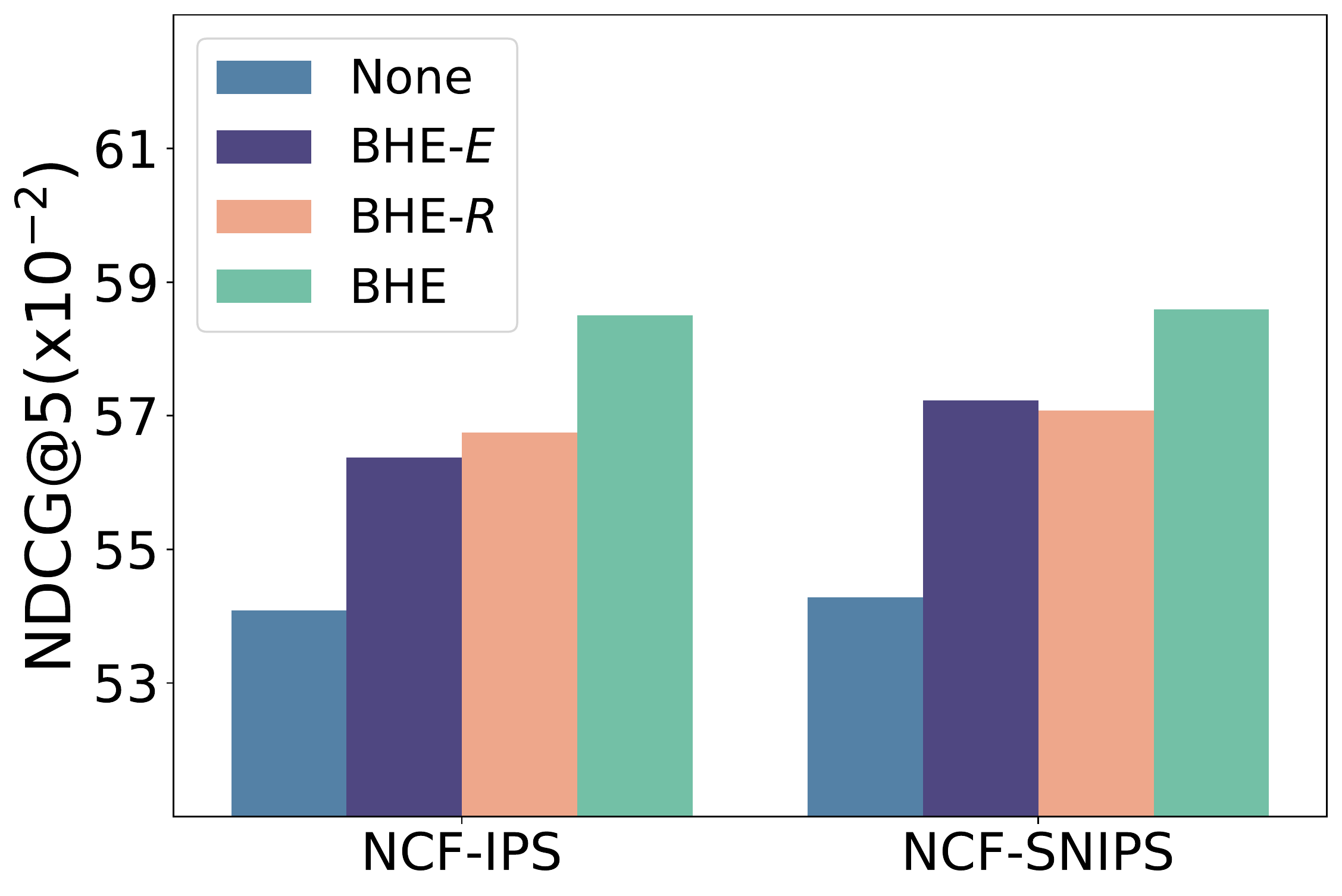}
 }
  \subfigure[MF as backbone on Coat] {
     \includegraphics[width=0.475\linewidth]{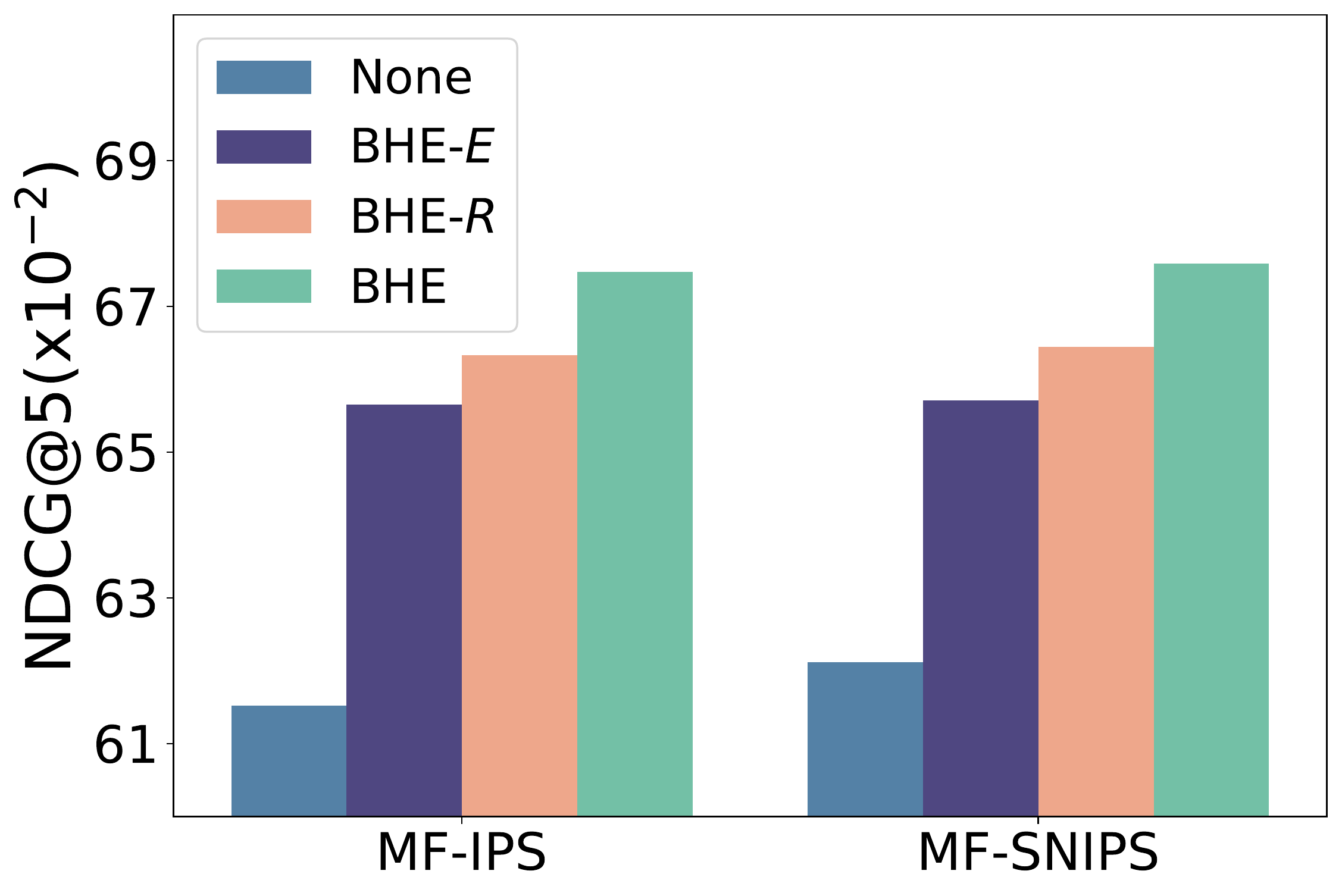}
 }
 \subfigure[NCF as backbone on Coat] {
     \includegraphics[width=0.475\linewidth]{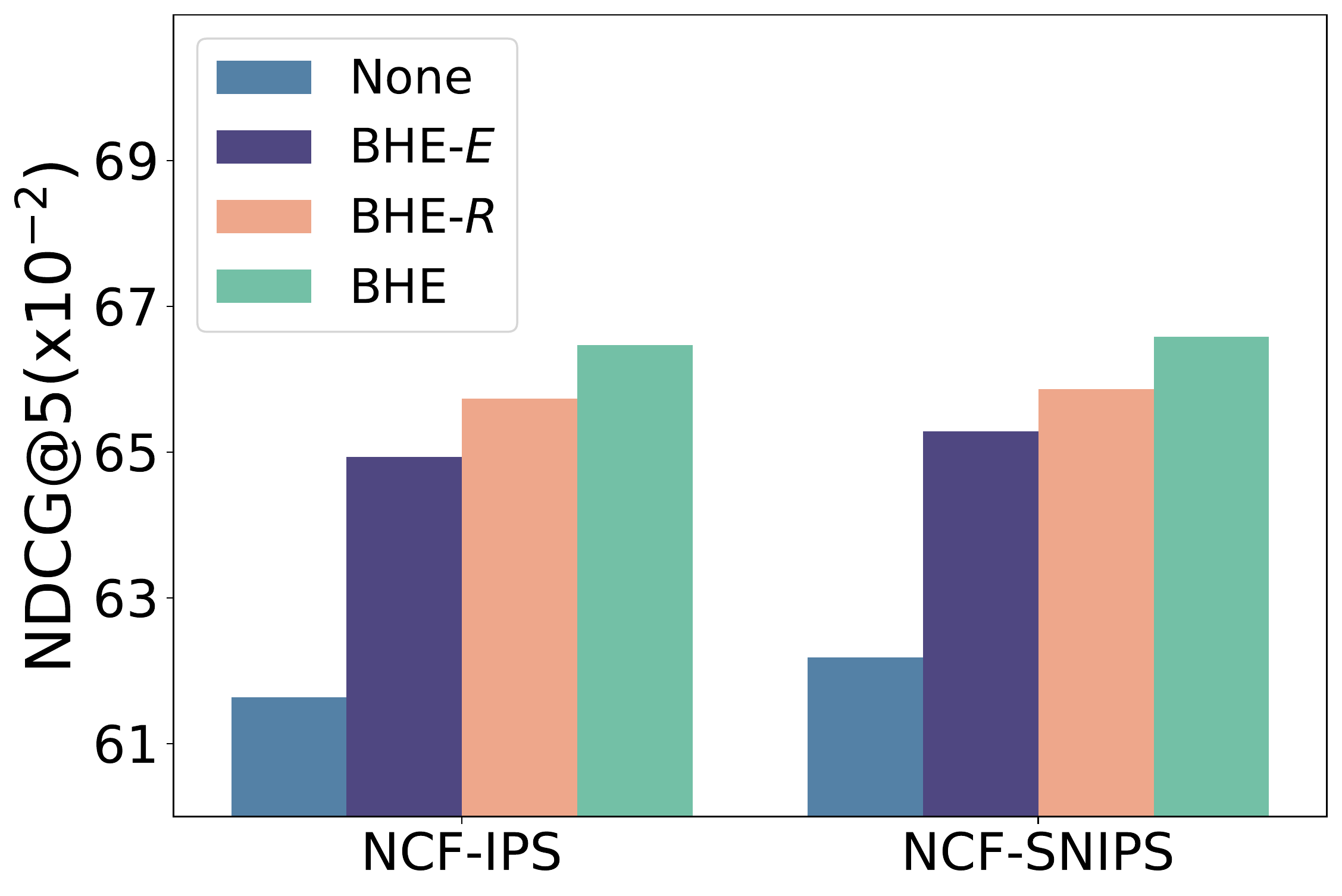}
 }
 \vskip -0.2in
 \caption{Performance of exploiting heterogeneity of $\Eupper$ or $\Rupper$ for debiasing. Both \OURS-$\Eupper$ and \OURS-$\Rupper$ improve compared to original IPS-based methods but are not as good as \OURS.}
 \label{debias_ablation_fig}
 \vskip -0.2in
\end{figure}

\subsection{RQ2: Exploiting Heterogeneity Promotes Generalization and Sub-populational Robustness}
\label{exp_better_generalization_and_sub_popu}

We evaluate the improvement of explicitly exploiting prediction mechanism heterogeneity introduced in Section~\ref{exploit_multi}.
We first examine the overall performance of the methods. In addition to \OURS~ and backbones, we also compare the methods training multiple sub-models for sub-populations obtained by following strategies:
\begin{itemize}[leftmargin=0.3cm]
    \item cluster-user/item. We categorized the samples into multiple environments according to the important feature of the user/item. In Yelp, we select the fans number/item popularity as the important feature for categorizing the samples, while in MovieLens-1M we select the age/number of type labels. 
    \item raw feature/embedding. We perform traditional clustering(e.g., k-means) on the raw features or the well-trained embeddings of samples to categorize samples into sub-populations. These methods explicitly exploit the heterogeneity of covariate distribution.
\end{itemize}
The main difference between the baselines and \OURS~ is that they exploit different heterogeneity. It should be emphasized that all methods, including \OURS, do not use the ground truth to categorize samples in the test phase. More details are in Appendix~\ref{app_baselines}.

\textbf{Overall Performance.}
The overall performance results are shown in Table\ref{labeling_func_table}. We can observe that:
\begin{itemize}[leftmargin=0.3cm]
    \item Compared to other baselines, \OURS~achieves the best performance
    with a remarkable improvement for all the metrics in each case, demonstrating the superiority of our method.
    \item Compared to backbones(FM and NFM), cluster-user/item can bring about some benefits. However, there is still a significant gap between them with \OURS. The reason is that heterogeneity defined by features can represent prediction mechanism heterogeneity to some extent but cannot accurately describe and model it. 
    \item The methods which explicitly exploit heterogeneity of covariate distribution do not bring a boost and are even worse. This indicates that there are differences between the heterogeneity of covariate distribution and prediction mechanism. Misuse of heterogeneity could cause unintended damage.
\end{itemize}

\textbf{Sub-populational Robustness.}
In addition to the overall performance, we also pay attention to sub-populational robustness. For ease of presentation and understanding, we artificially define several sub-populations on test data according to the important features of user and item. Specifically, in MovieLens-1M we use the item popularity and number of user fans respectively, while in Yelp we use item popularity and age respectively. The performances of each sub-population are shown in Figure\ref{minor_group}, where \textit{popular}, \textit{young}, and \textit{influencer} are major sub-populations, while others are minor sub-populations. From the results, we observe that:

\begin{itemize}[leftmargin=0.3cm]
    \item The backbones pay more attention to the major sub-populations and neglect the minor sub-populations. This could seriously damage the interests of minor sub-populations. Such imbalanced performances are harmful to the long-term development of the platform.

    \item Compared with backbones, \OURS~ has achieved improvement in each sub population, especially in the minor sub populations. It shows that appropriate exploitation of prediction mechanism heterogeneity can promote recommenders better capture the rating patterns of each sub-population, and prevent minor sub-populations from being dominated by the major to some extent.


    \item Our method which explores and exploits heterogeneity can bring improvement of performance generally to different sub-populations regardless of the division.
    
\end{itemize}

In summary, appropriate exploitation of recommendation data heterogeneity can improve performance by unveiling the unique statistical characteristics of each sub-population, with the most significant improvement in minor sub-populations.

\begin{figure} [t]
 \subfigure[FM-\OURS~on MovieLens-1M] {
     \includegraphics[width=0.475\linewidth]{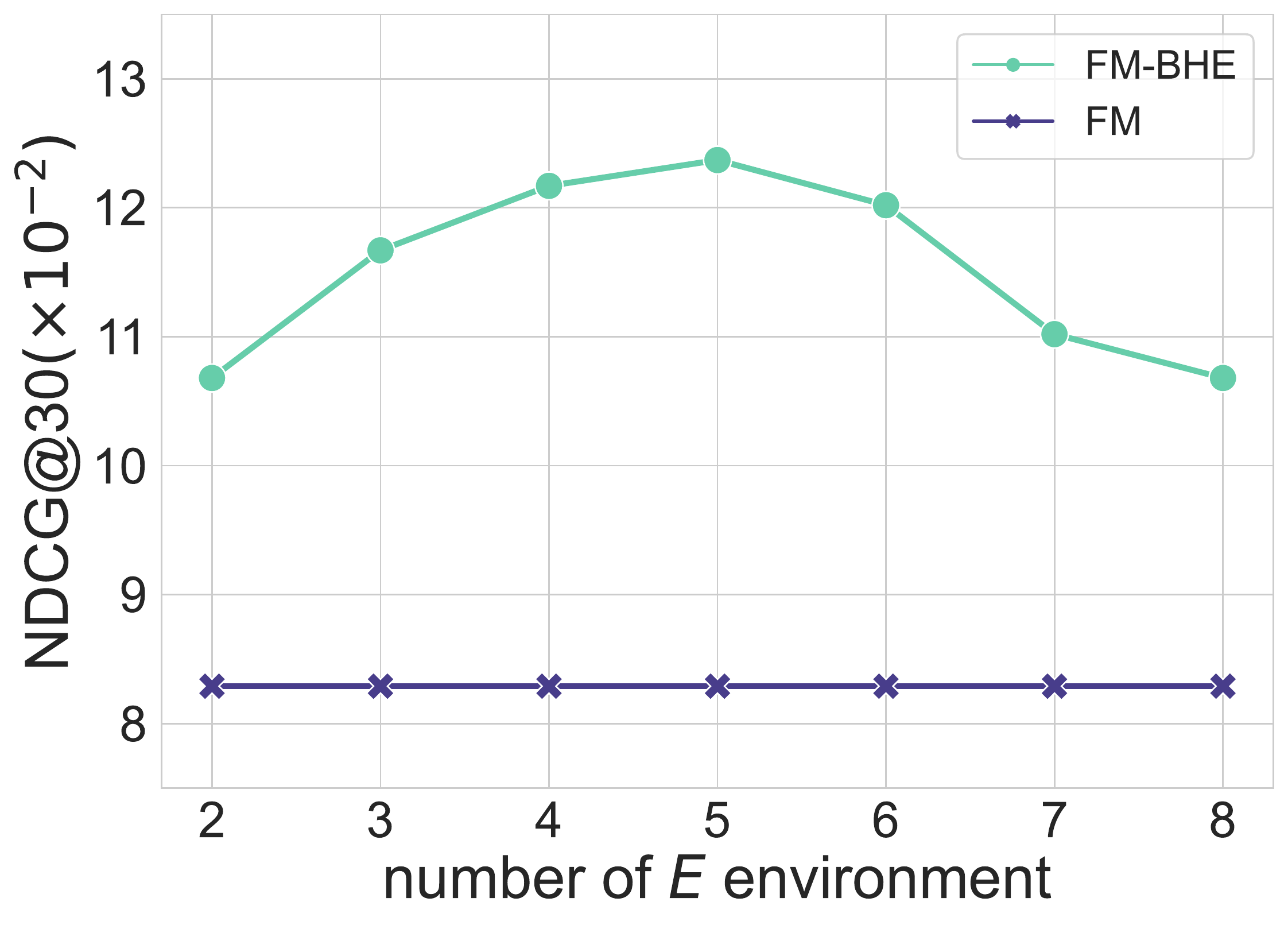}
 }
 \subfigure[NFM-\OURS~on MovieLens-1M] {
     \includegraphics[width=0.475\linewidth]{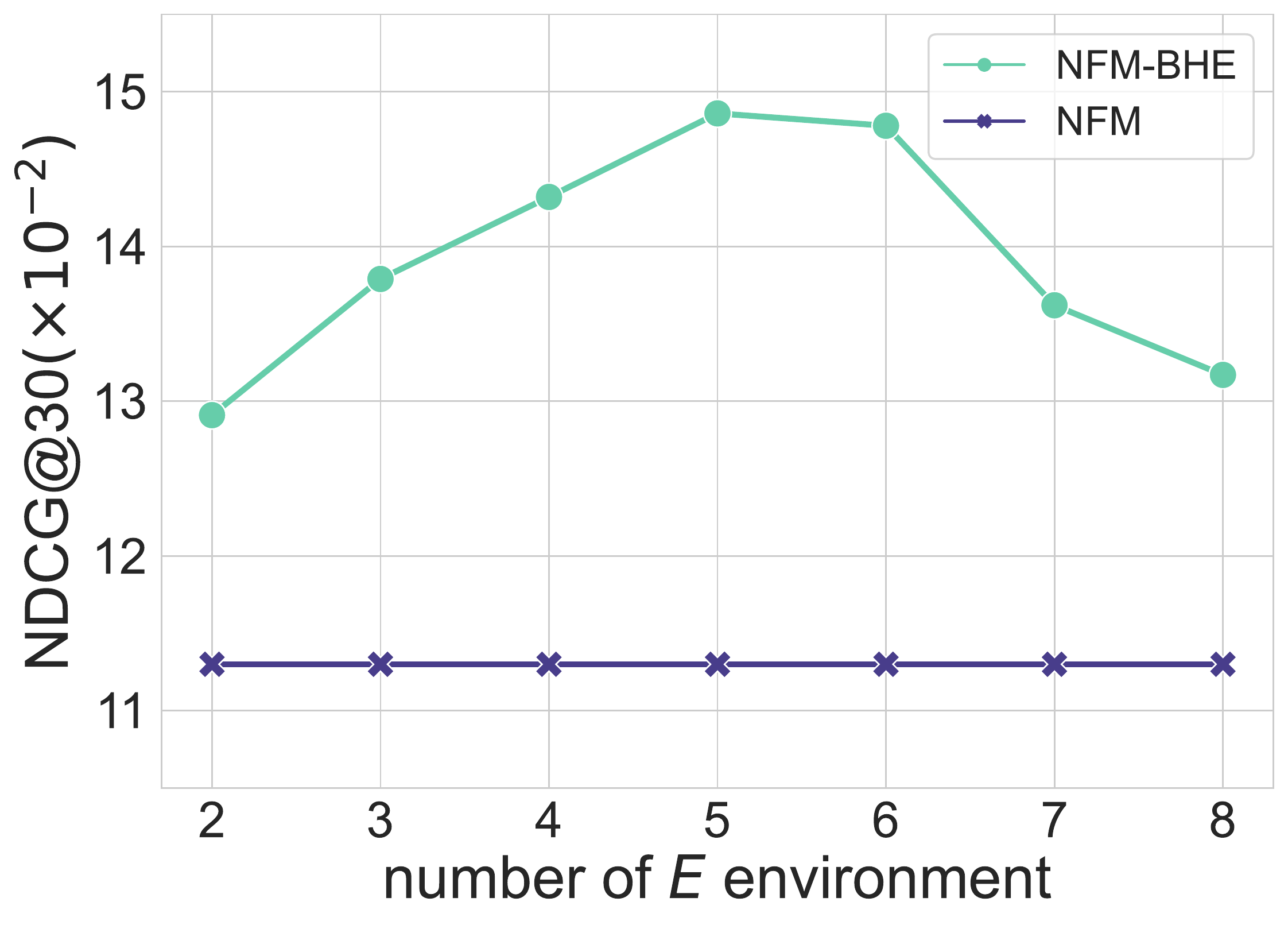}
 }
 \vskip -0.2in
 \caption{The influence of $\Eupper$ environment number on better generalization(settings in Section~\ref{exp_better_generalization_and_sub_popu}). We also show the performance of backbones as a comparison. Since backbones do not exploit heterogeneity, they are not influenced by the number of environments.}
 \label{env_num_ml}
 \vskip -0.25in
\end{figure}

\subsection{RQ3: Exploiting Heterogeneity for Better Debiasing}
\label{exp_better_debias}

\OURS~ can be combined with mainstream IPS-based methods. In this subsection we combine it with IPS\cite{schnabel2016recommendations} and SNIPS\cite{schnabel2016recommendations}. The baselines we consider include two backbones(MF and NCF) and IPS-based debiasing methods(IPS and SNIPS) implemented based on them. More details are in Appendix~\ref{app_baselines}. It is worth to mention that MF and NCF can not directly combine with \OURS~ for debiasing tasks.
In Figure\ref{debias_main_fig}, we observe that:
\begin{itemize}[leftmargin=0.3cm]
    \item \OURS~outperforms all baselines in each case, which demonstrates the superiority of exploiting the heterogeneity unveiled by \OURS.
    \item Although IPS and SNIPS has improved compared to backbones, there is a significant gap with \OURS. This is because \OURS~ exploits heterogeneity information for estimating propensity score more accurately than traditional methods. More accurate propensity score makes \OURS~ promote traditional debiasing methods.
\end{itemize}

In addition, we also conduct an ablation study on how $\Eupper$ heterogeneity and $\Rupper$ heterogeneity influence the performance of \OURS~ in the debiasing task. We compared \OURS~ with \OURS-$\Eupper$ and \OURS-$\Rupper$ which exploit only $\Eupper$ and $\Rupper$ respectively. According to the result shown in Figure\ref{debias_ablation_fig}:

\begin{itemize}[leftmargin=0.3cm]
    \item \OURS-$\Eupper$ and \OURS-$\Rupper$ outperform IPS and SNIPS in most cases. This demonstrates that heterogeneity of both $\Eupper$ and $\Rupper$ can improve debiasing. 
    \item However, both \OURS-$\Eupper$ and \OURS-$\Rupper$ can not cannot perform as well as \OURS. 
    It shows that neither $\Eupper$ nor $\Rupper$ alone are sufficient to fully characterize data heterogeneity useful in debiasing tasks.
\end{itemize}

\subsection{RQ4: Environment Numbers Influence \OURS}
\begin{figure} [t]
 \vskip -0.2in
 \subfigure[MF-IPS-\OURS~on Yahoo] {
     \includegraphics[width=0.475\linewidth]{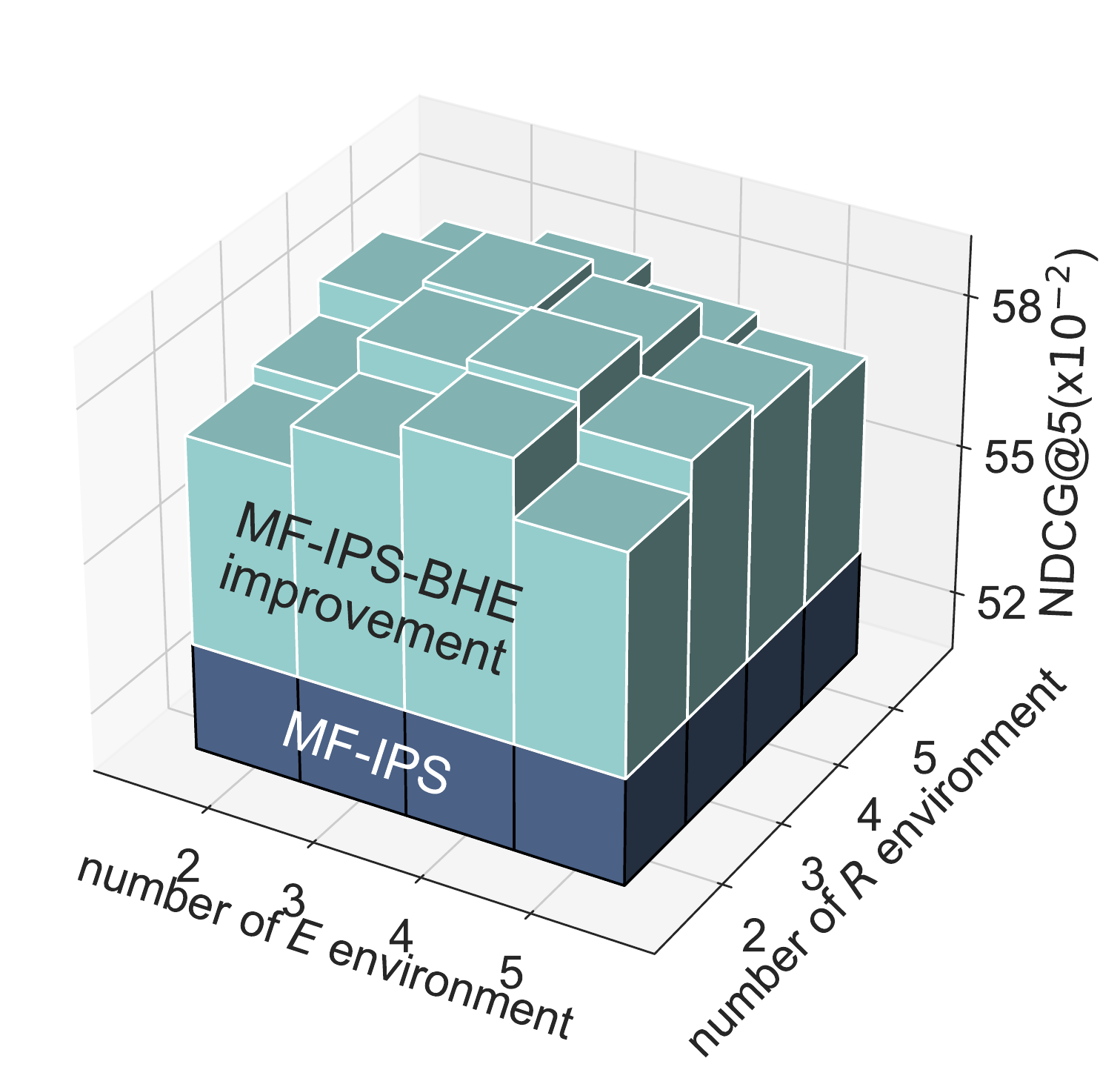}
 }
 \subfigure[MF-SNIPS-\OURS~on Yahoo] {
     \includegraphics[width=0.475\linewidth]{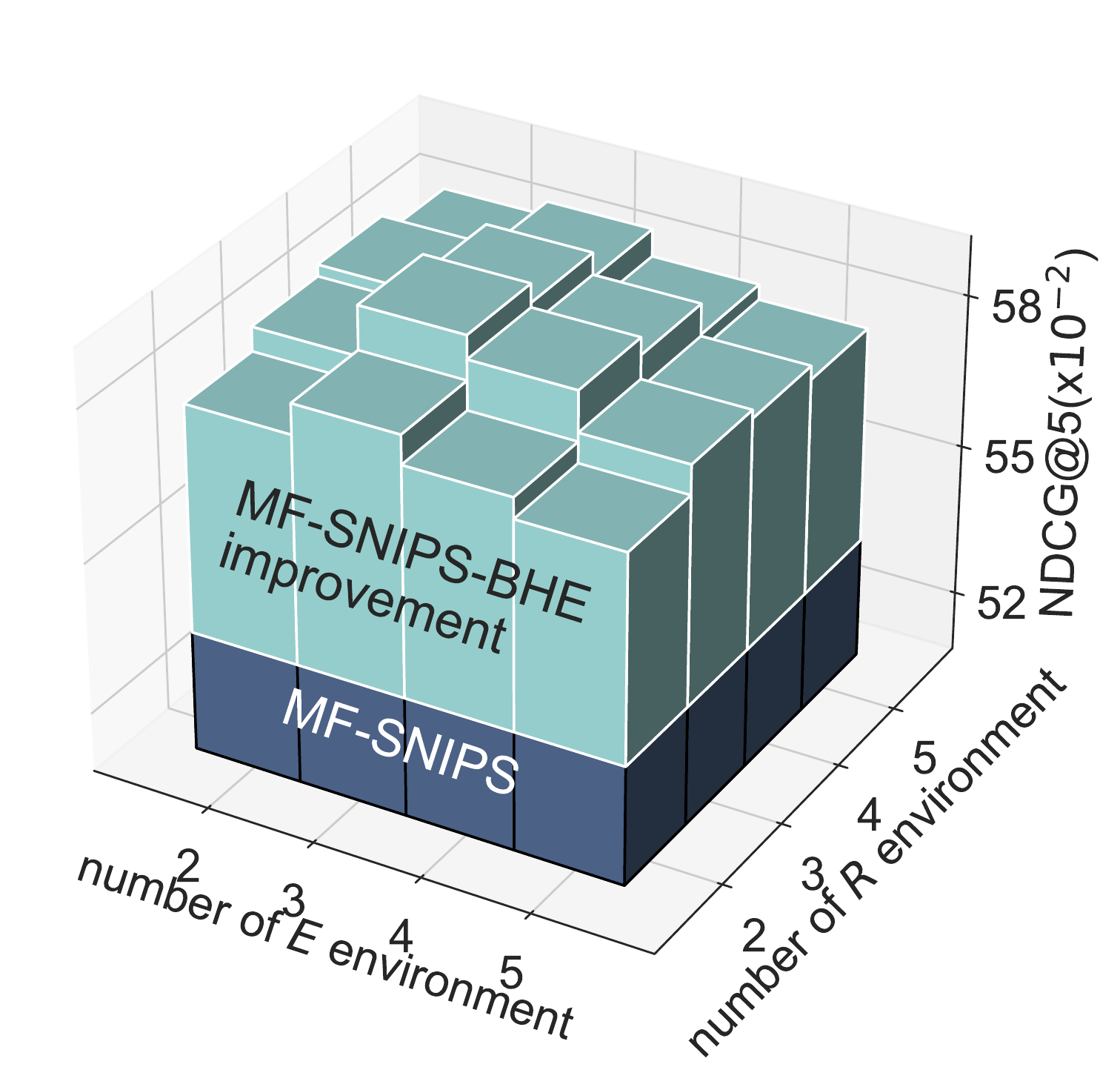}
 }

  \subfigure[NCF-IPS-\OURS~on Yahoo] {
     \includegraphics[width=0.475\linewidth]{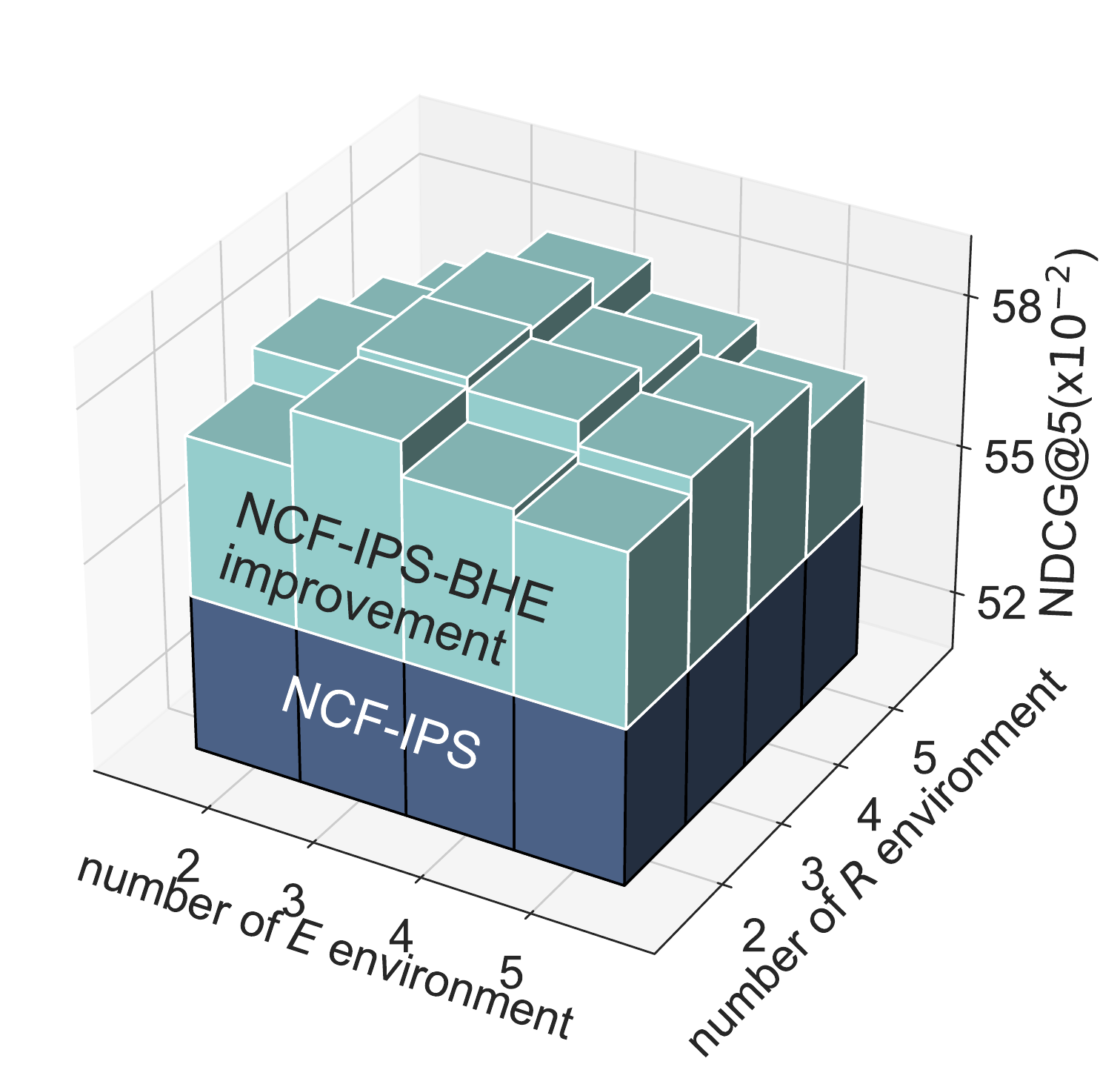}
 }
 \subfigure[NCF-SNIPS-\OURS~on Yahoo] {
     \includegraphics[width=0.475\linewidth]{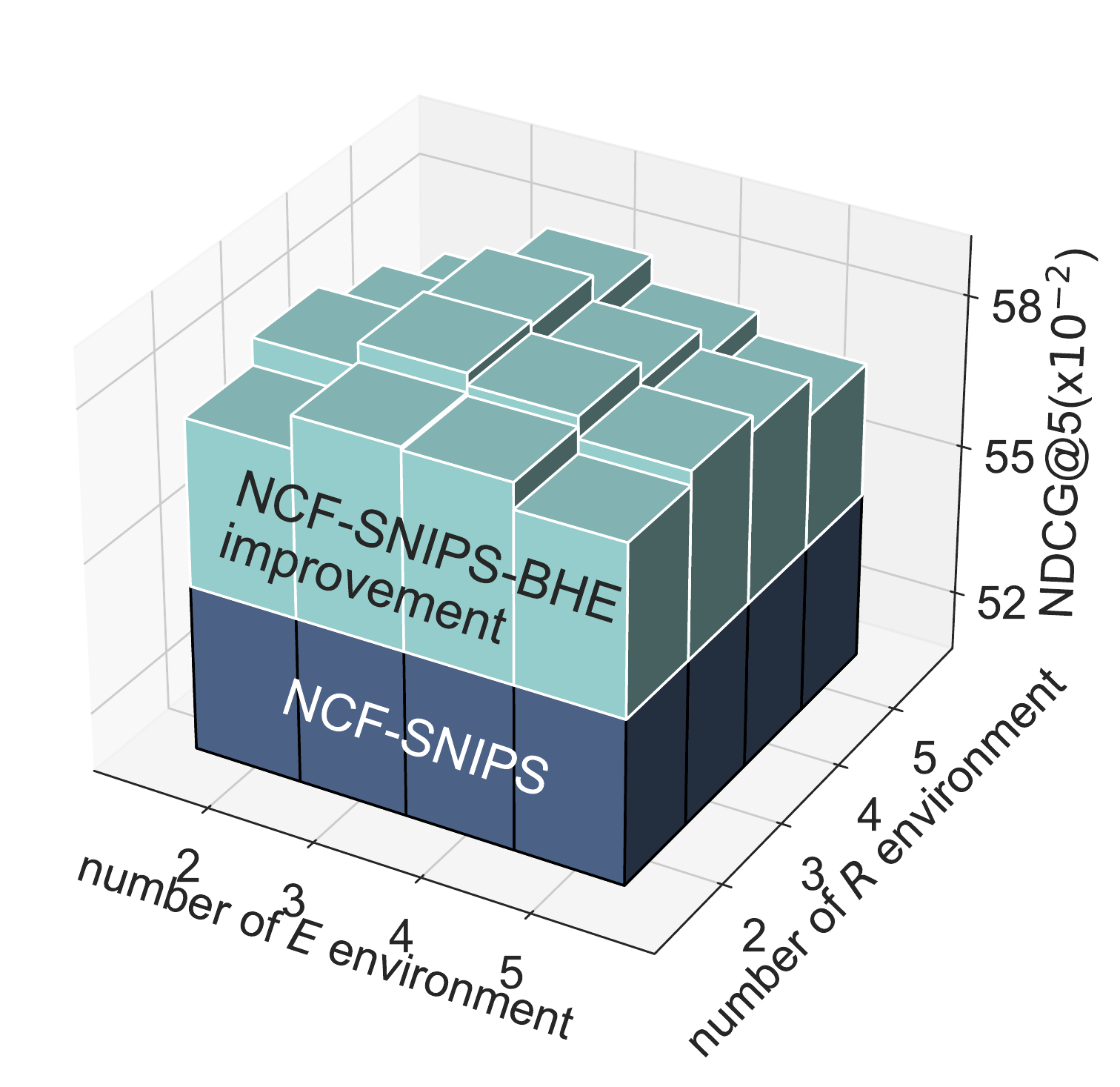}
 }
 
 \vskip -0.2in
 \caption{The influence of $\Eupper$ and $\Rupper$ environment numbers on better debiasing(settings in Section~\ref{exp_better_debias}). We also show the performance of backbones as a comparison(the blue pillars). Since backbones do not exploit heterogeneity, they are not influenced by the number of environments.}
 \label{env_num_yahoo}
 \vskip -0.1in
\end{figure}

In this subsection, we study how the environment numbers($|\mathcal{E}|$ and $|\mathcal{R}|$), which are hyper-parameters influence \OURS. In Figure\ref{env_num_ml}-\ref{env_num_yahoo}:
\begin{itemize}[leftmargin=0.3cm]
    \item The performance of \OURS~ gets better as the numbers of environments increases. This suggests that more environments are needed to model the heterogeneity of the data.
    \item However, if the numbers of environments is too large, data sparsity can be exacerbated and hurt performance.
    \item In most cases, despite fluctuations in the performance of \OURS, there is still a significant improvement over backbones.
\end{itemize}

On Yelp and Coat, we can also get consistent conclusions, see the Appendix~\ref{app_env_num} for details.

\section{conclusion}

Massive amounts of data from different sub-populations lead to widespread data heterogeneity in recommendation since each sub-population could exhibit some unique characteristics and patterns. Ignoring the heterogeneity in recommendation data could hurt the generalization of recommendation models, damage the sub-populational robustness, and make the recommendation models misled by biases. To study heterogeneity, we intrinsically define and model the heterogeneity in recommendation data. Since real-world data often lack accurate and practical sub-population labels, we propose \OURS~ to explore the heterogeneity in recommendation data. To exploit the heterogeneity explored, we propose two approaches also. We conduct extensive experiments on real-world datasets to validate the basic assumption of our approaches and the significant impacts of heterogeneity on recommendation models. Also, the evaluation on \OURS~ indicates that appropriately exploiting data heterogeneity in recommendation could improve generalization, promote sub-populational robustness, and support debiasing.
\clearpage
\bibliographystyle{ACM-Reference-Format}
\balance
\bibliography{ref}


\begin{thebibliography}{37}


\ifx \showCODEN    \undefined \def \showCODEN     #1{\unskip}     \fi
\ifx \showDOI      \undefined \def \showDOI       #1{#1}\fi
\ifx \showISBNx    \undefined \def \showISBNx     #1{\unskip}     \fi
\ifx \showISBNxiii \undefined \def \showISBNxiii  #1{\unskip}     \fi
\ifx \showISSN     \undefined \def \showISSN      #1{\unskip}     \fi
\ifx \showLCCN     \undefined \def \showLCCN      #1{\unskip}     \fi
\ifx \shownote     \undefined \def \shownote      #1{#1}          \fi
\ifx \showarticletitle \undefined \def \showarticletitle #1{#1}   \fi
\ifx \showURL      \undefined \def \showURL       {\relax}        \fi
\providecommand\bibfield[2]{#2}
\providecommand\bibinfo[2]{#2}
\providecommand\natexlab[1]{#1}
\providecommand\showeprint[2][]{arXiv:#2}

\bibitem[Abdollahpouri and Mansoury(2020)]%
        {abdollahpouri2020multi}
\bibfield{author}{\bibinfo{person}{Himan Abdollahpouri} {and}
  \bibinfo{person}{Masoud Mansoury}.} \bibinfo{year}{2020}\natexlab{}.
\newblock \showarticletitle{Multi-sided exposure bias in recommendation}.
\newblock \bibinfo{journal}{\emph{arXiv preprint arXiv:2006.15772}}
  (\bibinfo{year}{2020}).
\newblock


\bibitem[Abdollahpouri et~al\mbox{.}(2019)]%
        {abdollahpouri2019unfairness}
\bibfield{author}{\bibinfo{person}{Himan Abdollahpouri},
  \bibinfo{person}{Masoud Mansoury}, \bibinfo{person}{Robin Burke}, {and}
  \bibinfo{person}{Bamshad Mobasher}.} \bibinfo{year}{2019}\natexlab{}.
\newblock \showarticletitle{The unfairness of popularity bias in
  recommendation}.
\newblock \bibinfo{journal}{\emph{arXiv preprint arXiv:1907.13286}}
  (\bibinfo{year}{2019}).
\newblock


\bibitem[Abdollahpouri et~al\mbox{.}(2020)]%
        {abdollahpouri2020connection}
\bibfield{author}{\bibinfo{person}{Himan Abdollahpouri},
  \bibinfo{person}{Masoud Mansoury}, \bibinfo{person}{Robin Burke}, {and}
  \bibinfo{person}{Bamshad Mobasher}.} \bibinfo{year}{2020}\natexlab{}.
\newblock \showarticletitle{The connection between popularity bias,
  calibration, and fairness in recommendation}. In
  \bibinfo{booktitle}{\emph{Fourteenth ACM conference on recommender systems}}.
  \bibinfo{pages}{726--731}.
\newblock


\bibitem[Arjovsky et~al\mbox{.}(2019)]%
        {arjovsky2019invariant}
\bibfield{author}{\bibinfo{person}{Martin Arjovsky}, \bibinfo{person}{L{\'e}on
  Bottou}, \bibinfo{person}{Ishaan Gulrajani}, {and} \bibinfo{person}{David
  Lopez-Paz}.} \bibinfo{year}{2019}\natexlab{}.
\newblock \showarticletitle{Invariant risk minimization}.
\newblock \bibinfo{journal}{\emph{arXiv preprint arXiv:1907.02893}}
  (\bibinfo{year}{2019}).
\newblock


\bibitem[Bonner and Vasile(2018)]%
        {bonner2018causal}
\bibfield{author}{\bibinfo{person}{Stephen Bonner} {and}
  \bibinfo{person}{Flavian Vasile}.} \bibinfo{year}{2018}\natexlab{}.
\newblock \showarticletitle{Causal embeddings for recommendation}. In
  \bibinfo{booktitle}{\emph{Proceedings of the 12th ACM conference on
  recommender systems}}. \bibinfo{pages}{104--112}.
\newblock


\bibitem[Breitenstein et~al\mbox{.}(2020)]%
        {breitenstein2020systematization}
\bibfield{author}{\bibinfo{person}{Jasmin Breitenstein},
  \bibinfo{person}{Jan-Aike Term{\"o}hlen}, \bibinfo{person}{Daniel Lipinski},
  {and} \bibinfo{person}{Tim Fingscheidt}.} \bibinfo{year}{2020}\natexlab{}.
\newblock \showarticletitle{Systematization of corner cases for visual
  perception in automated driving}. In \bibinfo{booktitle}{\emph{2020 IEEE
  Intelligent Vehicles Symposium (IV)}}. IEEE, \bibinfo{pages}{1257--1264}.
\newblock


\bibitem[Cali{\'n}ski and Harabasz(1974)]%
        {calinski1974dendrite}
\bibfield{author}{\bibinfo{person}{Tadeusz Cali{\'n}ski} {and}
  \bibinfo{person}{Jerzy Harabasz}.} \bibinfo{year}{1974}\natexlab{}.
\newblock \showarticletitle{A dendrite method for cluster analysis}.
\newblock \bibinfo{journal}{\emph{Communications in Statistics-theory and
  Methods}} \bibinfo{volume}{3}, \bibinfo{number}{1} (\bibinfo{year}{1974}),
  \bibinfo{pages}{1--27}.
\newblock


\bibitem[Challen et~al\mbox{.}(2019)]%
        {challen2019artificial}
\bibfield{author}{\bibinfo{person}{Robert Challen}, \bibinfo{person}{Joshua
  Denny}, \bibinfo{person}{Martin Pitt}, \bibinfo{person}{Luke Gompels},
  \bibinfo{person}{Tom Edwards}, {and} \bibinfo{person}{Krasimira
  Tsaneva-Atanasova}.} \bibinfo{year}{2019}\natexlab{}.
\newblock \showarticletitle{Artificial intelligence, bias and clinical safety}.
\newblock \bibinfo{journal}{\emph{BMJ Quality \& Safety}} \bibinfo{volume}{28},
  \bibinfo{number}{3} (\bibinfo{year}{2019}), \bibinfo{pages}{231--237}.
\newblock


\bibitem[Chen et~al\mbox{.}(2021)]%
        {chen2021autodebias}
\bibfield{author}{\bibinfo{person}{Jiawei Chen}, \bibinfo{person}{Hande Dong},
  \bibinfo{person}{Yang Qiu}, \bibinfo{person}{Xiangnan He},
  \bibinfo{person}{Xin Xin}, \bibinfo{person}{Liang Chen},
  \bibinfo{person}{Guli Lin}, {and} \bibinfo{person}{Keping Yang}.}
  \bibinfo{year}{2021}\natexlab{}.
\newblock \showarticletitle{AutoDebias: Learning to debias for recommendation}.
  In \bibinfo{booktitle}{\emph{Proceedings of the 44th International ACM SIGIR
  Conference on Research and Development in Information Retrieval}}.
  \bibinfo{pages}{21--30}.
\newblock


\bibitem[Chen et~al\mbox{.}(2020)]%
        {chen2020bias}
\bibfield{author}{\bibinfo{person}{Jiawei Chen}, \bibinfo{person}{Hande Dong},
  \bibinfo{person}{Xiang Wang}, \bibinfo{person}{Fuli Feng},
  \bibinfo{person}{Meng Wang}, {and} \bibinfo{person}{Xiangnan He}.}
  \bibinfo{year}{2020}\natexlab{}.
\newblock \showarticletitle{Bias and debias in recommender system: A survey and
  future directions}.
\newblock \bibinfo{journal}{\emph{arXiv preprint arXiv:2010.03240}}
  (\bibinfo{year}{2020}).
\newblock


\bibitem[Chen et~al\mbox{.}(2018)]%
        {chen2018modeling}
\bibfield{author}{\bibinfo{person}{Jiawei Chen}, \bibinfo{person}{Yan Feng},
  \bibinfo{person}{Martin Ester}, \bibinfo{person}{Sheng Zhou},
  \bibinfo{person}{Chun Chen}, {and} \bibinfo{person}{Can Wang}.}
  \bibinfo{year}{2018}\natexlab{}.
\newblock \showarticletitle{Modeling Users' Exposure with Social Knowledge
  Influence and Consumption Influence for Recommendation}. In
  \bibinfo{booktitle}{\emph{Proceedings of the 27th ACM International
  Conference on Information and Knowledge Management}}.
  \bibinfo{pages}{953--962}.
\newblock


\bibitem[Collins et~al\mbox{.}(2018)]%
        {collins2018study}
\bibfield{author}{\bibinfo{person}{Andrew Collins}, \bibinfo{person}{Dominika
  Tkaczyk}, \bibinfo{person}{Akiko Aizawa}, {and} \bibinfo{person}{Joeran
  Beel}.} \bibinfo{year}{2018}\natexlab{}.
\newblock \showarticletitle{A study of position bias in digital library
  recommender systems}.
\newblock \bibinfo{journal}{\emph{arXiv preprint arXiv:1802.06565}}
  (\bibinfo{year}{2018}).
\newblock


\bibitem[Duchi and Namkoong(2018)]%
        {duchi2018learning}
\bibfield{author}{\bibinfo{person}{John Duchi} {and} \bibinfo{person}{Hongseok
  Namkoong}.} \bibinfo{year}{2018}\natexlab{}.
\newblock \showarticletitle{Learning models with uniform performance via
  distributionally robust optimization}.
\newblock \bibinfo{journal}{\emph{arXiv preprint arXiv:1810.08750}}
  (\bibinfo{year}{2018}).
\newblock


\bibitem[Dzobo et~al\mbox{.}(2018)]%
        {dzobo2018not}
\bibfield{author}{\bibinfo{person}{Kevin Dzobo},
  \bibinfo{person}{Dimakatso~Alice Senthebane}, \bibinfo{person}{Nicholas~Ekow
  Thomford}, \bibinfo{person}{Arielle Rowe}, \bibinfo{person}{Collet Dandara},
  {and} \bibinfo{person}{M~Iqbal Parker}.} \bibinfo{year}{2018}\natexlab{}.
\newblock \showarticletitle{Not everyone fits the mold: Intratumor and
  intertumor heterogeneity and innovative cancer drug design and development}.
\newblock \bibinfo{journal}{\emph{Omics: a journal of integrative biology}}
  \bibinfo{volume}{22}, \bibinfo{number}{1} (\bibinfo{year}{2018}),
  \bibinfo{pages}{17--34}.
\newblock


\bibitem[Fan et~al\mbox{.}(2014)]%
        {fan2014challenges}
\bibfield{author}{\bibinfo{person}{Jianqing Fan}, \bibinfo{person}{Fang Han},
  {and} \bibinfo{person}{Han Liu}.} \bibinfo{year}{2014}\natexlab{}.
\newblock \showarticletitle{Challenges of big data analysis}.
\newblock \bibinfo{journal}{\emph{National science review}}
  \bibinfo{volume}{1}, \bibinfo{number}{2} (\bibinfo{year}{2014}),
  \bibinfo{pages}{293--314}.
\newblock


\bibitem[Hamerly and Elkan(2003)]%
        {hamerly2003learning}
\bibfield{author}{\bibinfo{person}{Greg Hamerly} {and} \bibinfo{person}{Charles
  Elkan}.} \bibinfo{year}{2003}\natexlab{}.
\newblock \showarticletitle{Learning the k in k-means}.
\newblock \bibinfo{journal}{\emph{Advances in neural information processing
  systems}}  \bibinfo{volume}{16} (\bibinfo{year}{2003}).
\newblock


\bibitem[He(2017)]%
        {he2017learning}
\bibfield{author}{\bibinfo{person}{Jingrui He}.}
  \bibinfo{year}{2017}\natexlab{}.
\newblock \showarticletitle{Learning from Data Heterogeneity: Algorithms and
  Applications.}. In \bibinfo{booktitle}{\emph{IJCAI}}.
  \bibinfo{pages}{5126--5130}.
\newblock


\bibitem[He and Chua(2017)]%
        {he2017neural}
\bibfield{author}{\bibinfo{person}{Xiangnan He} {and} \bibinfo{person}{Tat-Seng
  Chua}.} \bibinfo{year}{2017}\natexlab{}.
\newblock \showarticletitle{Neural factorization machines for sparse predictive
  analytics}. In \bibinfo{booktitle}{\emph{Proceedings of the 40th
  International ACM SIGIR conference on Research and Development in Information
  Retrieval}}. \bibinfo{pages}{355--364}.
\newblock


\bibitem[He et~al\mbox{.}(2017)]%
        {he2017NCF}
\bibfield{author}{\bibinfo{person}{Xiangnan He}, \bibinfo{person}{Lizi Liao},
  \bibinfo{person}{Hanwang Zhang}, \bibinfo{person}{Liqiang Nie},
  \bibinfo{person}{Xia Hu}, {and} \bibinfo{person}{Tat-Seng Chua}.}
  \bibinfo{year}{2017}\natexlab{}.
\newblock \showarticletitle{Neural collaborative filtering}. In
  \bibinfo{booktitle}{\emph{Proceedings of the 26th international conference on
  world wide web}}. \bibinfo{pages}{173--182}.
\newblock


\bibitem[He et~al\mbox{.}(2022)]%
        {he2022causpref}
\bibfield{author}{\bibinfo{person}{Yue He}, \bibinfo{person}{Zimu Wang},
  \bibinfo{person}{Peng Cui}, \bibinfo{person}{Hao Zou},
  \bibinfo{person}{Yafeng Zhang}, \bibinfo{person}{Qiang Cui}, {and}
  \bibinfo{person}{Yong Jiang}.} \bibinfo{year}{2022}\natexlab{}.
\newblock \showarticletitle{CausPref: Causal Preference Learning for
  Out-of-Distribution Recommendation}. In \bibinfo{booktitle}{\emph{Proceedings
  of the ACM Web Conference 2022}}. \bibinfo{pages}{410--421}.
\newblock


\bibitem[Joachims et~al\mbox{.}(2007)]%
        {joachims2007evaluating}
\bibfield{author}{\bibinfo{person}{Thorsten Joachims}, \bibinfo{person}{Laura
  Granka}, \bibinfo{person}{Bing Pan}, \bibinfo{person}{Helene Hembrooke},
  \bibinfo{person}{Filip Radlinski}, {and} \bibinfo{person}{Geri Gay}.}
  \bibinfo{year}{2007}\natexlab{}.
\newblock \showarticletitle{Evaluating the accuracy of implicit feedback from
  clicks and query reformulations in web search}.
\newblock \bibinfo{journal}{\emph{ACM Transactions on Information Systems
  (TOIS)}} \bibinfo{volume}{25}, \bibinfo{number}{2} (\bibinfo{year}{2007}),
  \bibinfo{pages}{7--es}.
\newblock


\bibitem[Kearns et~al\mbox{.}(2018)]%
        {kearns2018preventing}
\bibfield{author}{\bibinfo{person}{Michael Kearns}, \bibinfo{person}{Seth
  Neel}, \bibinfo{person}{Aaron Roth}, {and} \bibinfo{person}{Zhiwei~Steven
  Wu}.} \bibinfo{year}{2018}\natexlab{}.
\newblock \showarticletitle{Preventing fairness gerrymandering: Auditing and
  learning for subgroup fairness}. In \bibinfo{booktitle}{\emph{International
  Conference on Machine Learning}}. PMLR, \bibinfo{pages}{2564--2572}.
\newblock


\bibitem[Kim and Wu(2022)]%
        {kim2022fedgpo}
\bibfield{author}{\bibinfo{person}{Young~Geun Kim} {and}
  \bibinfo{person}{Carole-Jean Wu}.} \bibinfo{year}{2022}\natexlab{}.
\newblock \showarticletitle{FedGPO: Heterogeneity-Aware Global Parameter
  Optimization for Efficient Federated Learning}. In
  \bibinfo{booktitle}{\emph{2022 IEEE International Symposium on Workload
  Characterization (IISWC)}}. IEEE, \bibinfo{pages}{117--129}.
\newblock


\bibitem[Koren(2008)]%
        {koren2008factorization}
\bibfield{author}{\bibinfo{person}{Yehuda Koren}.}
  \bibinfo{year}{2008}\natexlab{}.
\newblock \showarticletitle{Factorization meets the neighborhood: a
  multifaceted collaborative filtering model}. In
  \bibinfo{booktitle}{\emph{Proceedings of the 14th ACM SIGKDD international
  conference on Knowledge discovery and data mining}}.
  \bibinfo{pages}{426--434}.
\newblock


\bibitem[Li and Reynolds(1995)]%
        {li1995definition}
\bibfield{author}{\bibinfo{person}{H Li} {and} \bibinfo{person}{JF Reynolds}.}
  \bibinfo{year}{1995}\natexlab{}.
\newblock \showarticletitle{On definition and quantification of heterogeneity}.
\newblock \bibinfo{journal}{\emph{Oikos}} (\bibinfo{year}{1995}),
  \bibinfo{pages}{280--284}.
\newblock


\bibitem[Liu et~al\mbox{.}(2020)]%
        {liu2020general}
\bibfield{author}{\bibinfo{person}{Dugang Liu}, \bibinfo{person}{Pengxiang
  Cheng}, \bibinfo{person}{Zhenhua Dong}, \bibinfo{person}{Xiuqiang He},
  \bibinfo{person}{Weike Pan}, {and} \bibinfo{person}{Zhong Ming}.}
  \bibinfo{year}{2020}\natexlab{}.
\newblock \showarticletitle{A general knowledge distillation framework for
  counterfactual recommendation via uniform data}. In
  \bibinfo{booktitle}{\emph{Proceedings of the 43rd International ACM SIGIR
  Conference on Research and Development in Information Retrieval}}.
  \bibinfo{pages}{831--840}.
\newblock


\bibitem[Liu et~al\mbox{.}(2021a)]%
        {liu2021mitigating}
\bibfield{author}{\bibinfo{person}{Dugang Liu}, \bibinfo{person}{Pengxiang
  Cheng}, \bibinfo{person}{Hong Zhu}, \bibinfo{person}{Zhenhua Dong},
  \bibinfo{person}{Xiuqiang He}, \bibinfo{person}{Weike Pan}, {and}
  \bibinfo{person}{Zhong Ming}.} \bibinfo{year}{2021}\natexlab{a}.
\newblock \showarticletitle{Mitigating confounding bias in recommendation via
  information bottleneck}. In \bibinfo{booktitle}{\emph{Fifteenth ACM
  Conference on Recommender Systems}}. \bibinfo{pages}{351--360}.
\newblock


\bibitem[Liu et~al\mbox{.}(2021b)]%
        {liu2021heterogeneous}
\bibfield{author}{\bibinfo{person}{Jiashuo Liu}, \bibinfo{person}{Zheyuan Hu},
  \bibinfo{person}{Peng Cui}, \bibinfo{person}{Bo Li}, {and}
  \bibinfo{person}{Zheyan Shen}.} \bibinfo{year}{2021}\natexlab{b}.
\newblock \showarticletitle{Heterogeneous risk minimization}. In
  \bibinfo{booktitle}{\emph{International Conference on Machine Learning}}.
  PMLR, \bibinfo{pages}{6804--6814}.
\newblock


\bibitem[Liu et~al\mbox{.}(2021c)]%
        {liu2021kernelized}
\bibfield{author}{\bibinfo{person}{Jiashuo Liu}, \bibinfo{person}{Zheyuan Hu},
  \bibinfo{person}{Peng Cui}, \bibinfo{person}{Bo Li}, {and}
  \bibinfo{person}{Zheyan Shen}.} \bibinfo{year}{2021}\natexlab{c}.
\newblock \showarticletitle{Kernelized heterogeneous risk minimization}.
\newblock \bibinfo{journal}{\emph{arXiv preprint arXiv:2110.12425}}
  (\bibinfo{year}{2021}).
\newblock


\bibitem[Maeng et~al\mbox{.}(2022)]%
        {maeng2022towards}
\bibfield{author}{\bibinfo{person}{Kiwan Maeng}, \bibinfo{person}{Haiyu Lu},
  \bibinfo{person}{Luca Melis}, \bibinfo{person}{John Nguyen},
  \bibinfo{person}{Mike Rabbat}, {and} \bibinfo{person}{Carole-Jean Wu}.}
  \bibinfo{year}{2022}\natexlab{}.
\newblock \showarticletitle{Towards fair federated recommendation learning:
  Characterizing the inter-dependence of system and data heterogeneity}. In
  \bibinfo{booktitle}{\emph{Proceedings of the 16th ACM Conference on
  Recommender Systems}}. \bibinfo{pages}{156--167}.
\newblock


\bibitem[Rendle(2010)]%
        {rendle2010factorization}
\bibfield{author}{\bibinfo{person}{Steffen Rendle}.}
  \bibinfo{year}{2010}\natexlab{}.
\newblock \showarticletitle{Factorization machines}. In
  \bibinfo{booktitle}{\emph{2010 IEEE International conference on data
  mining}}. IEEE, \bibinfo{pages}{995--1000}.
\newblock


\bibitem[Rosenbaum(2005)]%
        {rosenbaum2005heterogeneity}
\bibfield{author}{\bibinfo{person}{Paul~R Rosenbaum}.}
  \bibinfo{year}{2005}\natexlab{}.
\newblock \showarticletitle{Heterogeneity and causality: Unit heterogeneity and
  design sensitivity in observational studies}.
\newblock \bibinfo{journal}{\emph{The American Statistician}}
  \bibinfo{volume}{59}, \bibinfo{number}{2} (\bibinfo{year}{2005}),
  \bibinfo{pages}{147--152}.
\newblock


\bibitem[Saito et~al\mbox{.}(2022)]%
        {saito2022unbiased}
\bibfield{author}{\bibinfo{person}{Yuta Saito}, \bibinfo{person}{Suguru
  Yaginuma}, \bibinfo{person}{Taketo Naito}, {and} \bibinfo{person}{Kazuhide
  Nakata}.} \bibinfo{year}{2022}\natexlab{}.
\newblock \showarticletitle{Unbiased Recommender Learning from Biased Graded
  Implicit Feedback}.
\newblock \bibinfo{journal}{\emph{WSDM 2022 Workshop on Decision Making for
  Modern Information Retrieval System}} (\bibinfo{year}{2022}).
\newblock


\bibitem[Schnabel et~al\mbox{.}(2016)]%
        {schnabel2016recommendations}
\bibfield{author}{\bibinfo{person}{Tobias Schnabel}, \bibinfo{person}{Adith
  Swaminathan}, \bibinfo{person}{Ashudeep Singh}, \bibinfo{person}{Navin
  Chandak}, {and} \bibinfo{person}{Thorsten Joachims}.}
  \bibinfo{year}{2016}\natexlab{}.
\newblock \showarticletitle{Recommendations as treatments: Debiasing learning
  and evaluation}. In \bibinfo{booktitle}{\emph{international conference on
  machine learning}}. PMLR, \bibinfo{pages}{1670--1679}.
\newblock


\bibitem[Wagner(1982)]%
        {wagner1982simpson}
\bibfield{author}{\bibinfo{person}{Clifford~H Wagner}.}
  \bibinfo{year}{1982}\natexlab{}.
\newblock \showarticletitle{Simpson's paradox in real life}.
\newblock \bibinfo{journal}{\emph{The American Statistician}}
  \bibinfo{volume}{36}, \bibinfo{number}{1} (\bibinfo{year}{1982}),
  \bibinfo{pages}{46--48}.
\newblock


\bibitem[Wang et~al\mbox{.}(2020)]%
        {wang2020information}
\bibfield{author}{\bibinfo{person}{Zifeng Wang}, \bibinfo{person}{Xi Chen},
  \bibinfo{person}{Rui Wen}, \bibinfo{person}{Shao-Lun Huang},
  \bibinfo{person}{Ercan~E Kuruoglu}, {and} \bibinfo{person}{Yefeng Zheng}.}
  \bibinfo{year}{2020}\natexlab{}.
\newblock \showarticletitle{Information Theoretic Counterfactual Learning from
  Missing-Not-At-Random Feedback}. In \bibinfo{booktitle}{\emph{Neural
  Information Processing Systems (NeurIPS)}}.
\newblock


\bibitem[Wang et~al\mbox{.}(2022)]%
        {wang2022invariant}
\bibfield{author}{\bibinfo{person}{Zimu Wang}, \bibinfo{person}{Yue He},
  \bibinfo{person}{Jiashuo Liu}, \bibinfo{person}{Wenchao Zou},
  \bibinfo{person}{Philip~S Yu}, {and} \bibinfo{person}{Peng Cui}.}
  \bibinfo{year}{2022}\natexlab{}.
\newblock \showarticletitle{Invariant Preference Learning for General Debiasing
  in Recommendation}. In \bibinfo{booktitle}{\emph{Proceedings of the 28th ACM
  SIGKDD Conference on Knowledge Discovery and Data Mining}}.
  \bibinfo{pages}{1969--1978}.
\newblock


\end{thebibliography}
\clearpage
\nobalance
\section*{Appendix}

\appendix

\begin{figure*}[t!]
 \subfigure[Showcase on Yelp] {
 \label{cross_case_yelp_nfm}
     \includegraphics[width=0.235\linewidth]{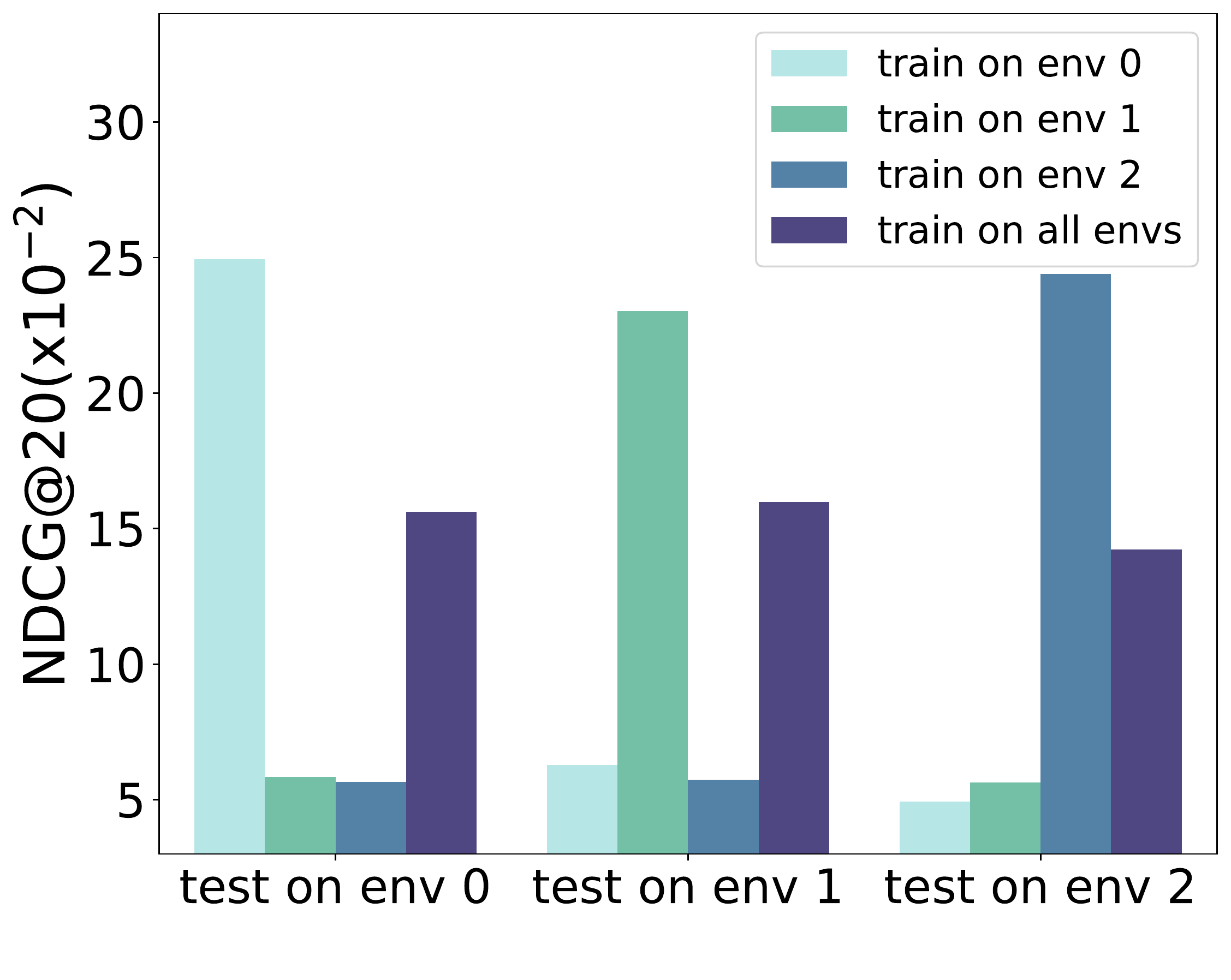}
 }
  \subfigure[Average on Yelp] {
 \label{cross_mean_yelp_nfm}
     \includegraphics[width=0.235\linewidth]{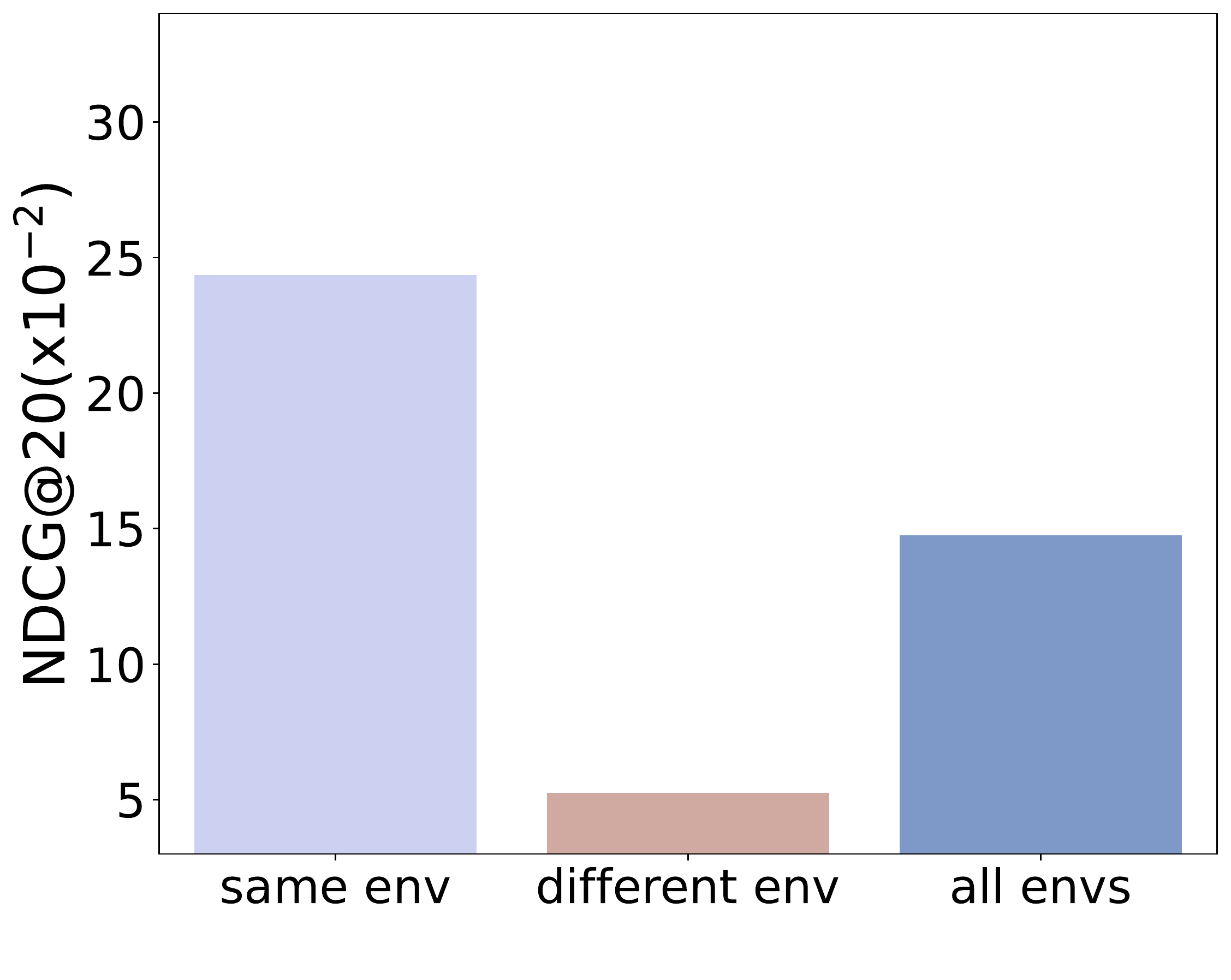}
 }
 \subfigure[Showcase on MovieLens-1M] {
 \label{cross_case_ml_nfm}
     \includegraphics[width=0.235\linewidth]{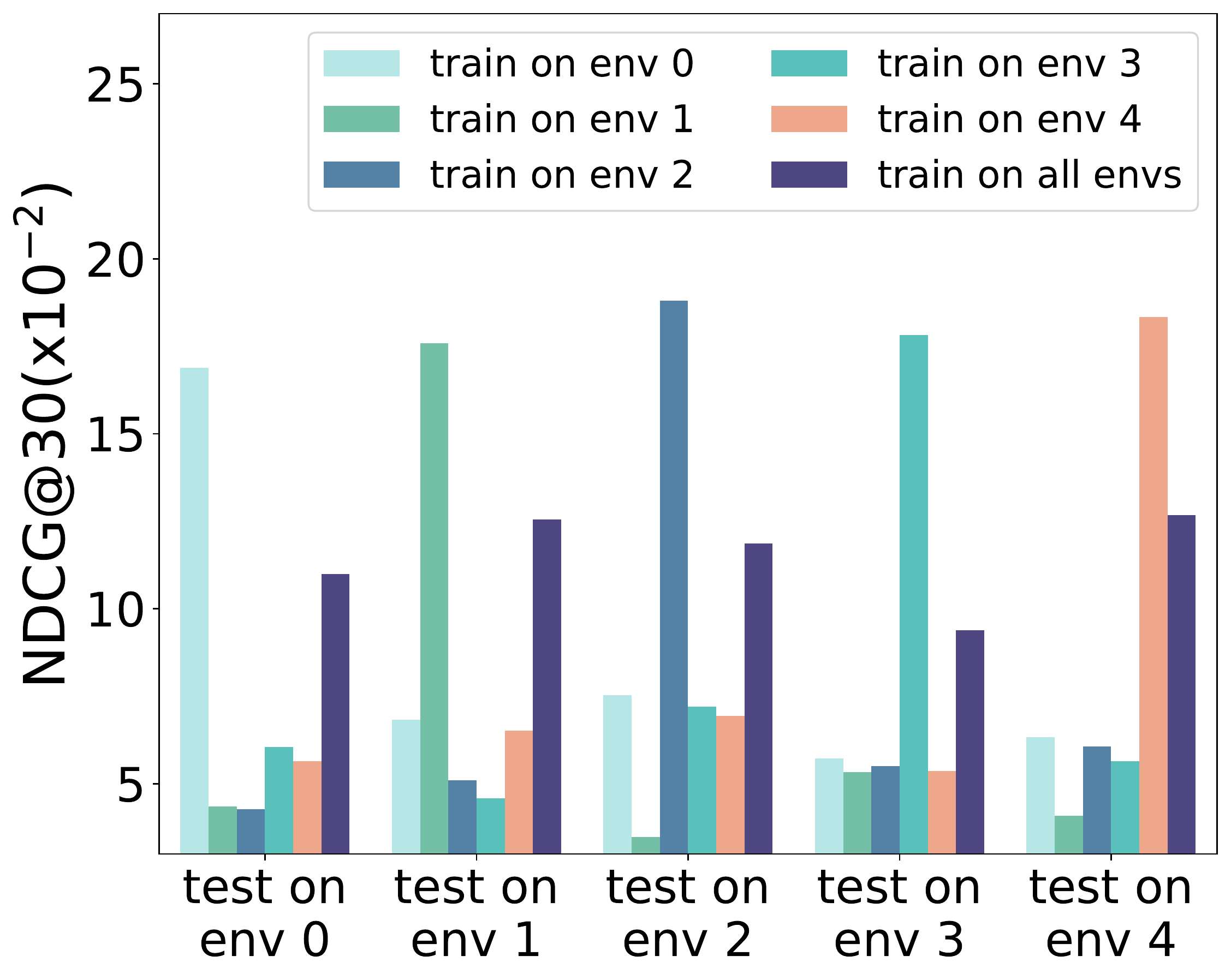}
 }
 \subfigure[Average on MovieLens-1M] {
 \label{cross_mean_ml_nfm}
     \includegraphics[width=0.235\linewidth]{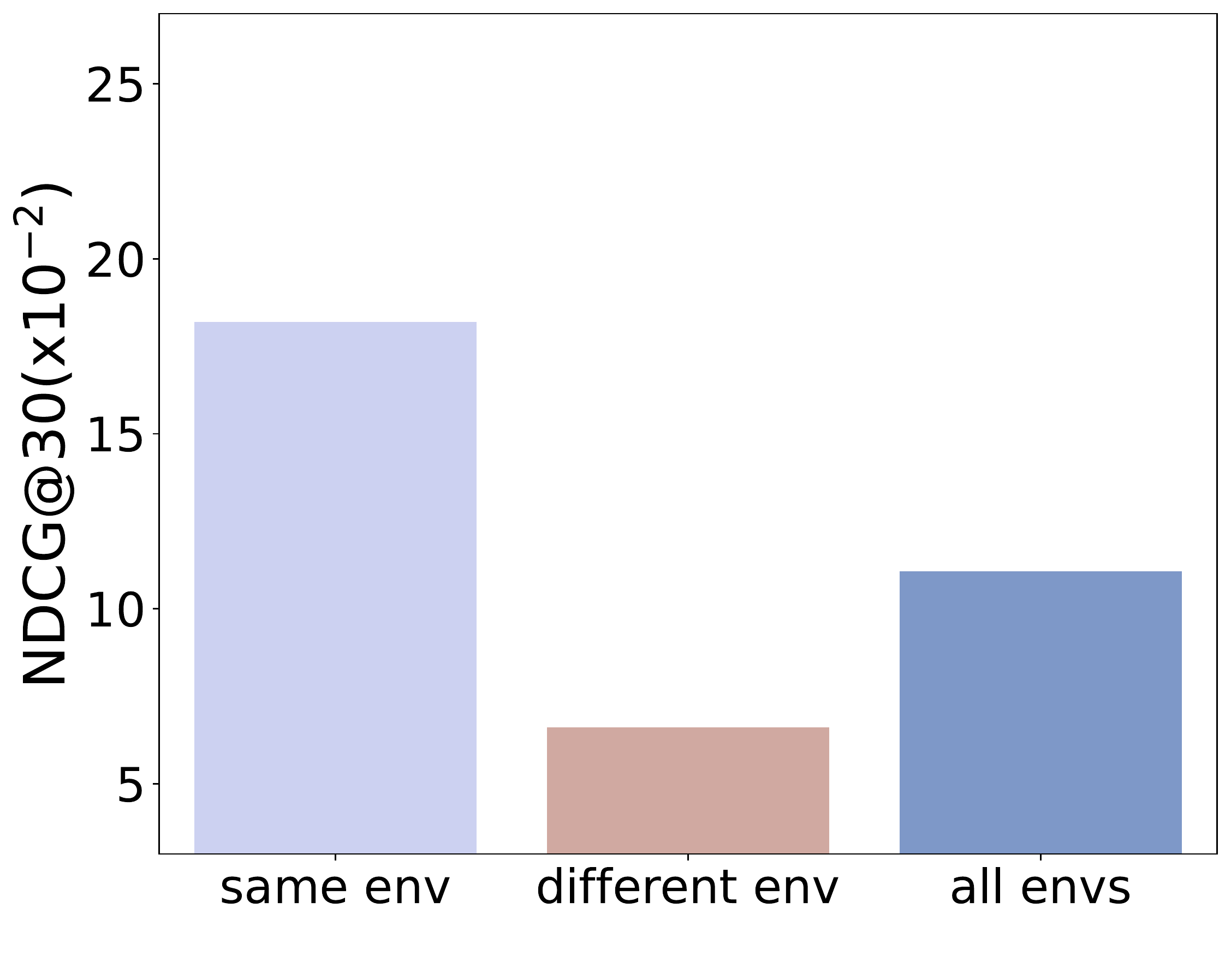}
 }
 \vskip -0.1in
 \caption{The heterogeneity of recommendation data is consistent with cognition and significantly impacts the recommenders. When the model is trained and evaluated in different environments, the performance will drop sharply. The backbone is NFM.}
 \label{cross_fig_nfm}
\end{figure*}

\begin{figure*}[t]
 \subfigure[Sub-population 0] {
     \includegraphics[width=0.17\linewidth]{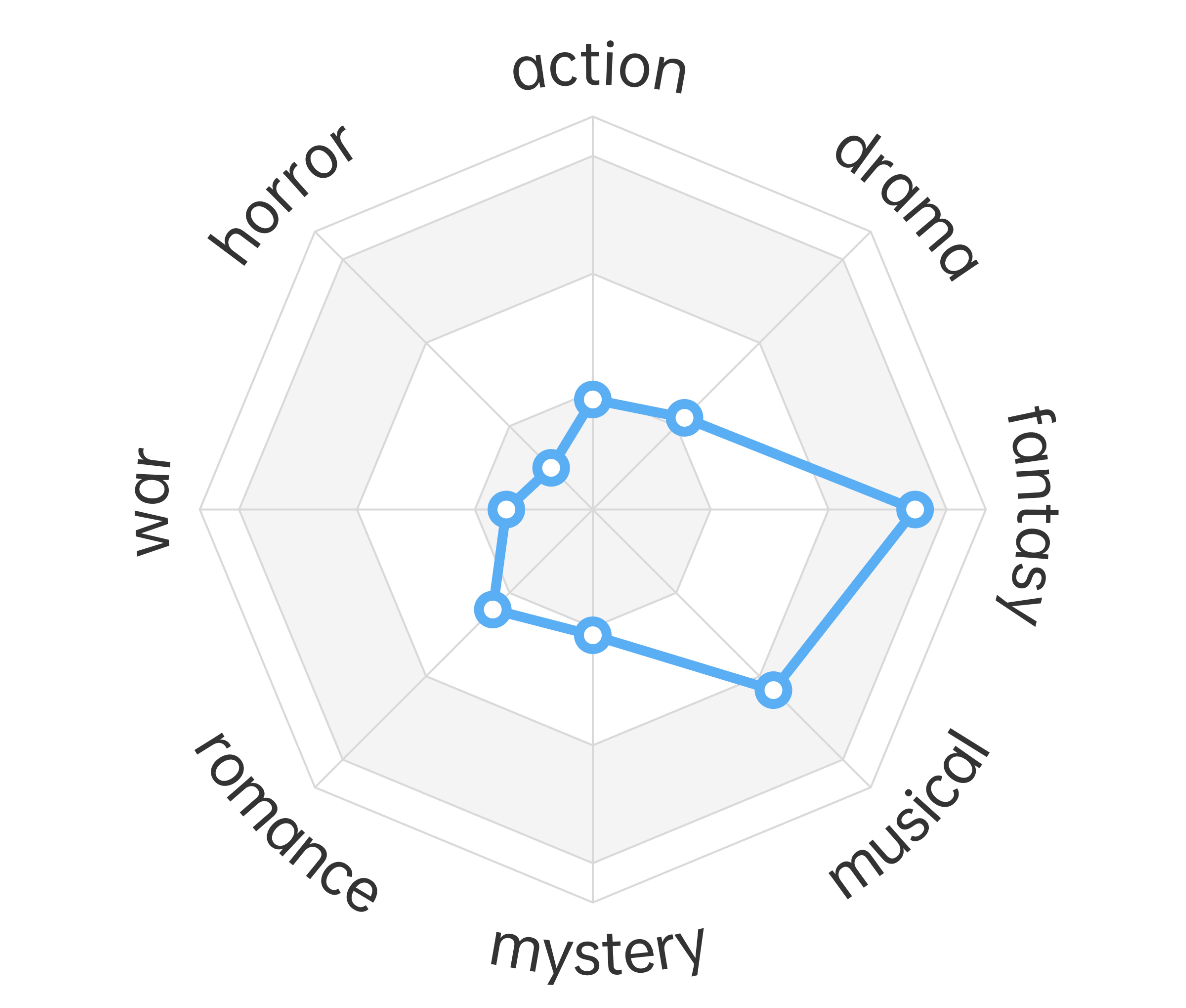}
 }
 \subfigure[Sub-population 1] {
     \includegraphics[width=0.17\linewidth]{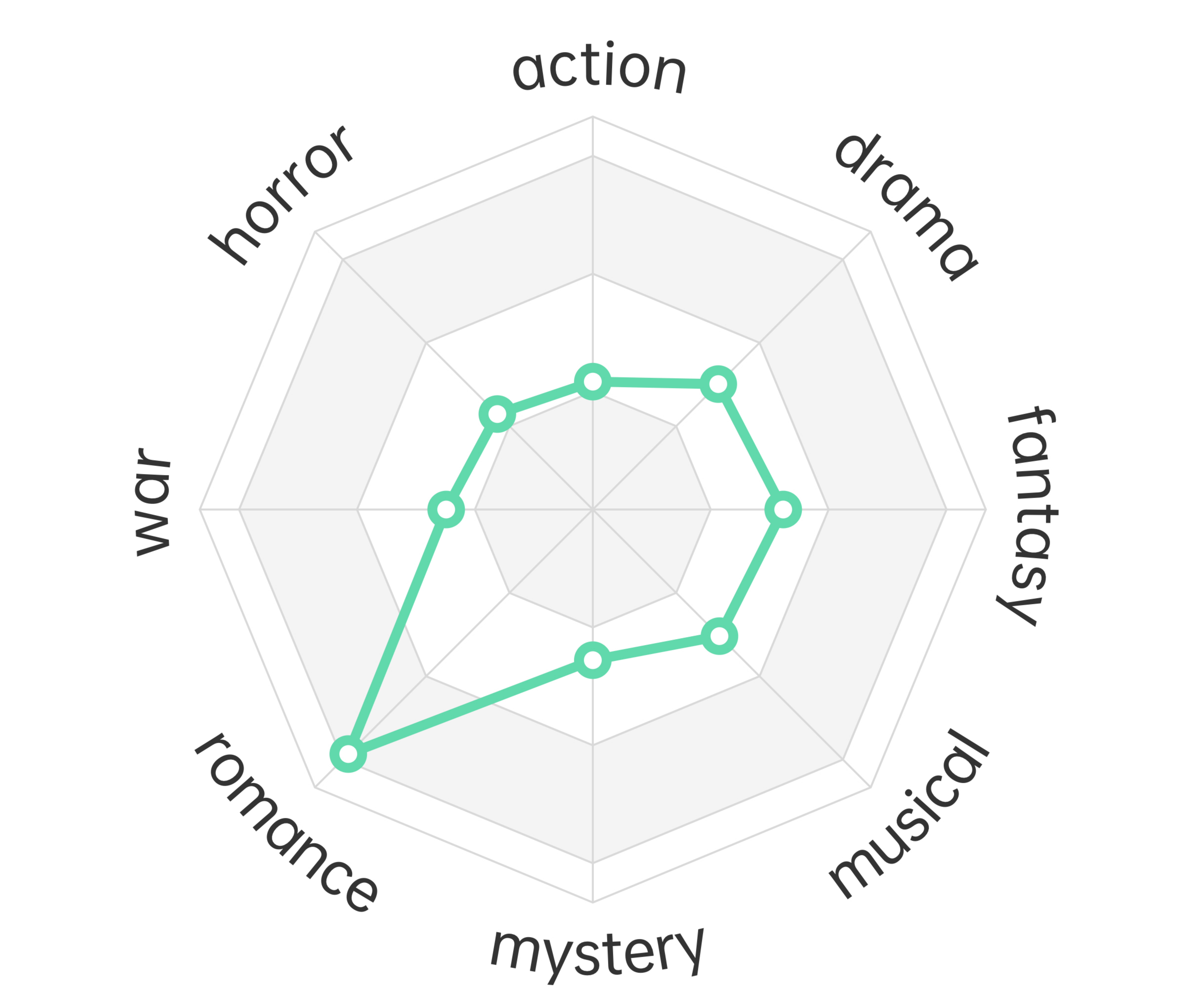}
 }
  \subfigure[Sub-population 2] {
     \includegraphics[width=0.17\linewidth]{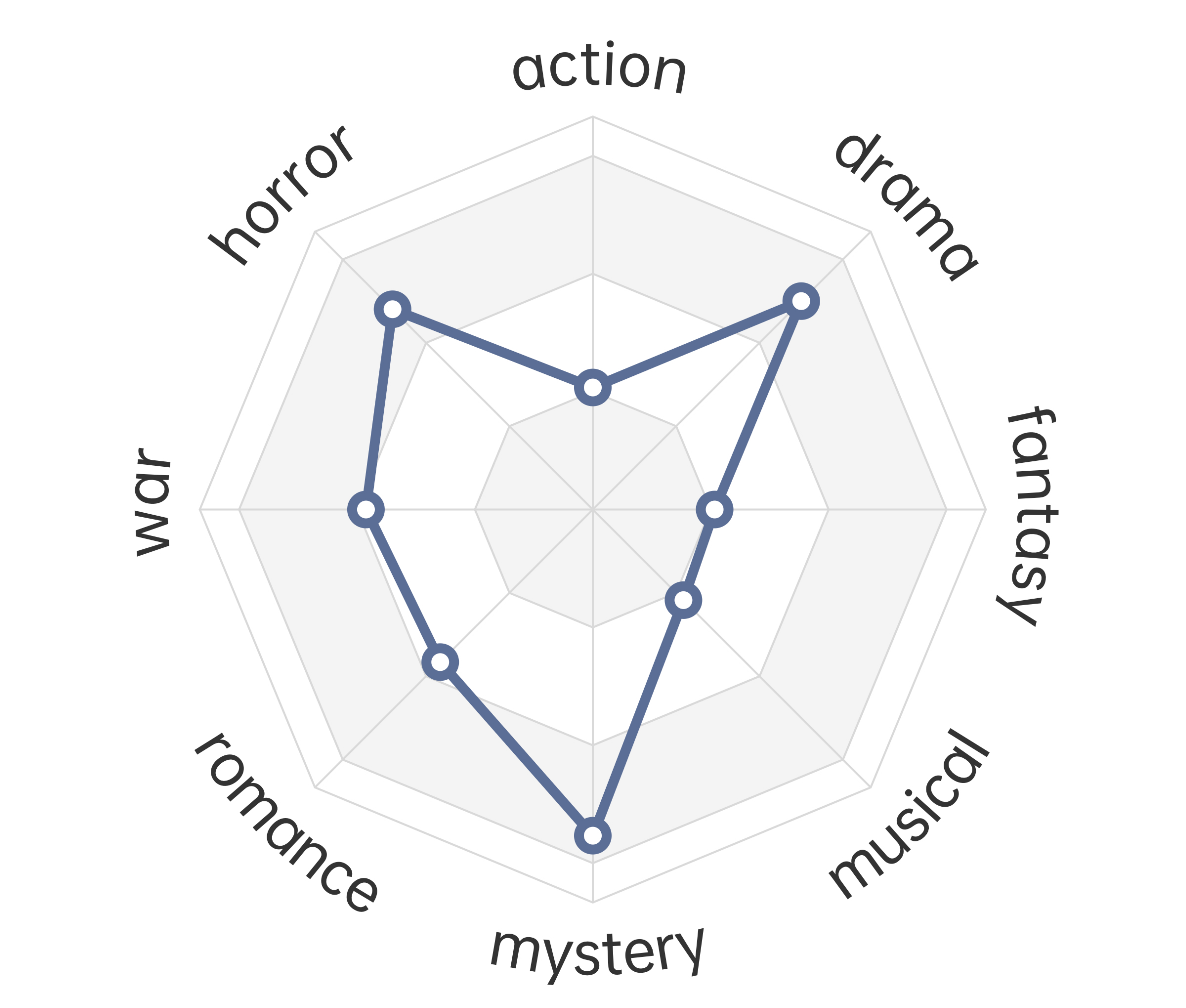}
 }
  \subfigure[Sub-population 3] {
     \includegraphics[width=0.17\linewidth]{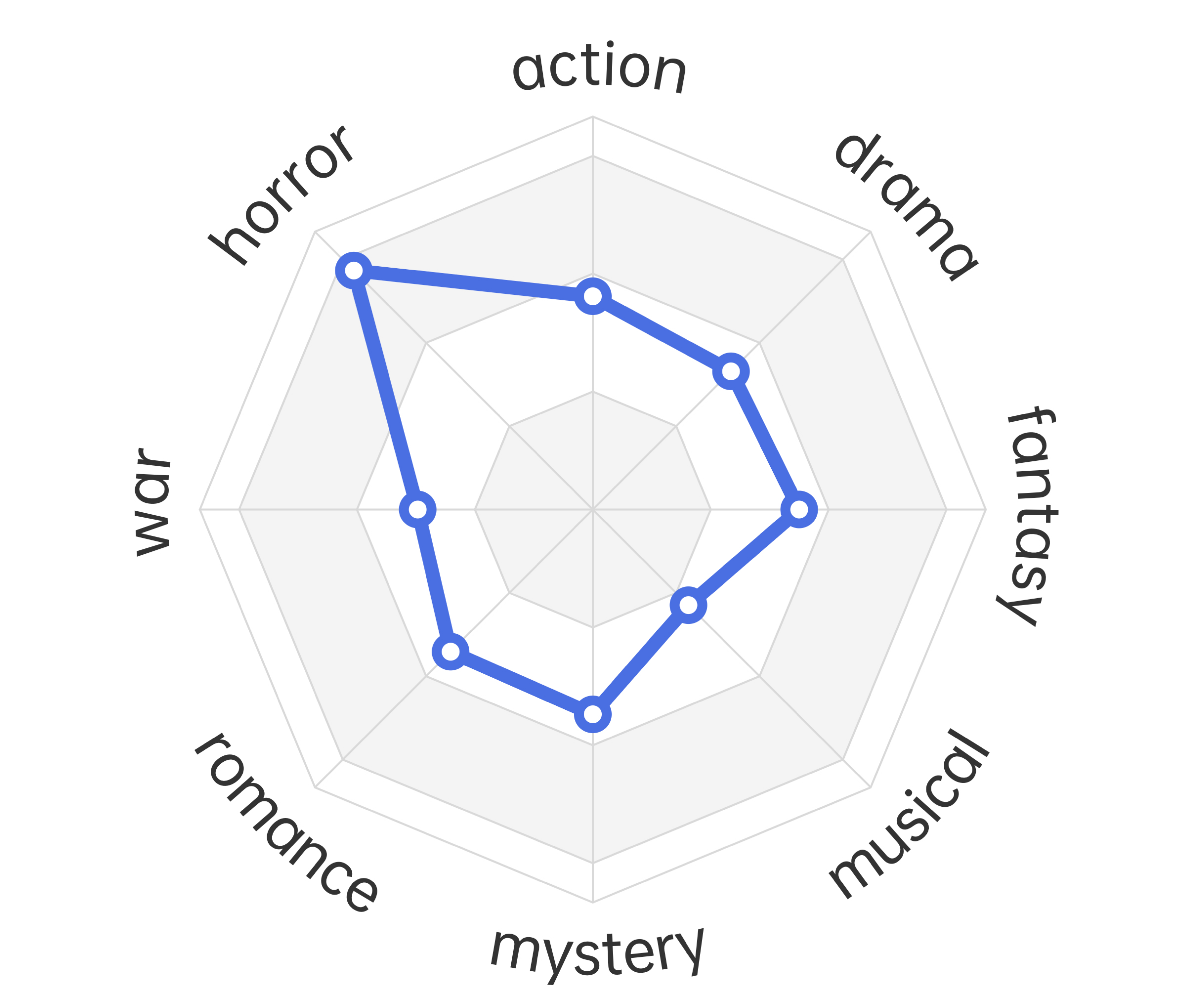}
 }
  \subfigure[Sub-population 4] {
     \includegraphics[width=0.17\linewidth]{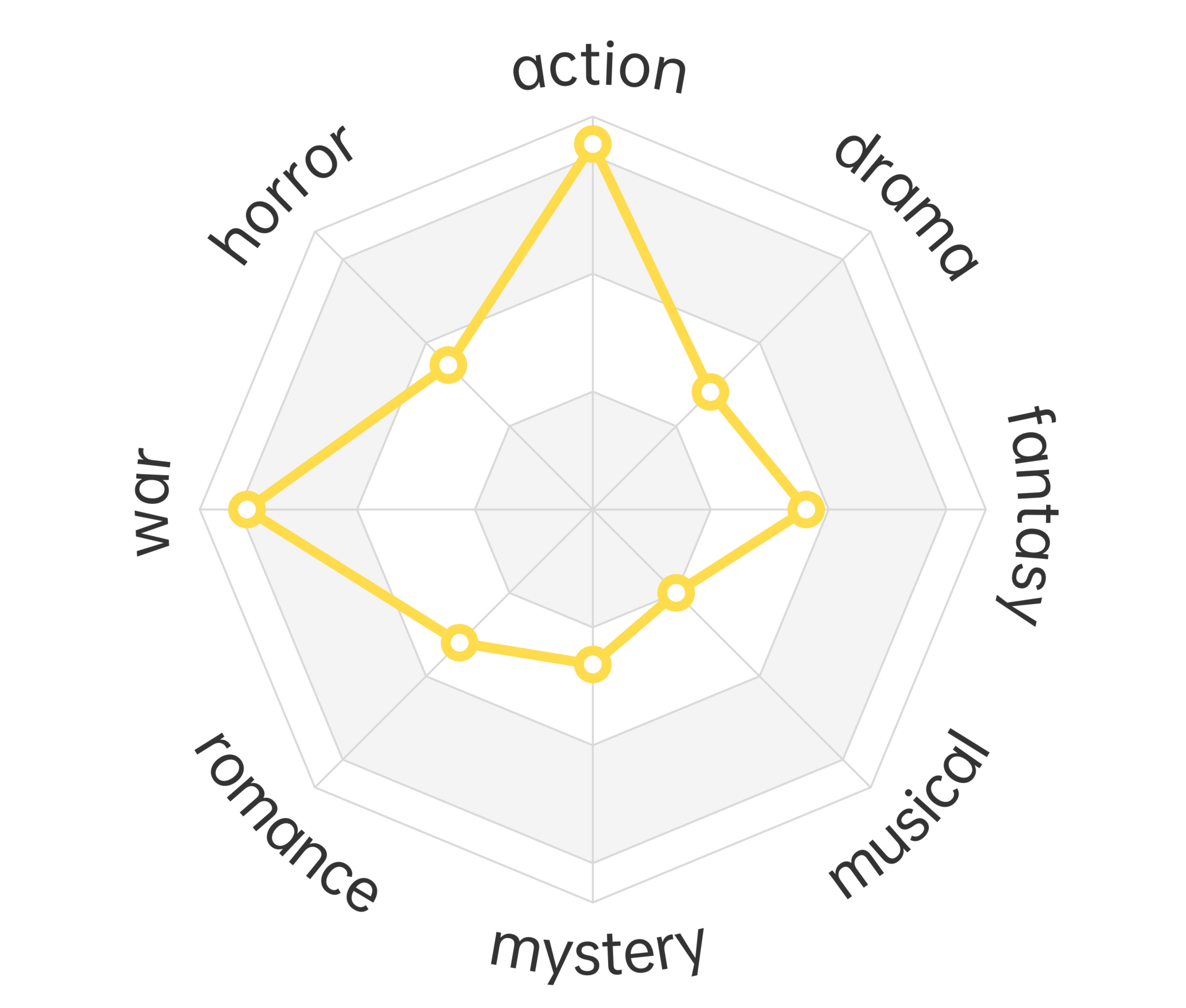}
 }
 
 \vskip -0.15in
 \caption{Five sub-populations explored by \OURS. Each dimension of radar chart is normalized.}
 \label{ml_explainable}
 
 \vskip -0.1in
\end{figure*}

\section{Metric Details}
\label{app_metrics}

Recommenders are evaluated by following.
\quad \\
We use the following metrics to measure the performance of the model under implicit feedback.

$\bm{NDCG@K}$ measures the quality of recommendation through discounted weight based on position.
\begin{equation} \begin{aligned} DCG_u@K &= \sum_{(u,v)\in D_{test}} \frac{I(\hat{j}_{u,v} \leq K)}{\log(\hat{j}_{u,v} + 1)} \\
NDCG@K &= \frac{1}{|\mathcal{U}|} \sum_{u\in \mathcal{U}} \frac{DCG_u@K}{IDCG_u@K},
\nonumber
\end{aligned}\end{equation}
where $IDCG_u@K$ is the ideal $DCG_u@K$, $D_{test}$ is the ground truth of test data, and $\hat{j}_{u,v}$ is the position of item $v$ in the recommended rank for user $u$. 

$\bm{Recall@K}$ means the proportion of relevant items found in the top-k recommendations.
\begin{equation} \begin{aligned} 
Recall_u@K &=  \frac{\sum_{(u,v)\in D_{test}} I(\hat{j}_{u,v} \leq K)}{|D_{test}^u|} \\
Recall@K &= \frac{1}{|\mathcal{U}|} \sum_{u\in \mathcal{U}} Recall_u@K,
\nonumber
\end{aligned}\end{equation}
where $D_{test}^u$ is the set of all positive interactions of  $u$ in test data $D_{test}$.

$\bm{compactness(CP)}$ measures how compact the clusters are.

\begin{equation} \begin{aligned} 
CP_{i} &=  \frac{\sum_{x \in D_{c}} \parallel x - \bar{x} \parallel }{|D_{c}|} \\
CP &=  \frac{1}{|\mathcal{C}|} \sum_{c \in \mathcal{C}} CP_{i},
\nonumber
\end{aligned}\end{equation}
where $\mathcal{C}$ is the space of all clusters, $x$ is the covariates of a sample.

\section{Details of Baselines}
\label{app_baselines}

In Section \ref{imapct_and_explainable}, we train backbones in each environment by Eq(\ref{loss_sub_model}). The model trained in all envs means the weights of every samples $w^{tr}=1$.

In Section 
\ref{exp_better_generalization_and_sub_popu}, we experimentally compared \OURS~ with a number of methods for exploring heterogeneity. We train the proposed composed recommender $\recomendersub_{cm}$ based on the heterogeneity they explored as follows:
\begin{itemize}[leftmargin=0.3cm]
    \item Explore heterogeneity according to expert knowledge(user-feature and item-feature). Since these methods do not output the distances between the samples and the centers of the sub-populations, the sample weight cannot be calculated by Eq(\ref{sample_weight_e}). Therefore, for the environment $\Elower$, we uniformly assign higher weights($w^{tr} > 1$) to the samples belonging to it, and assign lower weights($0< w^{tr} < 1$) to the rest of the samples. In the test phase, as there are raw features of test samples as input, we can categorized the test samples to each environment $\Eupper$. So, it is not necessary to predict by Eq(\ref{final_pred}). 
    \item Explore heterogeneity by traditional clustering methods(raw feature and embedding).  These clustering methods output the distances between samples to centers of environments. Therefore, we can replace $d_{u,v,\Elower}$ in Eq(\ref{sample_weight_e}) with the distance to center of each environment. The rest of the details are consistent with \OURS.
\end{itemize}

In Section \ref{exp_better_debias}, we compared the methods only use $\Eupper$ or $\Rupper$. \OURS~ improves the IPS-based methods by estimating $P_{\Elower, \Rlower}(Y=y|O=1)$ for each sub-population(Eq(\ref{our_estimate_propensity_method})). The two methods are the same, except that $(\Elower, \Rlower)$ is replaced by $\Elower$ and $\Rlower$ respectively.

\section{Supplementary Experimental Results}
\label{app_exp_results}

\subsection{Huge Impact of Heterogeneity on Recommenders}
\label{app_cross_nfm}

In Section~\ref{imapct_and_explainable}, we verify the impact of heterogeneity on recommenders using FM as backbone. In this section, we supplement the results of using NFM as backbone on both Yelp and MovieLens-1M. The results are shown in Figure\ref{cross_fig_nfm}.

\begin{figure*}[th!]
 \subfigure[NFM-BHE on MovieLens-1M(item popularity)] {
 \label{minor_group_ml_item_nfm}
     \includegraphics[width=0.23\linewidth]{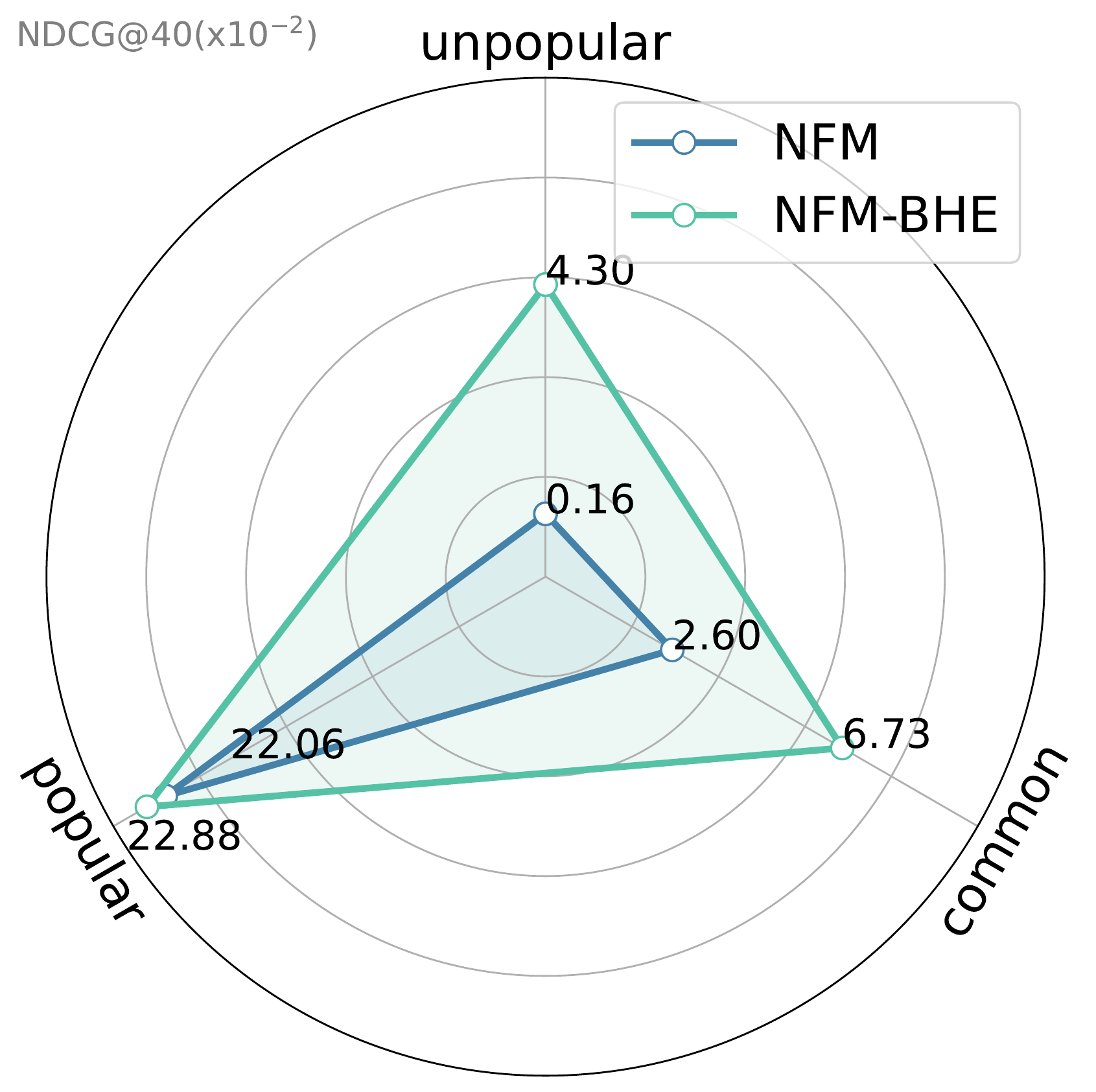}
 }
 \subfigure[NFM-BHE on MovieLens-1M(user age)] {
 \label{minor_group_ml_user_nfm}
     \includegraphics[width=0.23\linewidth]{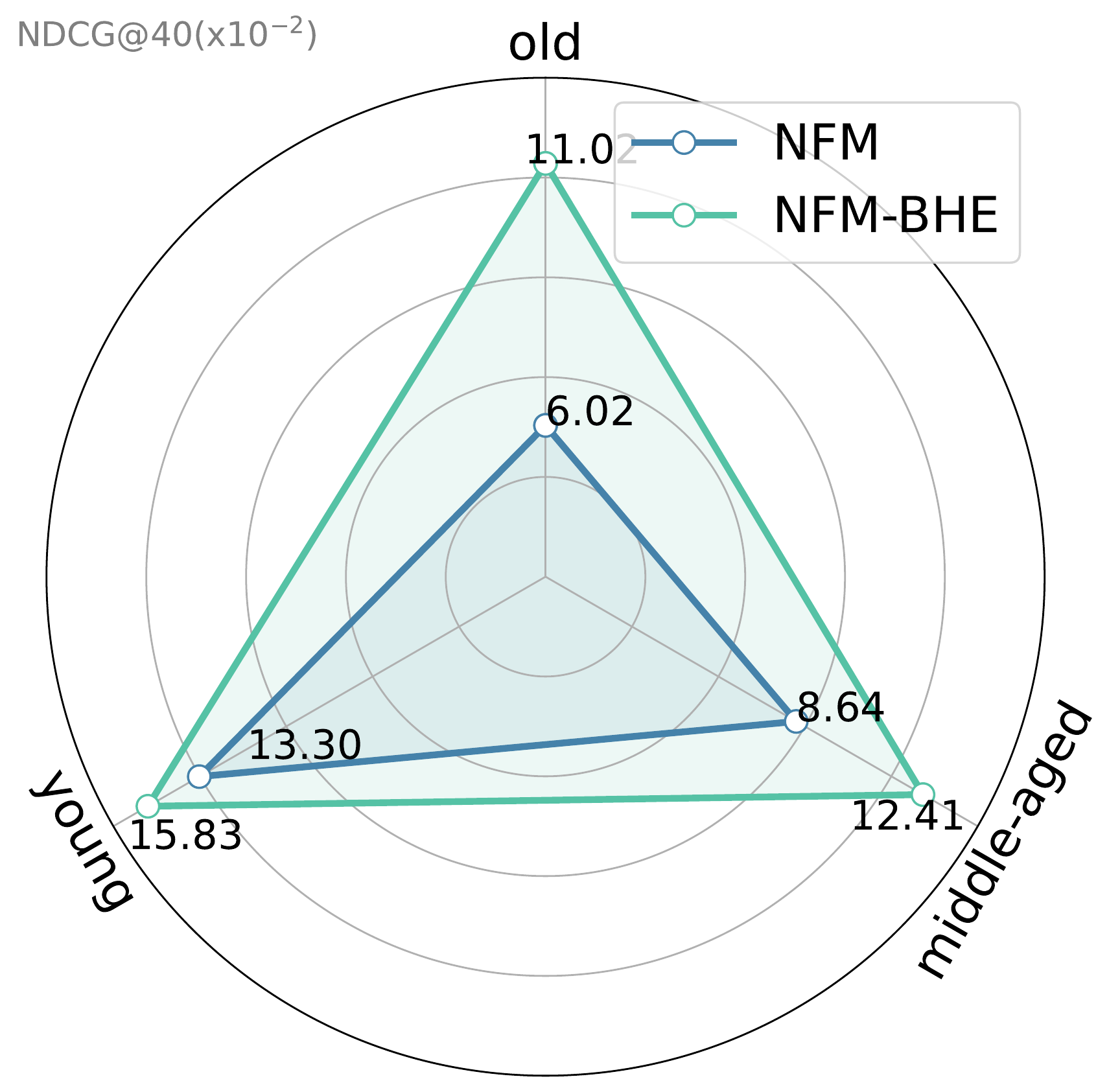}
 }
  \subfigure[NFM-BHE on Yelp(item popularity)] {
 \label{minor_group_yelp_item_nfm}
     \includegraphics[width=0.23\linewidth]{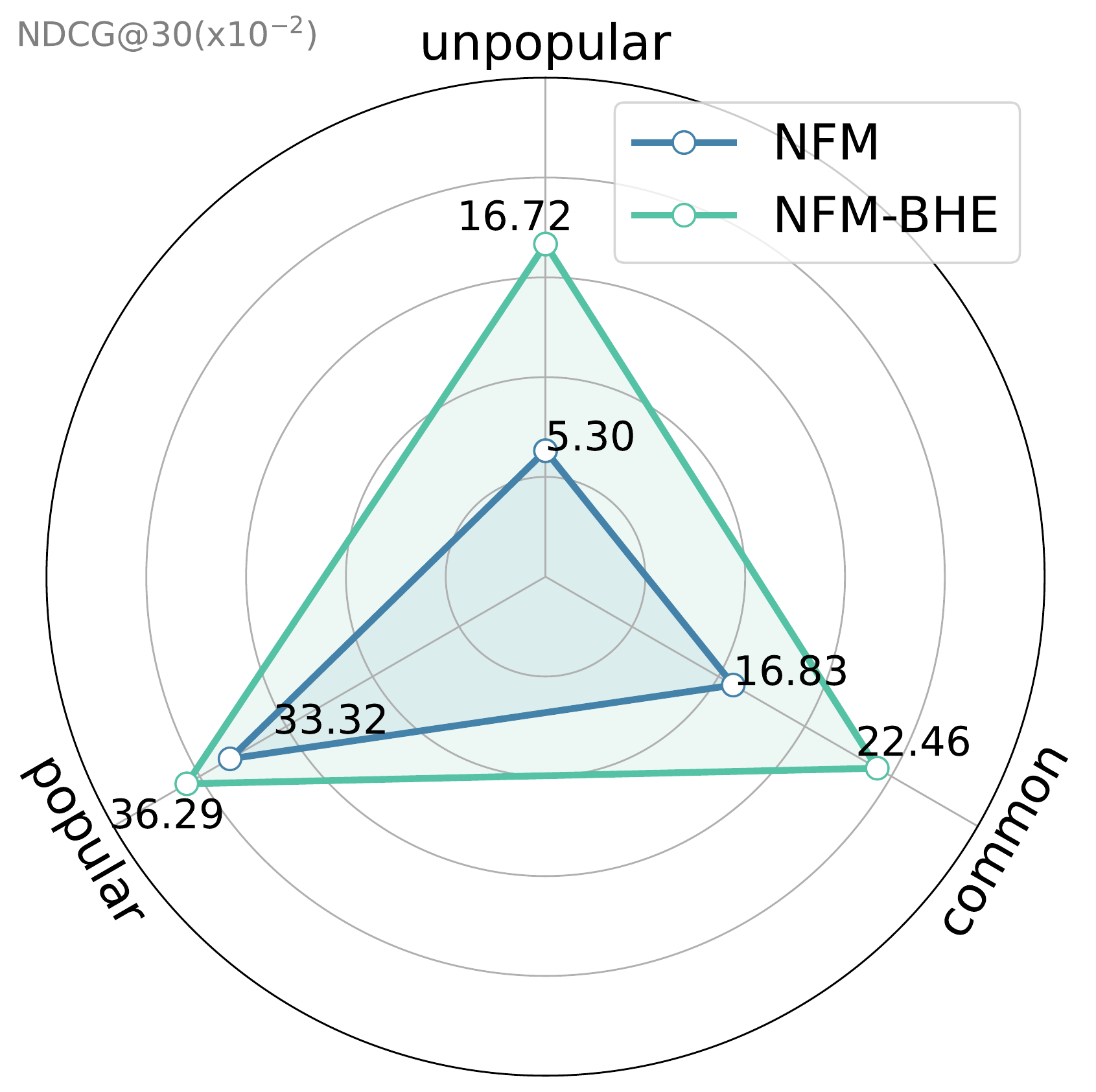}
 }
  \subfigure[NFM-BHE on Yelp(user fans number)] {
 \label{minor_group_yelp_user_nfm}
     \includegraphics[width=0.23\linewidth]{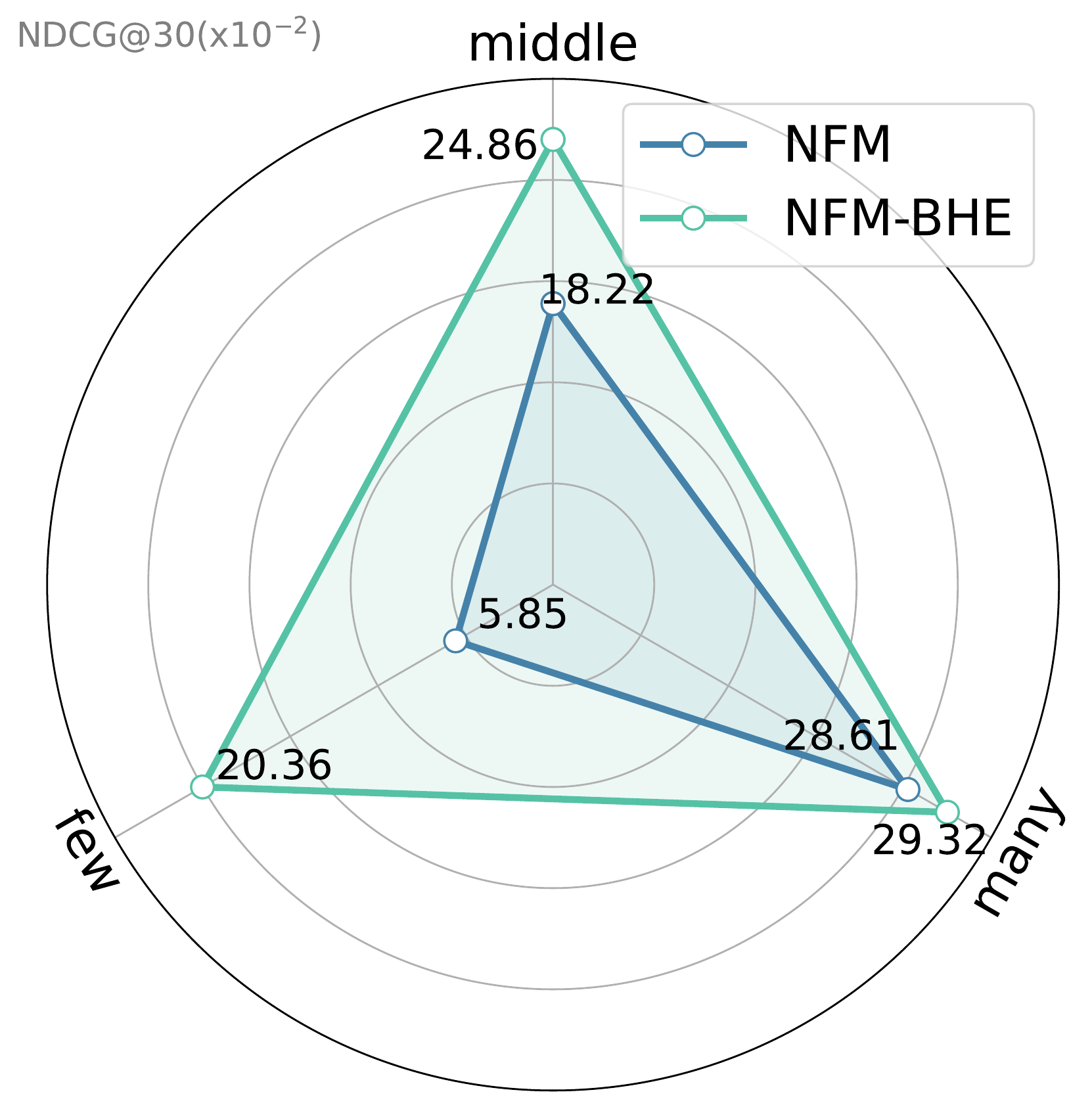}
 }
 
 \vskip -0.1in
 \caption{Exploiting heterogeneity explored by BHE makes the recommender better serve each sub-population. }
 \label{app_minor_group}
 
 \vskip -0.1in
\end{figure*}

\subsection{Explainable Sub-populations Explored by \OURS}
\label{app_explainable}
In this section, we show the explainable sub-populations explored by \OURS~ on MovieLens-1M mentioned in Section~\ref{imapct_and_explainable}. The results are shown in Figure\ref{ml_explainable}. Each sub-figure represents the preferences of a sub-population of data samples. And each dimension of the radar chart represents the degree of preference. For demonstration, we normalize across all dimensions. From the results:
\begin{itemize}[leftmargin=0.3cm]
    \item \textbf{Sub-population 0} prefers fantasy movies, and fantasy movies are often accompanied by beautiful music in a fantasy style. Therefore, they are quite fond of fantasy movies with musical theater styles such as "Frozen" and "Alice in Wonderland". On the other hand, They are interested in action movies and other movies that are too thrilling and exciting and think drama movies are boring. 
    \item \textbf{Sub-population 1} prefers romance movies. Since romance films are often accompanied by romantic music or take place in a fantasy world, they also favor musical films and fantasy films to a certain extent, such as Twilight. In addition, they have little interest in exciting movies such as action movies and horror movies.
    \item \textbf{Sub-population 2} prefers drama and mystery films, and they like the reasoning content in these films. At the same time, since mystery films often contain a certain degree of horror elements, they also show a certain preference for horror films, such as the horror cruise ship. Other movies lack the reasoning element that users of this genre love, so they don't like it.
    \item \textbf{Sub-population 3} prefers horror movies and enjoys the thrill of being frightened. Unlike sub-population 2, what they like is the horror element itself, not the reasoning element, so they are not interested in drama and mystery films. Thrilling escape stories and fantasy elements are often accompanied by horror elements (e.g., Tomb Raider and Mummy), so they also show favor for action movies and fantasy movies. 
    \item \textbf{Sub-population 4} especially prefers exciting scenes, so they are most fond of action movies and war movies. These movies are sometimes combined with fantasy elements (e.g., Lord of the Rings), so this sub-population also prefers fantasy movies. Additionally, they tend to find musicals and dramas too boring.
\end{itemize}

\begin{figure*}[t!]
 \subfigure[MF-IPS-\OURS~on Coat] {
     \includegraphics[width=0.235\linewidth]{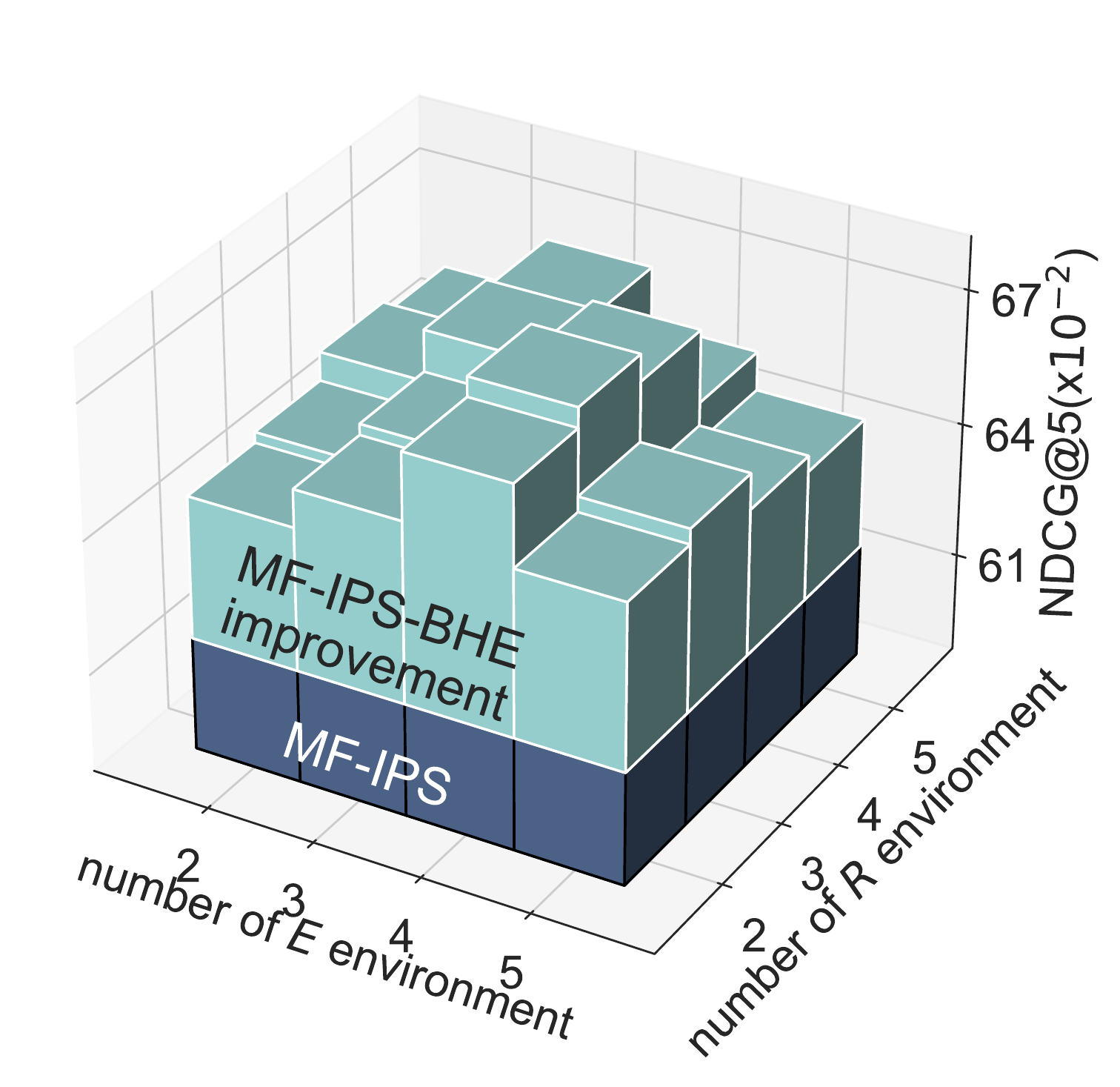}
 }
  \subfigure[MF-SNIPS-\OURS~on Coat] {
     \includegraphics[width=0.235\linewidth]{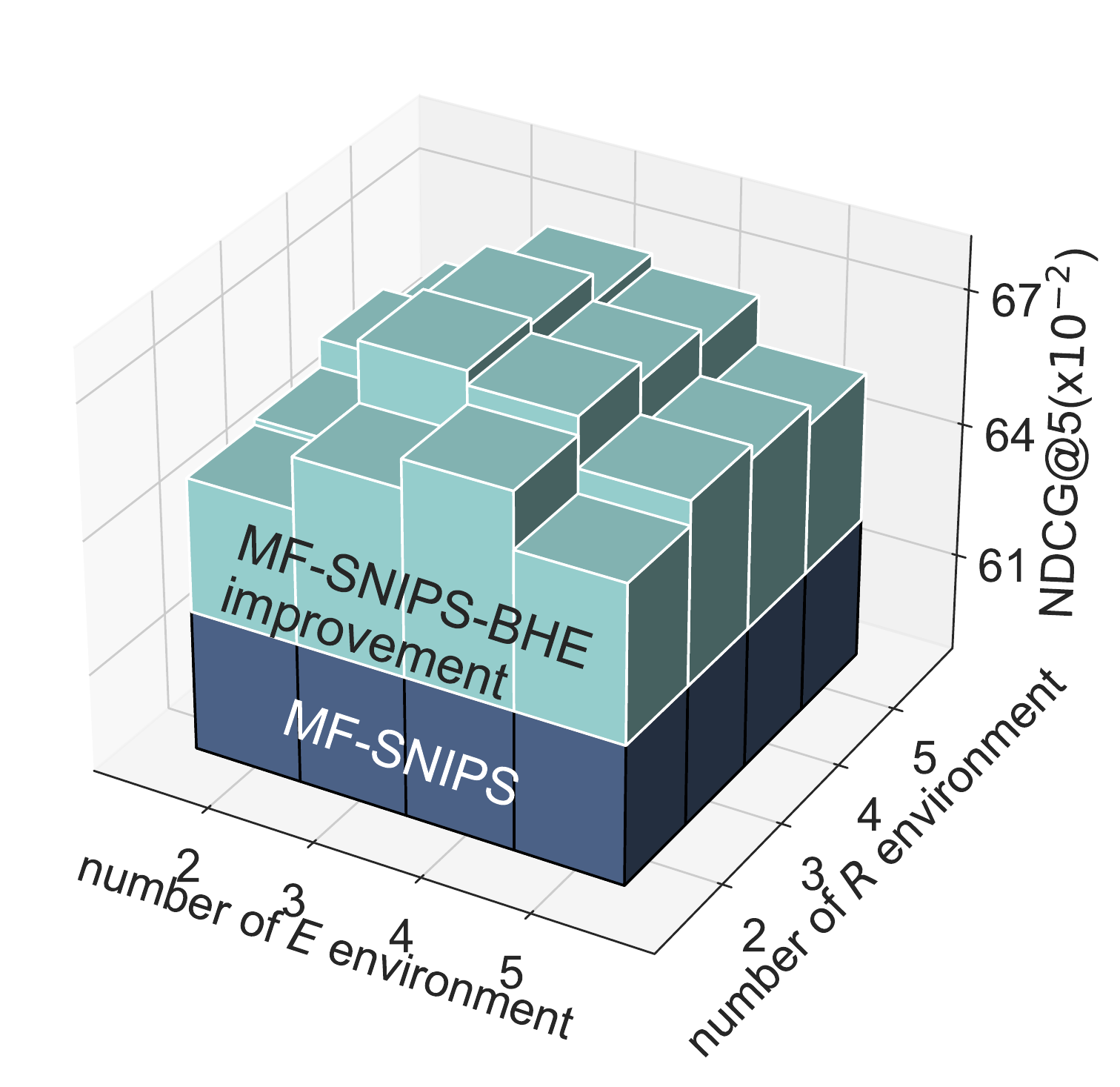}
 }
 \subfigure[NCF-IPS-\OURS~on Coat] {
     \includegraphics[width=0.235\linewidth]{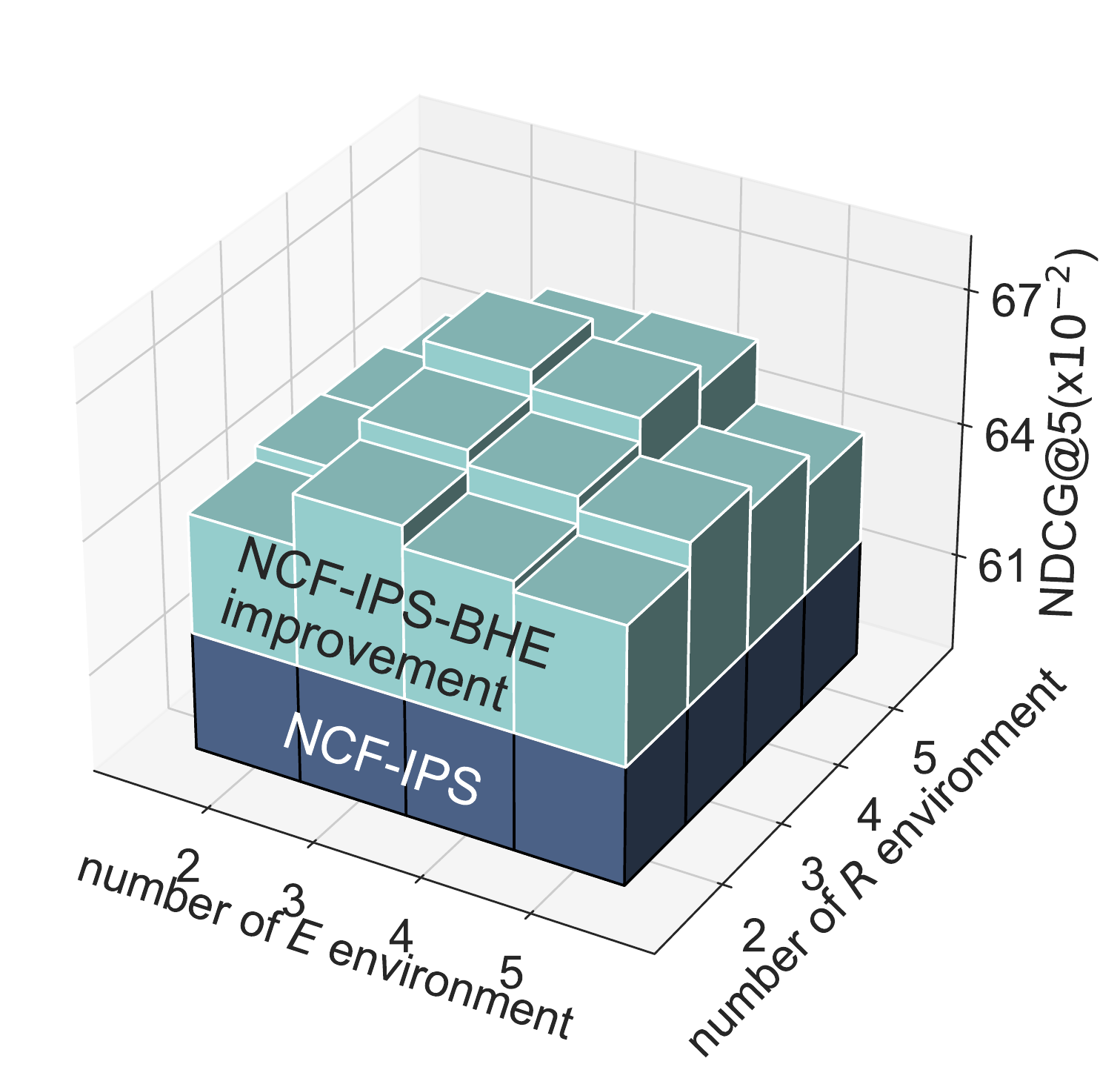}
 }
 \subfigure[NCF-SNIPS-\OURS~on Coat] {
     \includegraphics[width=0.235\linewidth]{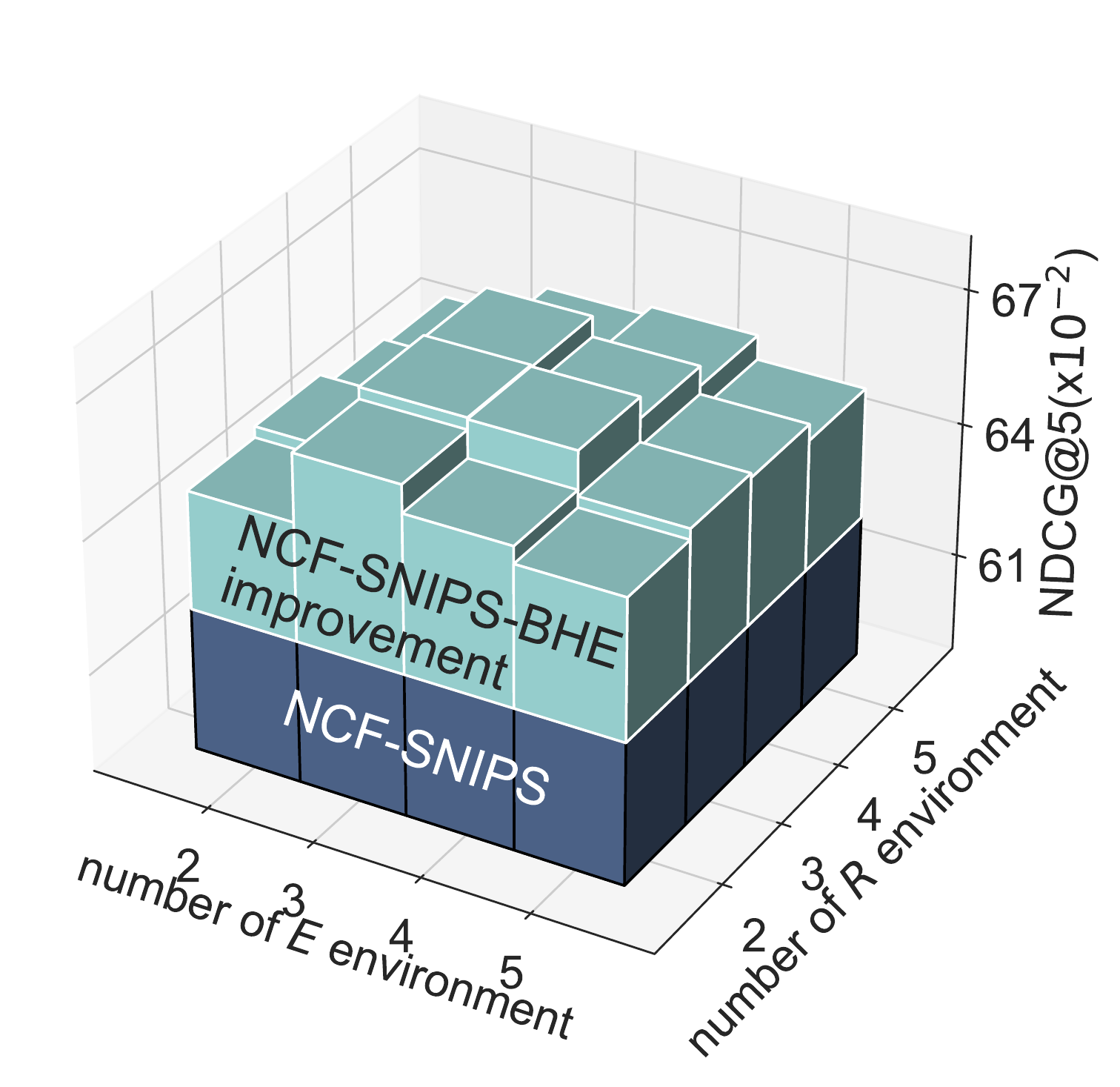}
 }
 \vskip -0.1in
 \caption{The influence of $\Eupper$ and $\Rupper$ environment numbers on better debiasing(settings in Section~\ref{exp_better_debias}). We also show the performance of backbones as a comparison(the blue pillars). Since backbones do not exploit heterogeneity, they are not influenced by the number of environments.}
 \label{env_num_coat}
\end{figure*}

\begin{figure*}[t!]
 \subfigure[FM-\OURS~on Yelp] {
     \includegraphics[width=0.235\linewidth]{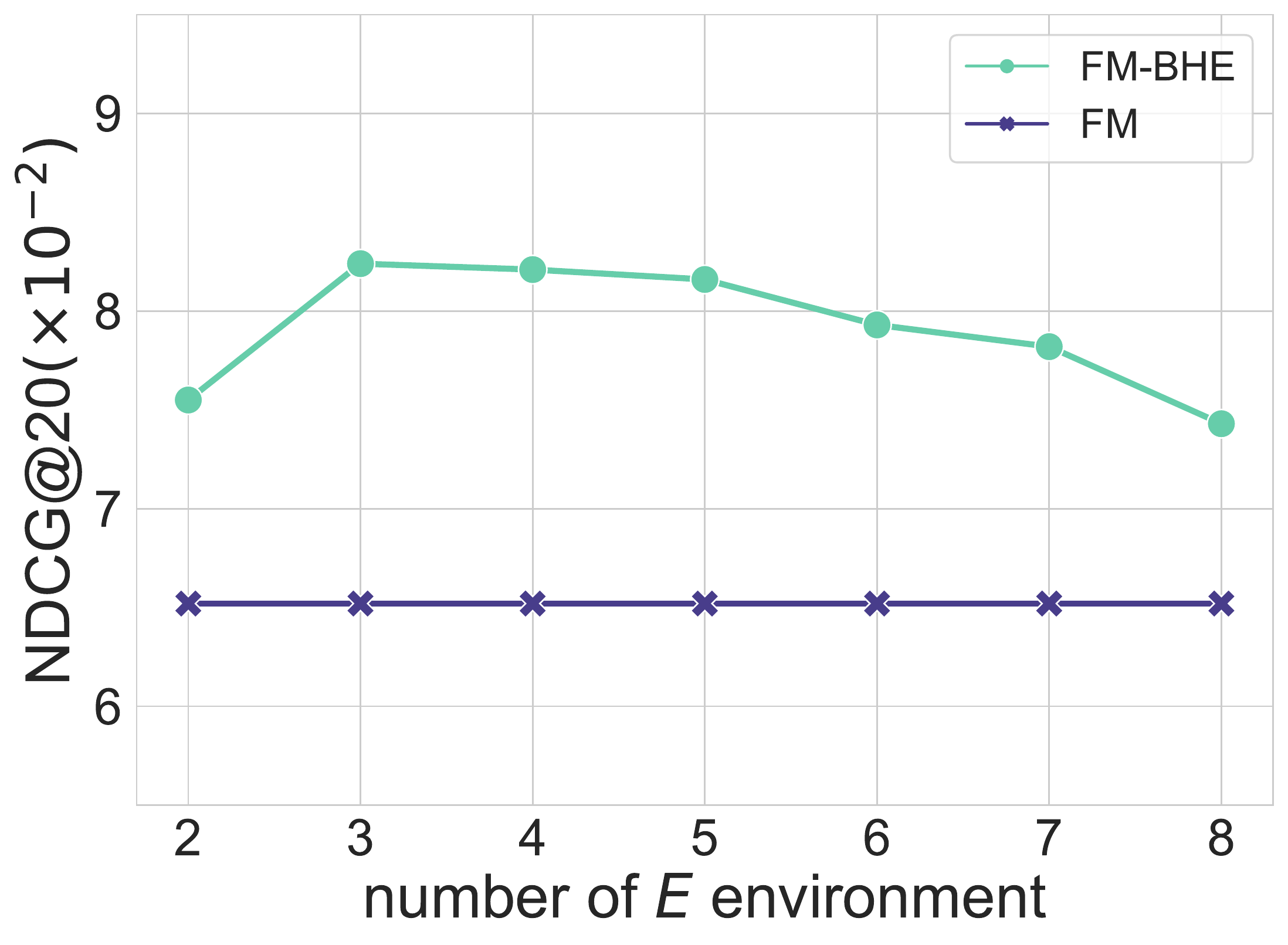}
 }
  \subfigure[NFM-\OURS~on Yelp] {
     \includegraphics[width=0.235\linewidth]{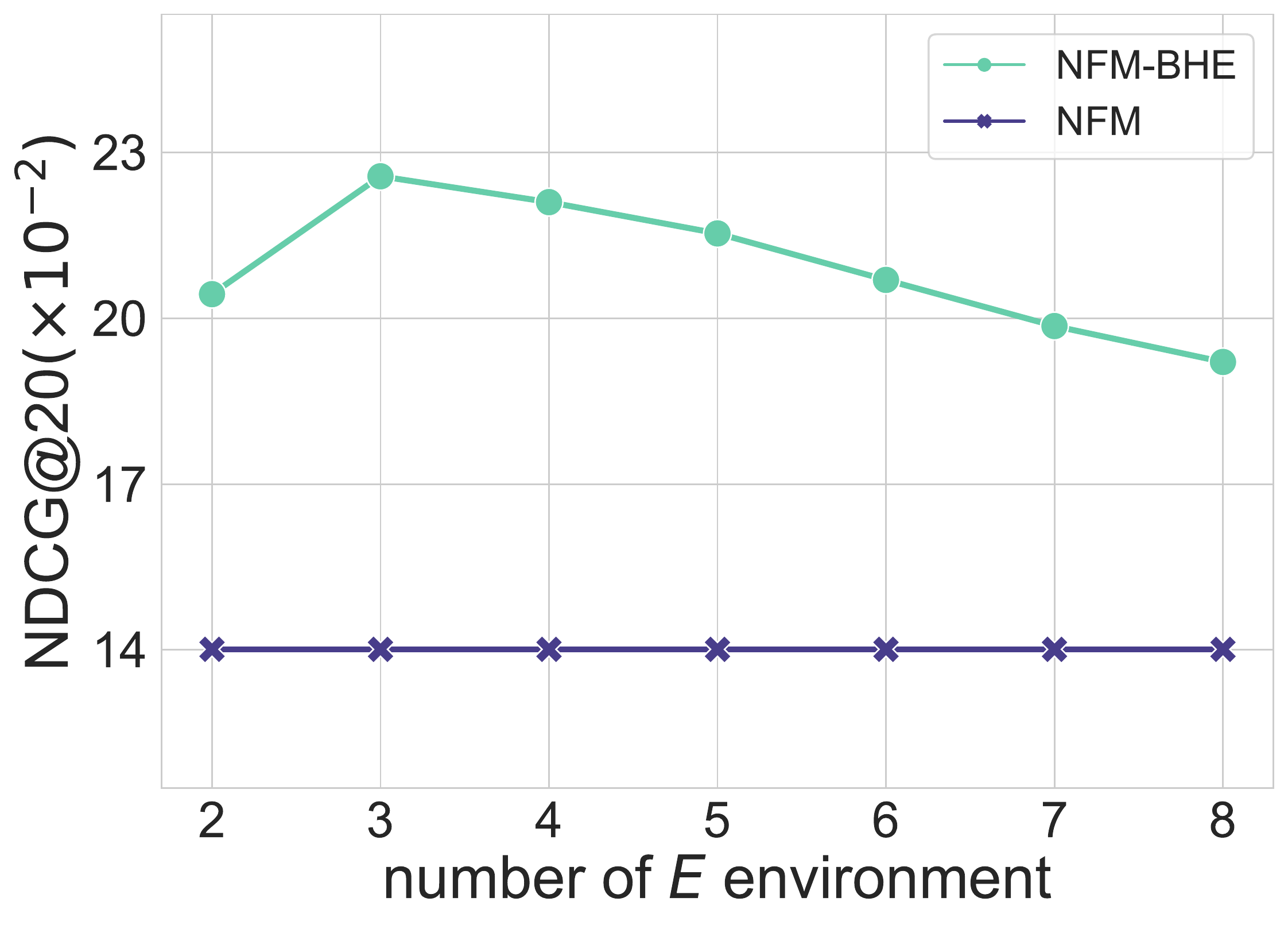}
 }
 \vskip -0.1in
 \caption{The influence of $\Eupper$ environment number on better generalization(settings in Section~\ref{exp_better_generalization_and_sub_popu}). We also show the performance of backbones as a comparison. Since backbones do not exploit heterogeneity, they are not influenced by the number of environments.}
 \label{env_num_yelp}
\end{figure*}


\subsection{Sub-populational Robustness of \OURS}
We demonstrate the results of sub-populational robustness of \OURS using NFM as backbone. The format of results is the same as Figure\ref{minor_group}  in Section~\ref{exp_better_generalization_and_sub_popu}. The results are shown in Figure\ref{app_minor_group}.

\subsection{Environments Numbers Influence \OURS}
\label{app_env_num}
In this section, we supplement the ablation experiments about the number of environments on both Yelp and Coat. The results on Coat are shown in Figure\ref{env_num_coat}, and The results on Yelp are shown in Figure\ref{env_num_yelp}.

\end{document}